\documentclass[runningheads]{llncs}
\usepackage[T1]{fontenc}
\usepackage{amsmath}
\usepackage{amssymb}
\usepackage{graphicx}
\usepackage{floatrow}
\usepackage{floatflt}
\usepackage{float}
\usepackage{amsmath}
\usepackage{amsbsy}
\usepackage{amssymb}
\usepackage{textcomp}
\usepackage{graphicx}
\usepackage{amsfonts}
\usepackage{multirow}
\usepackage{url}
\usepackage{longtable}
\usepackage{lscape}
\usepackage{hhline}
\usepackage[utf8]{inputenc}
\usepackage{colortbl,array} 
\usepackage{multirow,bigdelim}
\usepackage{booktabs}
\usepackage[linesnumbered,ruled]{algorithm2e}
\usepackage{xcolor,colortbl}
\usepackage[mathscr]{eucal}
\usepackage{mathtools, nccmath}
\usepackage{tikz}
\usepackage{pgfplots}
\usepackage{pgfplotstable}
\usepgfplotslibrary{groupplots}
\usetikzlibrary{pgfplots.groupplots}
\pgfplotsset{compat=1.3}
\usepackage{subcaption}
\usepackage{wrapfig}
\definecolor{Gray}{gray}{0.8}
\usepackage{float}
\floatstyle{plaintop}
\restylefloat{table}

\SetKwComment{Comment}{$\triangleright$~~}{}

\begin{document}
\title{Morphological Granulometric Analysis of Particle Imagery 
from Microgravity Experiments}
\titlerunning{Morphological Granulometry for Microgravity Experiments}
%

\author{Shima Shabani\inst{1}\and
Michael Breu\ss\inst{1}\and
Marvin Kahra\inst{1},\\ Jens Teiser\inst{2}, Gretha Swantje Völke\inst{2} \and Nico Wenders\inst{2}}
%
\authorrunning{S. Shabani et al.}
%
\institute{Institute for Mathematics, Brandenburg University of Technology\\
Platz der Deutschen Einheit 1, 03046 Cottbus, Germany,\\ 
\email{\{shima.shabani, breuss, Marvin.Kahra\}@b-tu.de}\\
\and
Faculty of Physics, University of Duisburg-Essen\\ Lotharstr. 1,
47057 Duisburg, Germany\\
\email{\{jens.teiser, gretha.voelke, nico.wenders\}@uni-due.de}}
\maketitle              
\begin{abstract}
The aim of our work is to analyze size distributions
of particles and their agglomerates in imagery from 
astrophysical microgravity experiments.
The data acquired in these experiments are given by
sequences consisting of several hundred images. 
It is desirable to establish an automated routine 
that helps to assess size distributions of important 
image structures and their dynamics in a statistical way.

The main technique we adopt to this end is the morphological granulometry.
After preprocessing steps that facilitate granulometric analysis, 
we show how to extract useful information on 
size of particle agglomerates as well as underlying dynamics.
At hand of the discussion of two different 
microgravity key experiments we demonstrate 
that the granulometric analysis enables to assess
important experimental aspects. 
We conjecture that our developments are a useful basis for the quantitative assessment of microgravity particle
experiments.
.
\end{abstract}
\section{Introduction}

Mathematical morphology is a subfield of image processing in which different filters are constructed to select and change the desired properties of an image, as detailed in \cite{ref_najman2010,ref_roerdink2011,ref_serra2012}. This involves moving a mask, which is defined by a structuring element ({\tt SE}) in terms of its shape, size, and centre, over the image. Corresponding morphological operations are performed on the resulting pixels, two of the most fundamental being dilation and erosion. These cause the structure in a binary image to grow or shrink according to the {\tt SE}.

Image segmentation as a part of mathematical morphology is based on a so-called scale space. Its basic idea is the gradual removal of structures without changing the contours of the objects. However, dilation and erosion are not sufficient in themselves to create such a scale space method, as they enlarge or reduce the contours \cite{ref_jackway1996}. For this reason, a combination of these operations is used to partially avoid this undesirable behaviour.
The application of an erosion to remove elements, followed by a dilation to partially reconstruct these structures, is referred to as opening. This process selectively removes structures smaller than the {\tt SE}, while preserving vertical edges with a high degree of fidelity. However, there is a potential for displacement of horizontal contours during this procedure. By performing openings with increasing {\tt SE}s, the image is progressively sifted. This process of sifting results in the removal of the next scale of structures at each step, thereby creating a scale space. Because of this sifting, the method is known as granulometry \cite{ref_goutsias2000,ref_soille1999}. Granulometry is one of the simplest and first morphological scale space approaches that were studied by Matheron \cite{ref_matheron1975} and Maragos \cite{ref_maragos1989}. Areas of application include medicine \cite{ref_theera2007}, biology \cite{ref_devaux2008} and geosciences \cite{ref_balagurunathan2001}.

This paper is about the application of granulometry to imagery from microgravity experiments. In a detailed discussion, we show how to make use of information obtained from the granulometry in order to obtain insights on the underlying physical processes that can be observed.
In doing this, we show here first steps in the development of discriminative features that help to assess the large amount of experimental data in an automated way.

\section{Microgravity Experiments and Data Sets}

Microgravity experiments are a common tool for understanding early phases of planet formation \cite{ref_wurm2021}. Planet formation occurs in protoplanetary disks, clouds of gas and dust surrounding a young star, with around 1\% of the mass in solids. It starts with dust grains in the micrometer range, which collide and form extremely porous agglomerates in hit-and-stick collisions \cite{ref_drazkowska2023}. With increasing size, the agglomerates decouple from the surrounding gas and the collision velocities increase, leading to the compaction of the particles \cite{ref_blum2008}. From here, it is an open debate on how planetary growth can proceed. With only surface forces acting, dust agglomerates and solid particles in the (sub) millimeter range bounce off each other rather than forming larger agglomerates, leading to the term bouncing barrier in the literature \cite{ref_zsom2010}. 

Once particles have a critical size, hydrodynamic processes in protoplanetary disks are an efficient mechanism to concentrate solid particles. Such processes are referred to as the streaming instability, the baroclinic instability, or vortices in general. All these mechanisms have in common that they require a critical minimum size of the dust agglomerates / solid particles, depending on the model in the decimeter to meter range \cite{ref_carrera2022,ref_squire2020}.

\begin{figure}[t]
    \begin{subfigure}[t]{.49\textwidth}
        \centering
        \includegraphics[width = 0.3\linewidth, height = 0.34\linewidth]{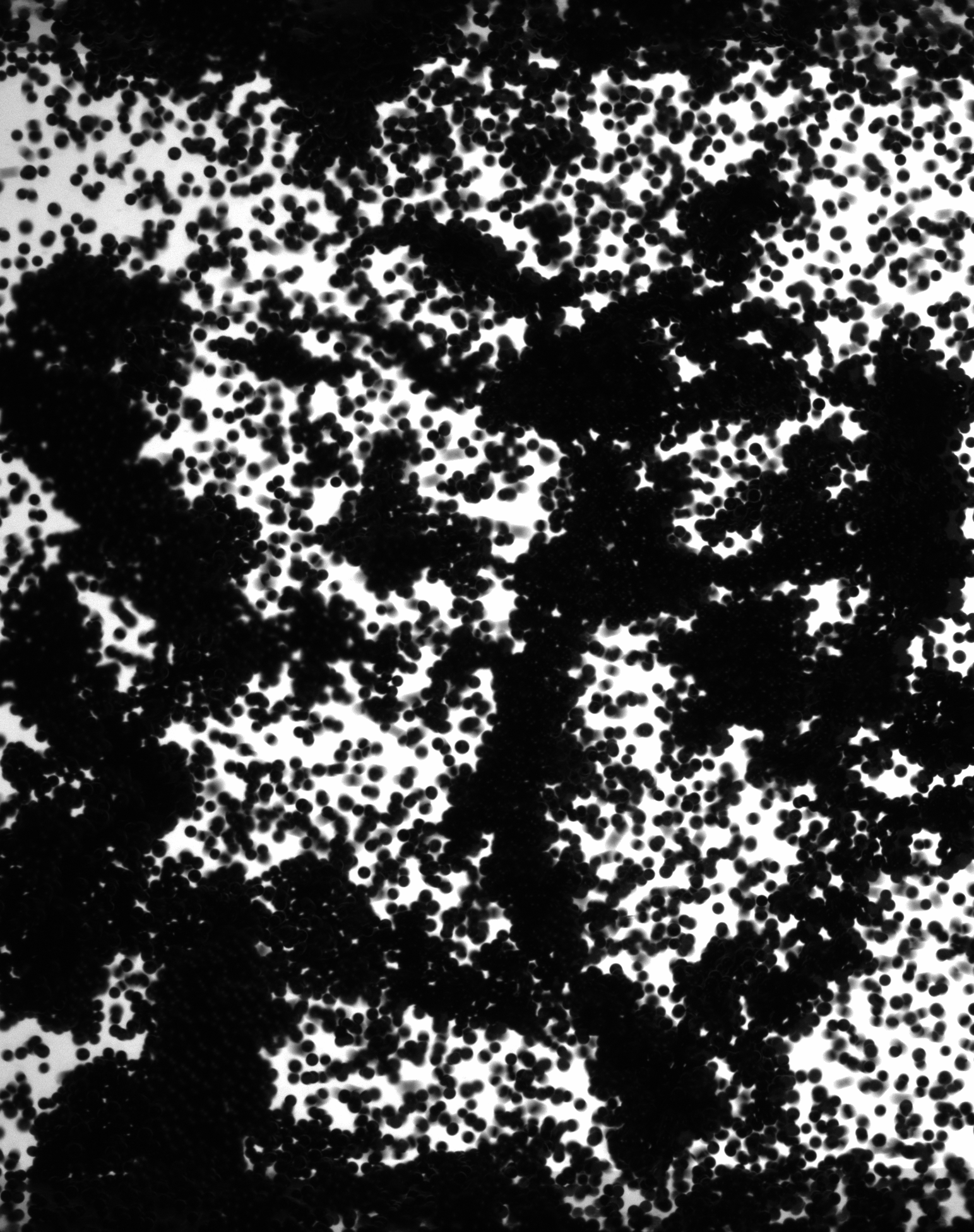}
        \includegraphics[width = 0.3\linewidth, height = 0.34\linewidth]{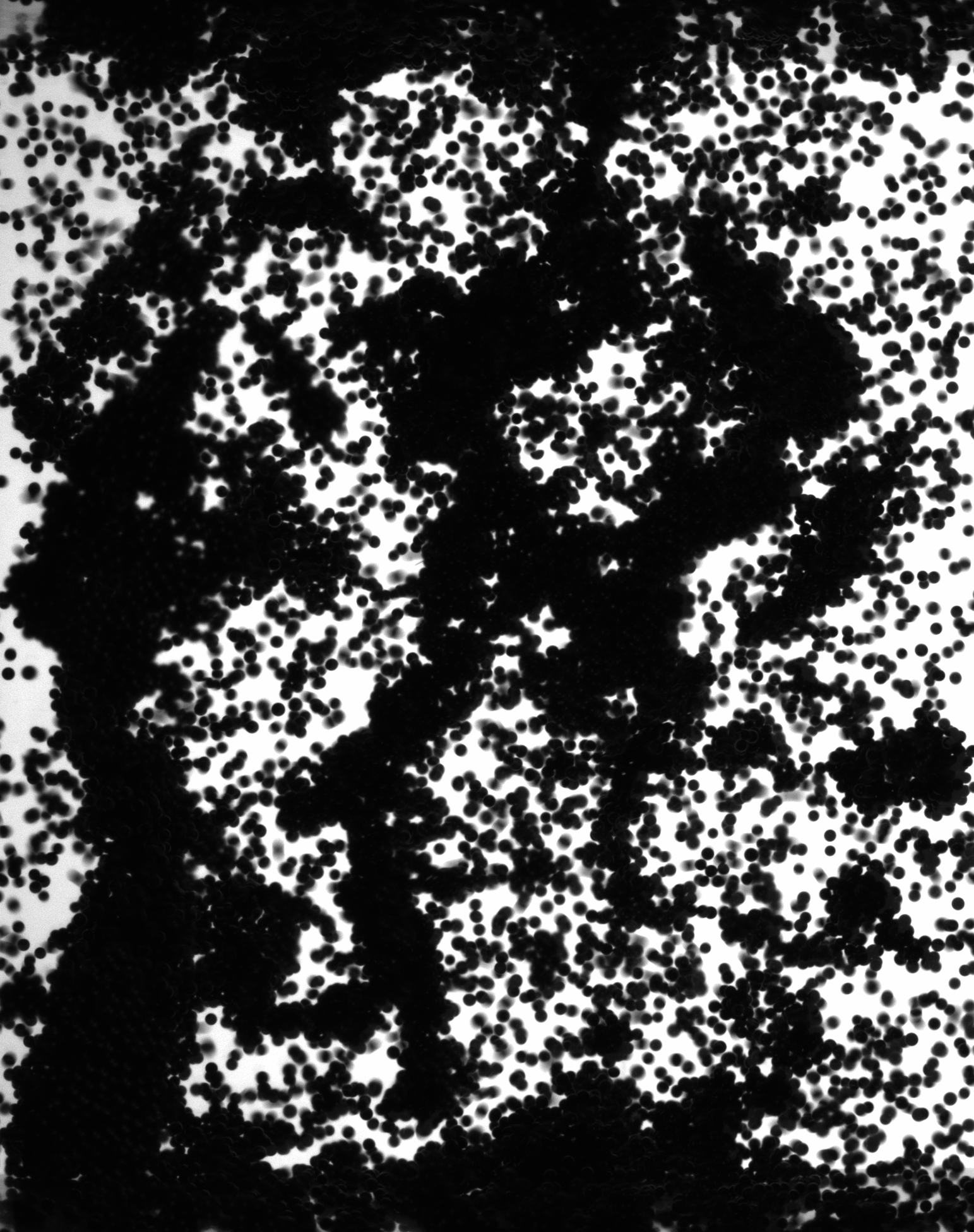} 
        \includegraphics[width = 0.3\linewidth, height = 0.34\linewidth]{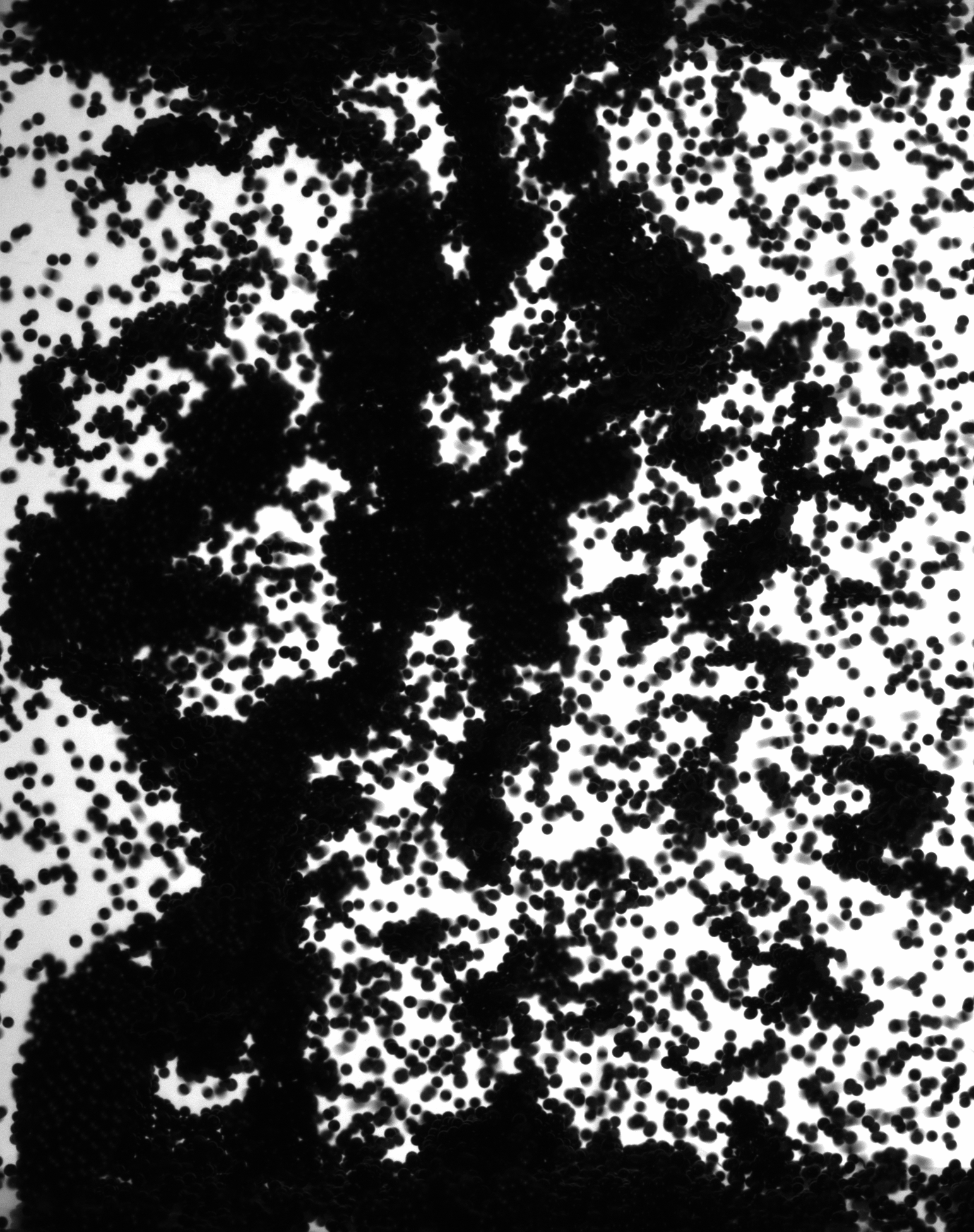} \\
    \hspace{0.01cm}$t= 1.875$\hspace{1.0cm}$t=2$\hspace{1.0cm}$t=2.125$\\
        \vspace{0.3em}
        \includegraphics[width = 0.3\linewidth, height = 0.34\linewidth]{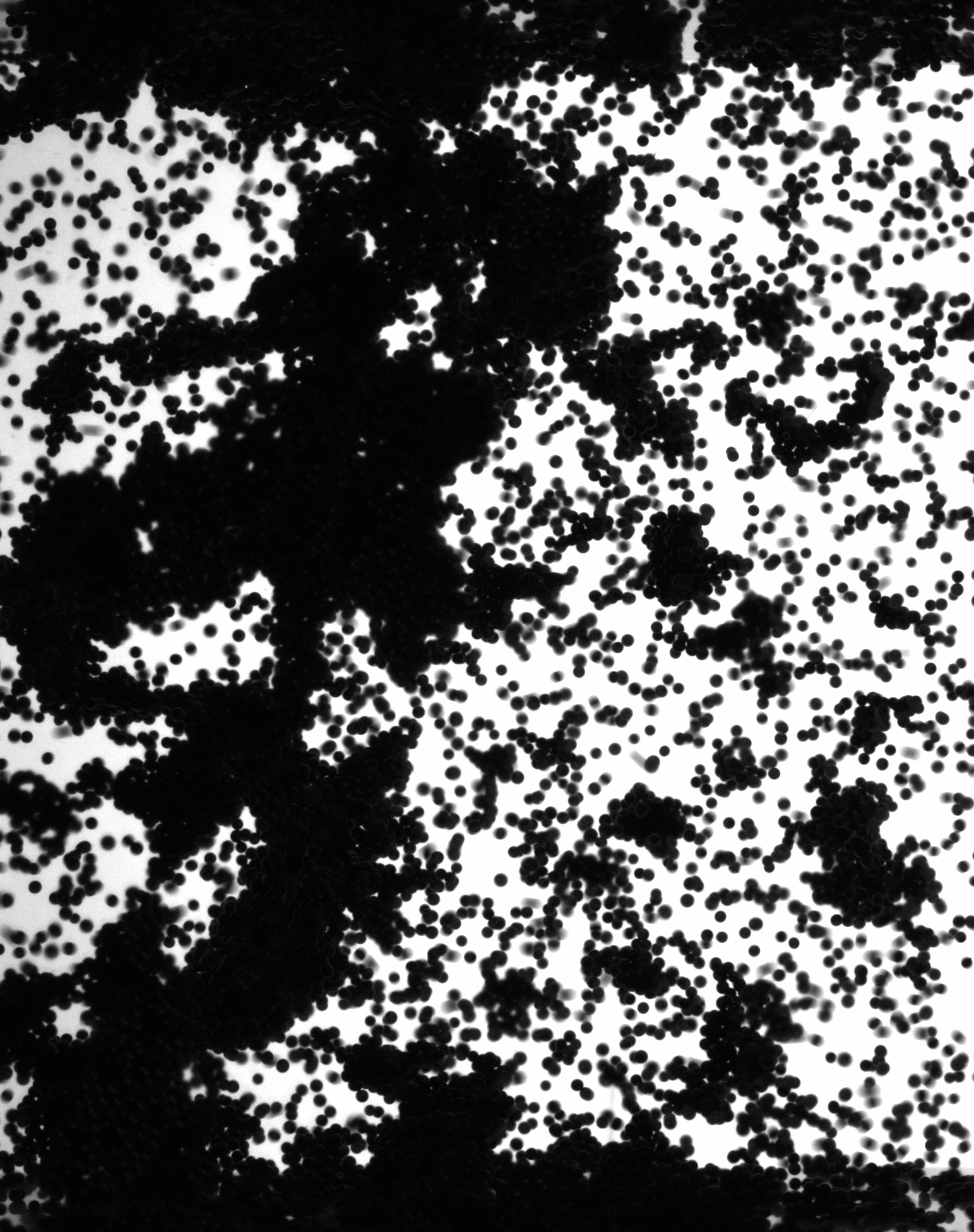}
        \includegraphics[width = 0.3\linewidth, height = 0.34\linewidth]{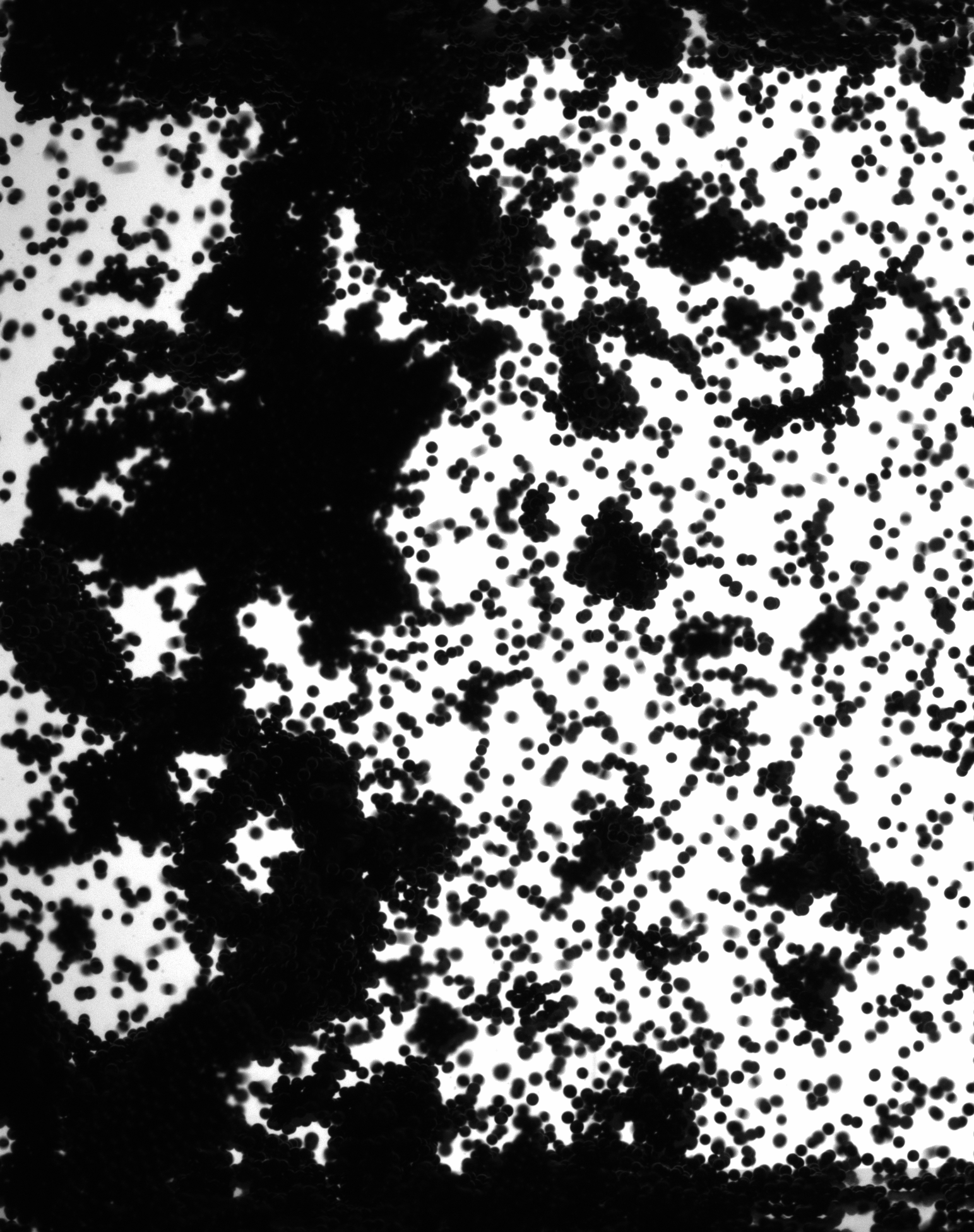} 
        \includegraphics[width = 0.3\linewidth, height = 0.34\linewidth]{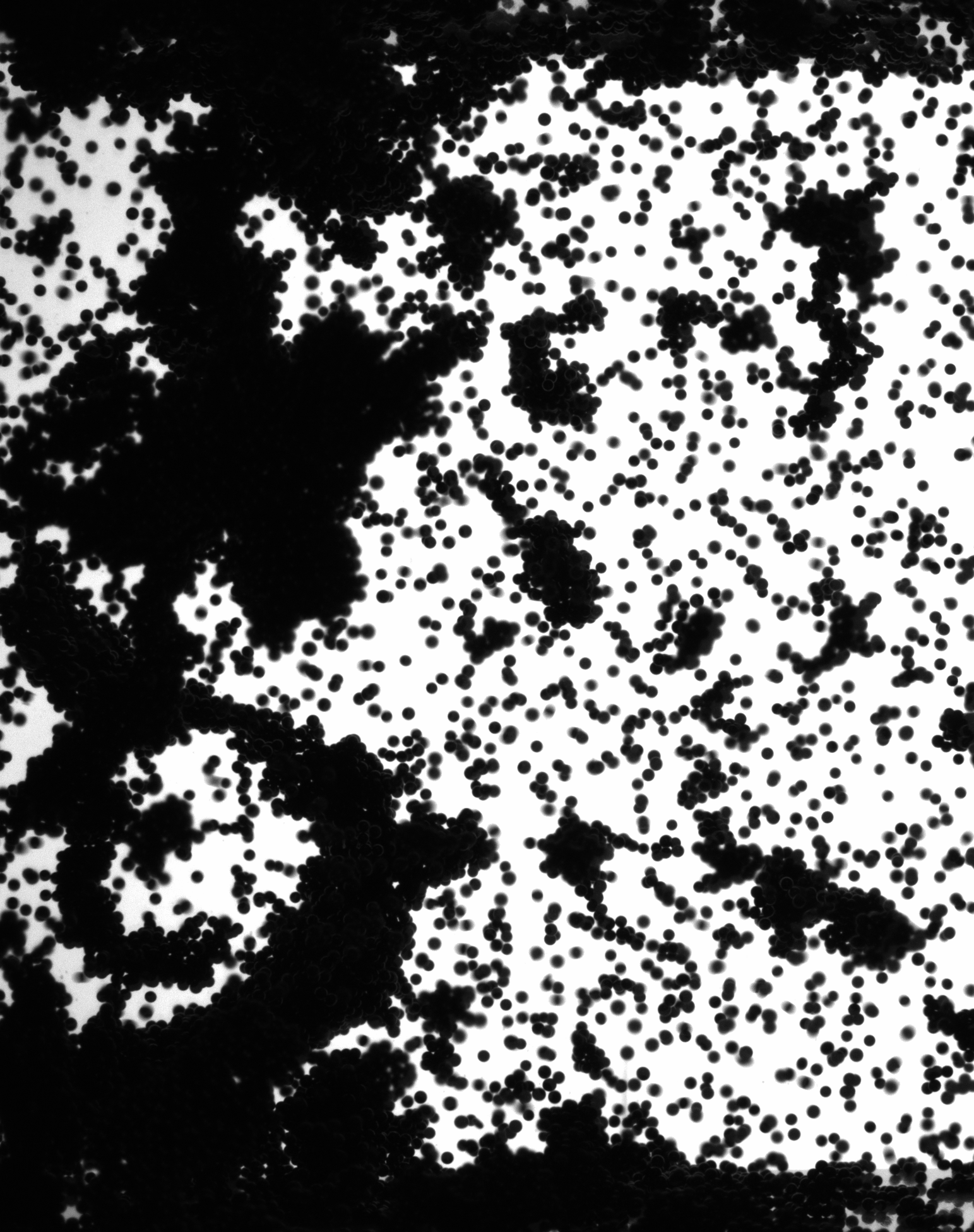}\\
        \hspace{0.01cm}$t= 2.250$\hspace{0.6cm}$t=2.375$\hspace{0.6cm}$t=2.500$\\
        \vspace{0.3em}
        \includegraphics[width = 0.3\linewidth, height = 0.34\linewidth]{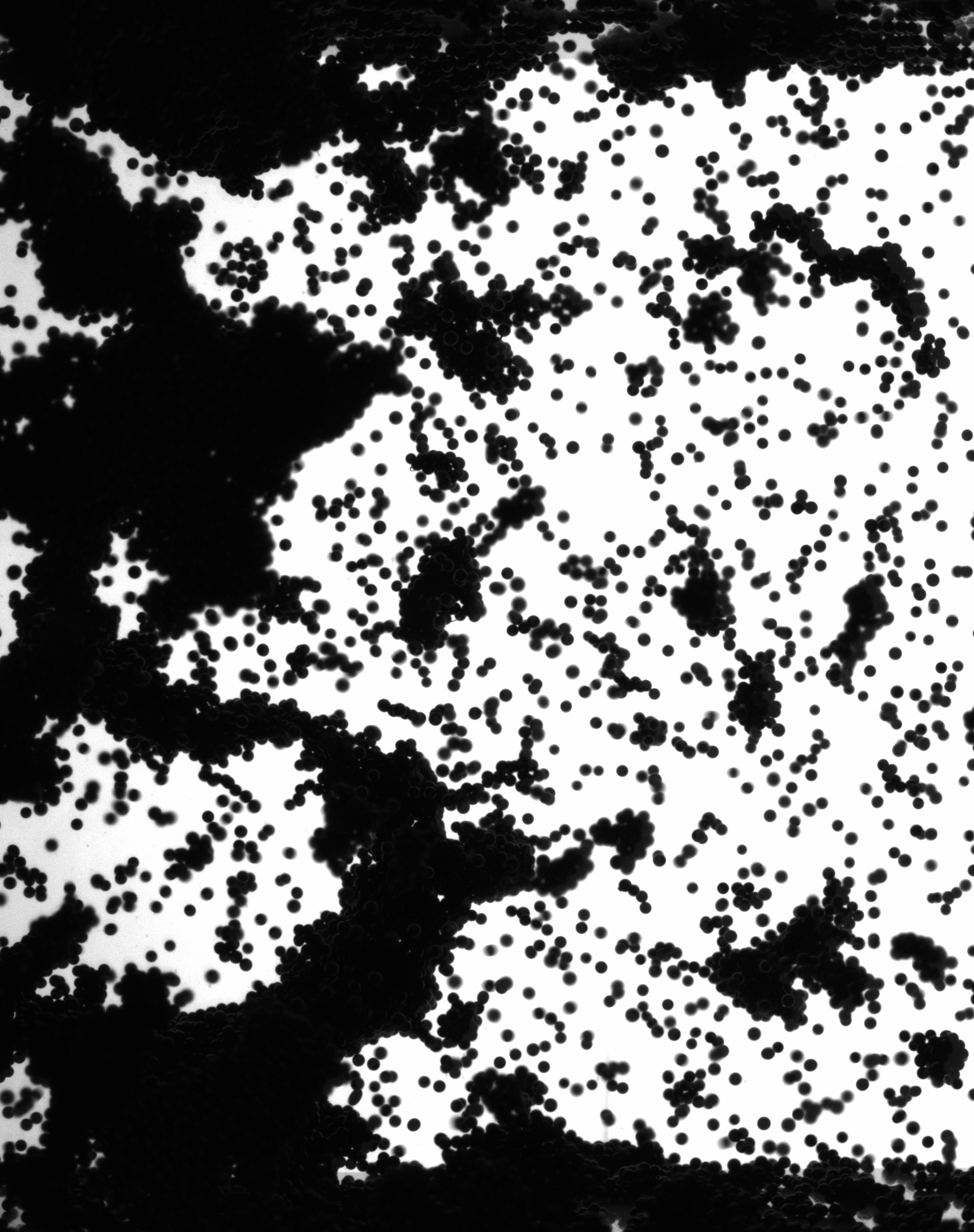}
        \includegraphics[width = 0.3\linewidth, height = 0.34\linewidth]{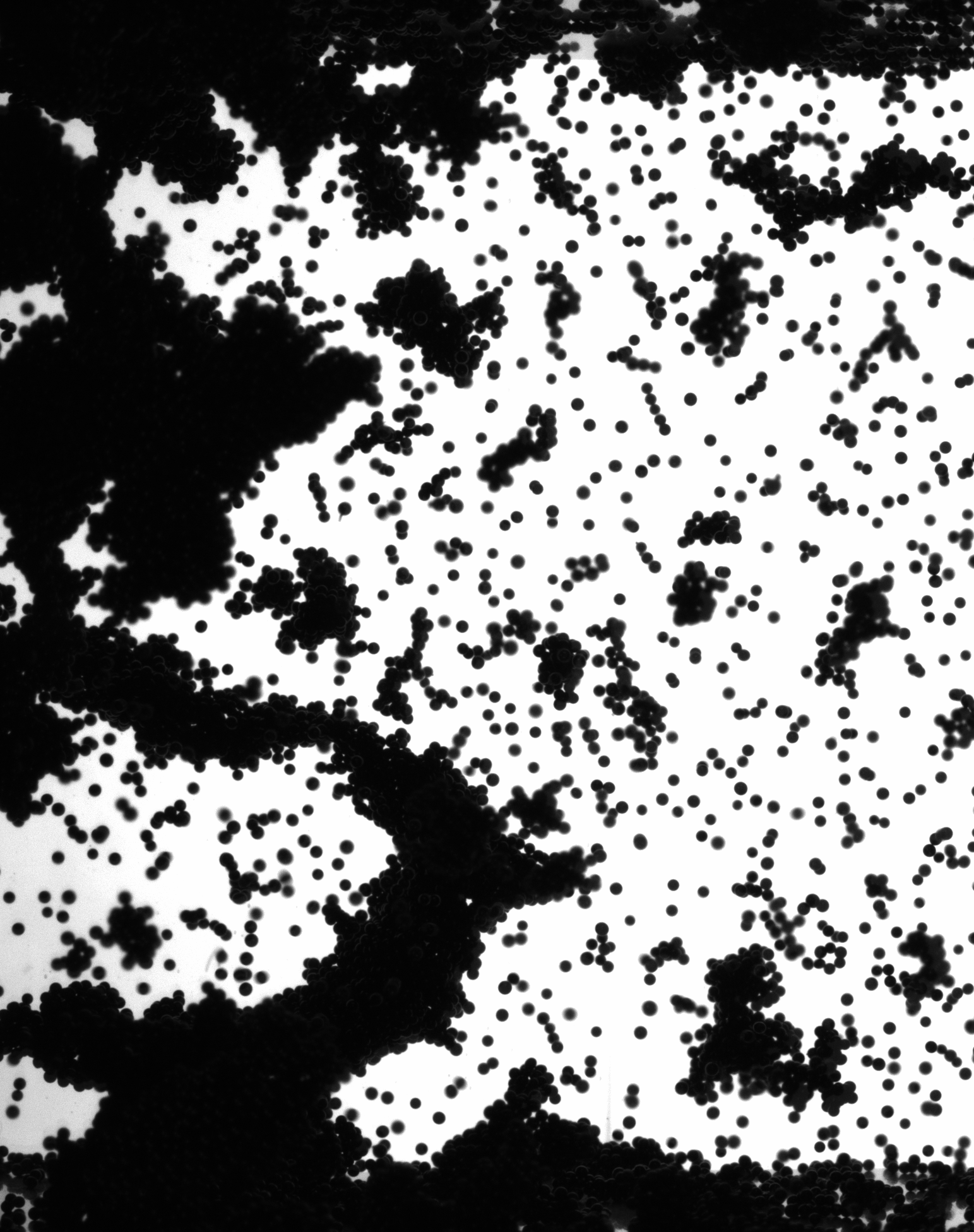}
        \includegraphics[width = 0.3\linewidth, height = 0.34\linewidth]{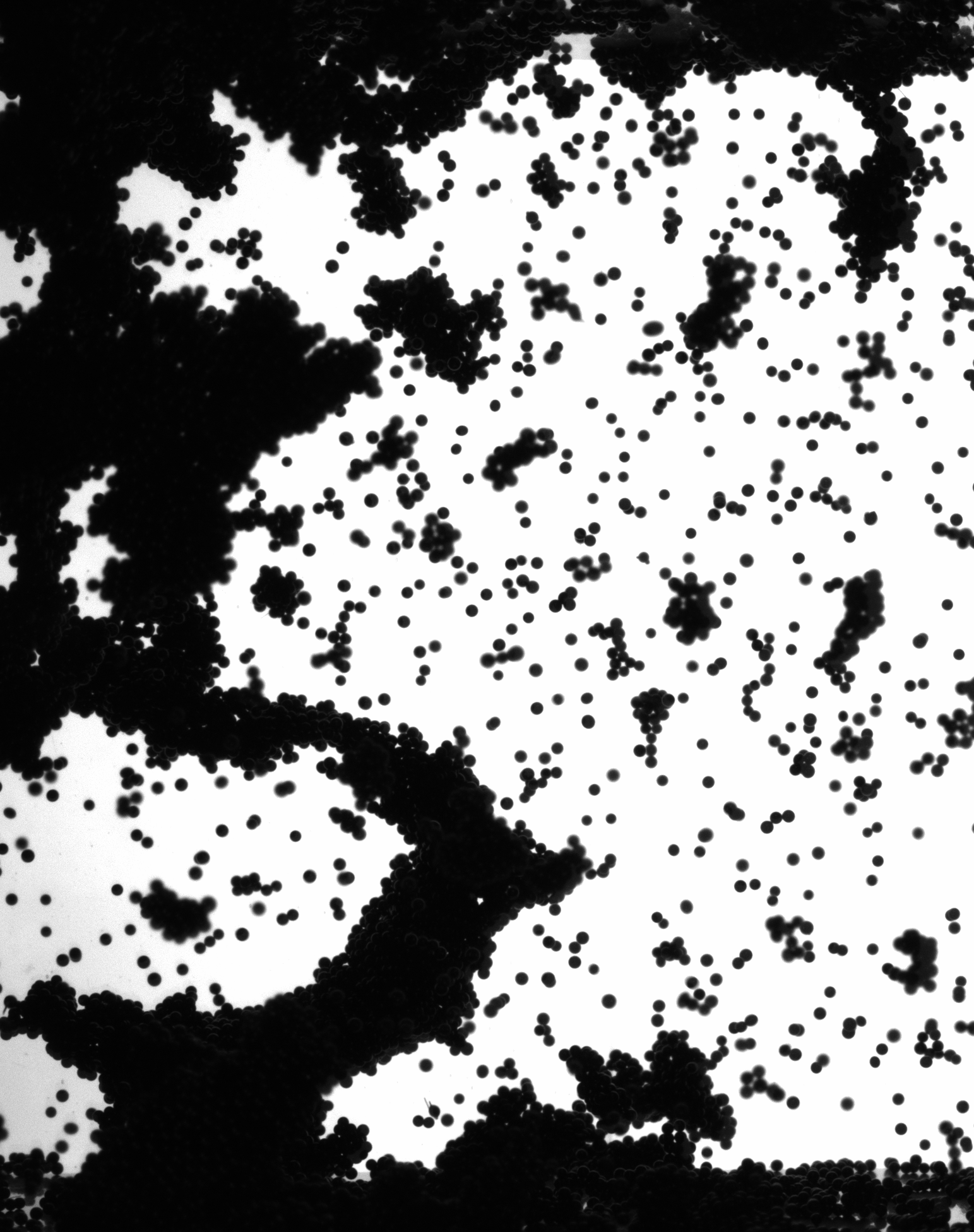} \\
        \hspace{0.01cm}$t= 2.750$\hspace{0.6cm}$t=3$\hspace{0.6cm}$t=3.500$\\
        \vspace{0.3em}
        \includegraphics[width = 0.3\linewidth, height = 0.34\linewidth]{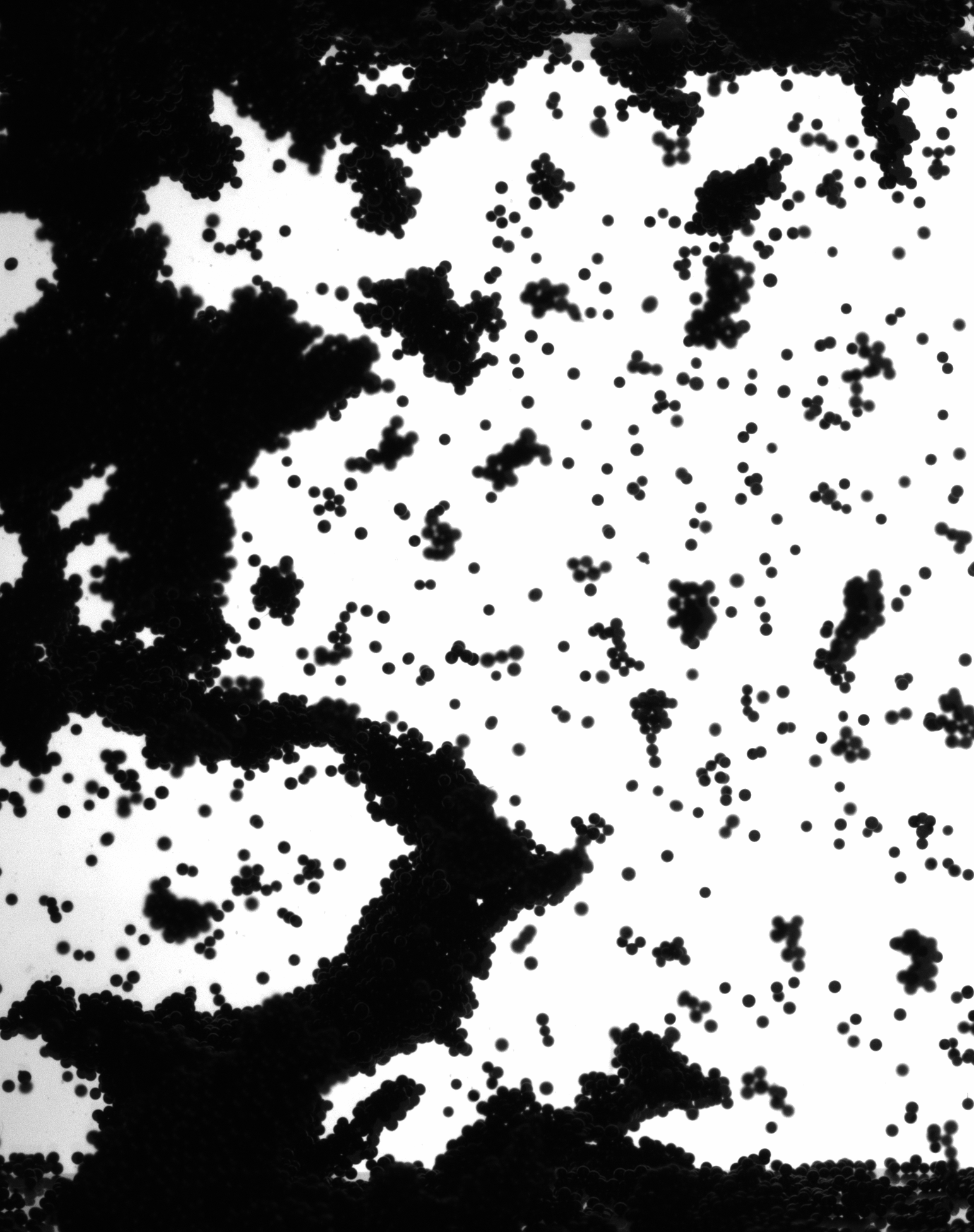}
        \includegraphics[width = 0.3\linewidth, height = 0.34\linewidth]{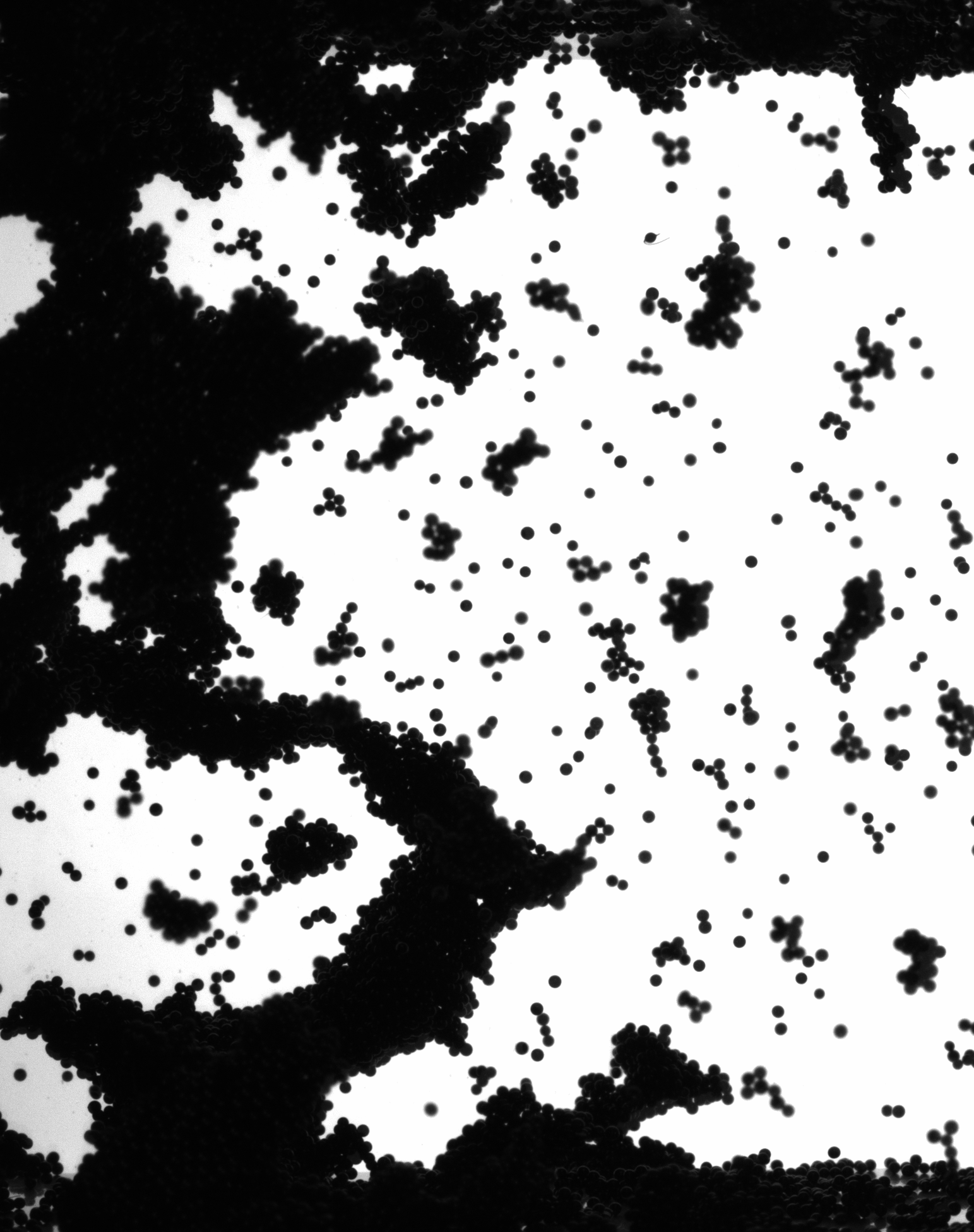} 
        \includegraphics[width = 0.3\linewidth, height = 0.34\linewidth]{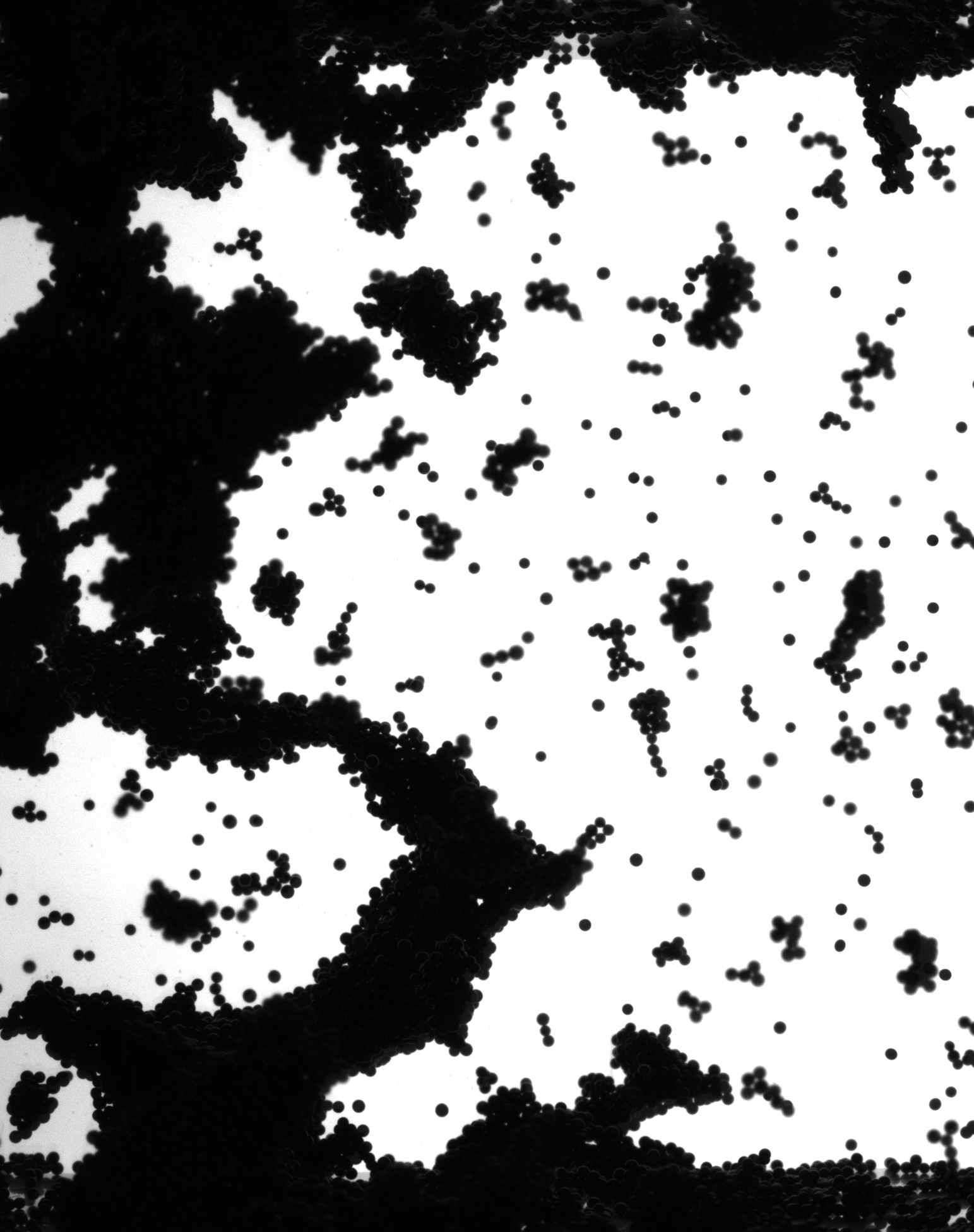}\\
        \hspace{0.01cm}$t= 4$\hspace{1.5cm}$t=5$\hspace{1.5cm}$t=6$\\
    \caption{\label{fig:Exp1}
    Drop tower experiment (\tt{Ser1})
    }
        
    \end{subfigure}
    \hfill
    \begin{subfigure}[t]{.49\textwidth}
        \centering
        \includegraphics[width = 0.3\linewidth, height = 0.34\linewidth]{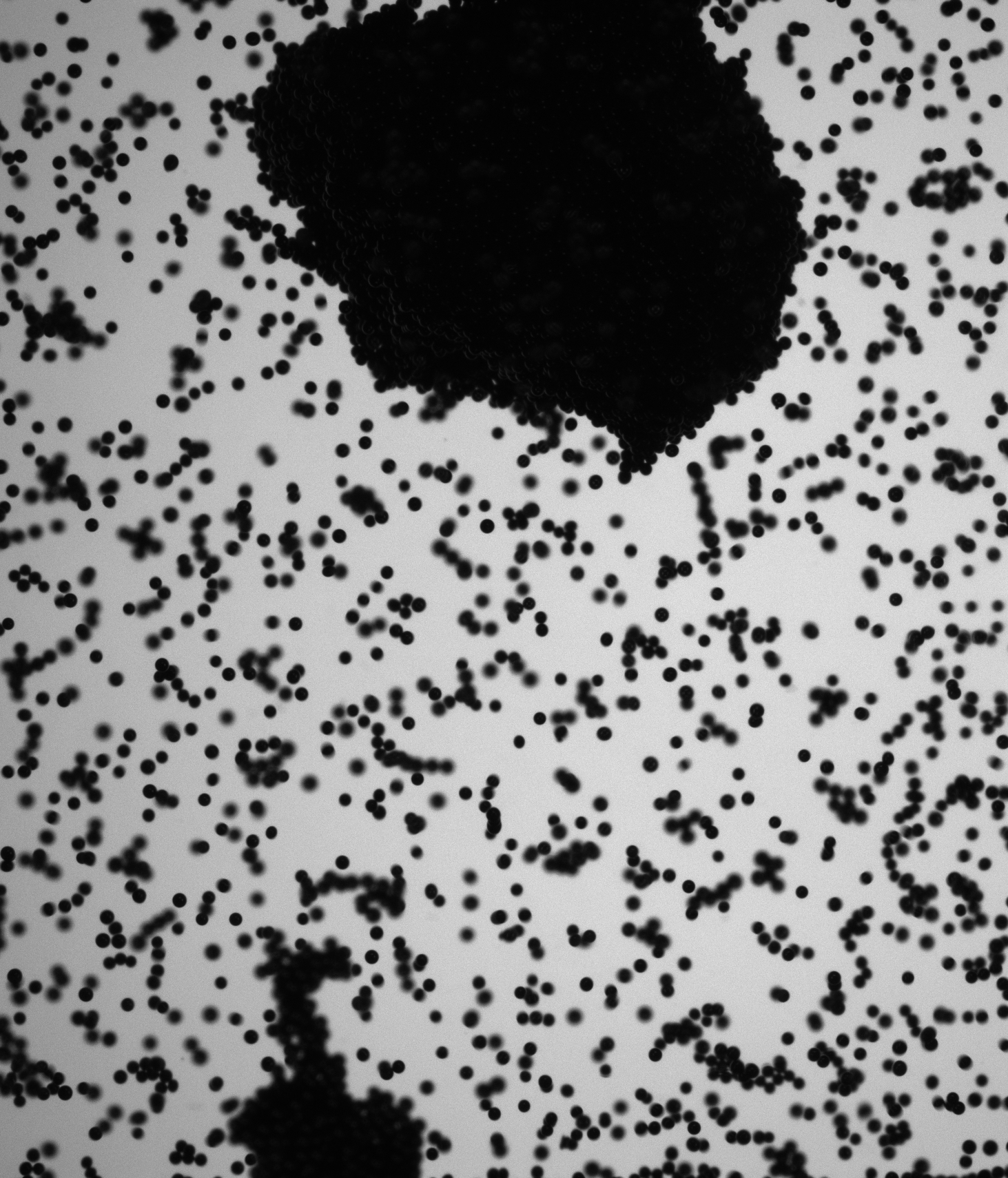}
        \includegraphics[width = 0.3\linewidth, height = 0.34\linewidth]{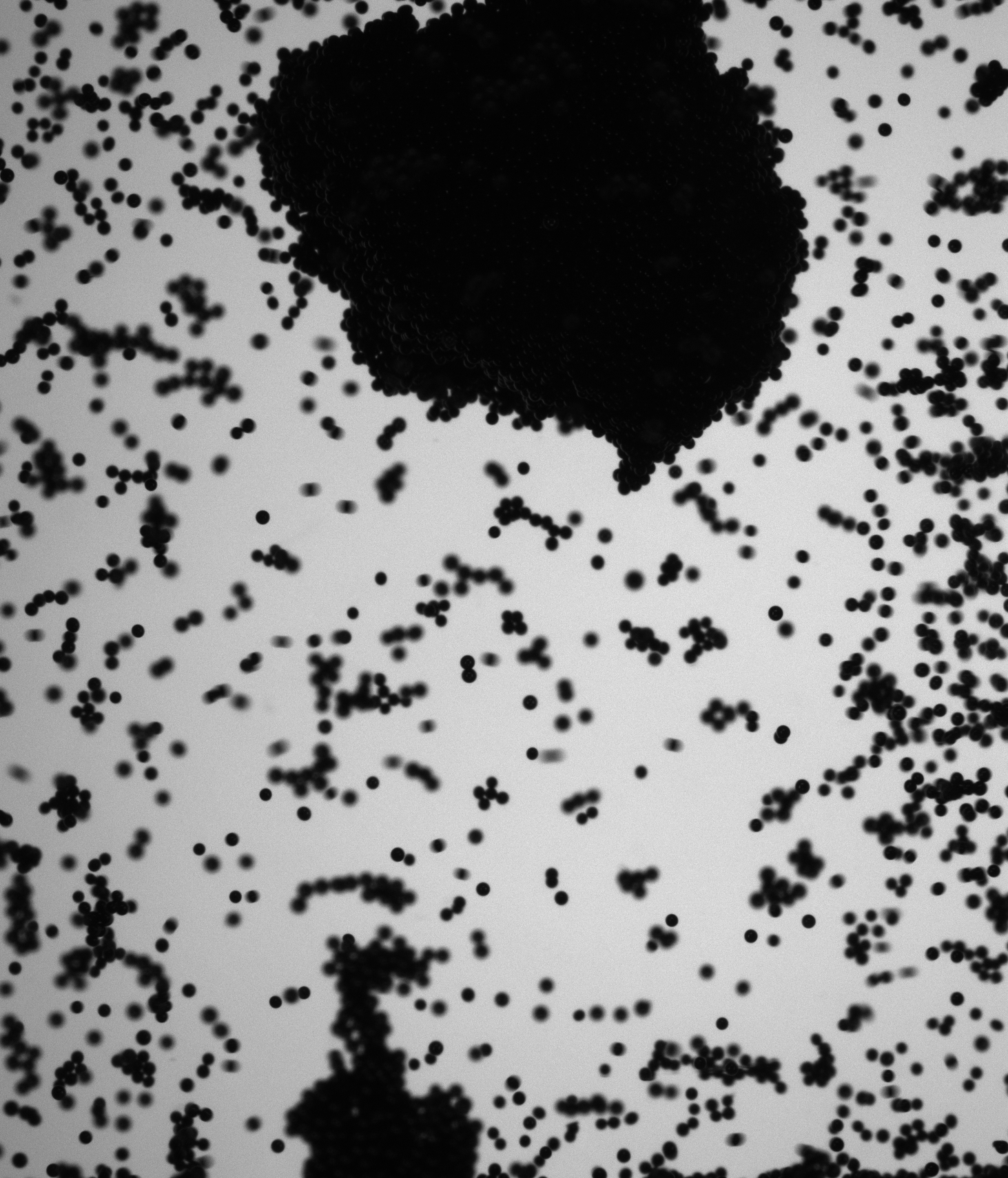} 
        \includegraphics[width = 0.3\linewidth, height = 0.34\linewidth]{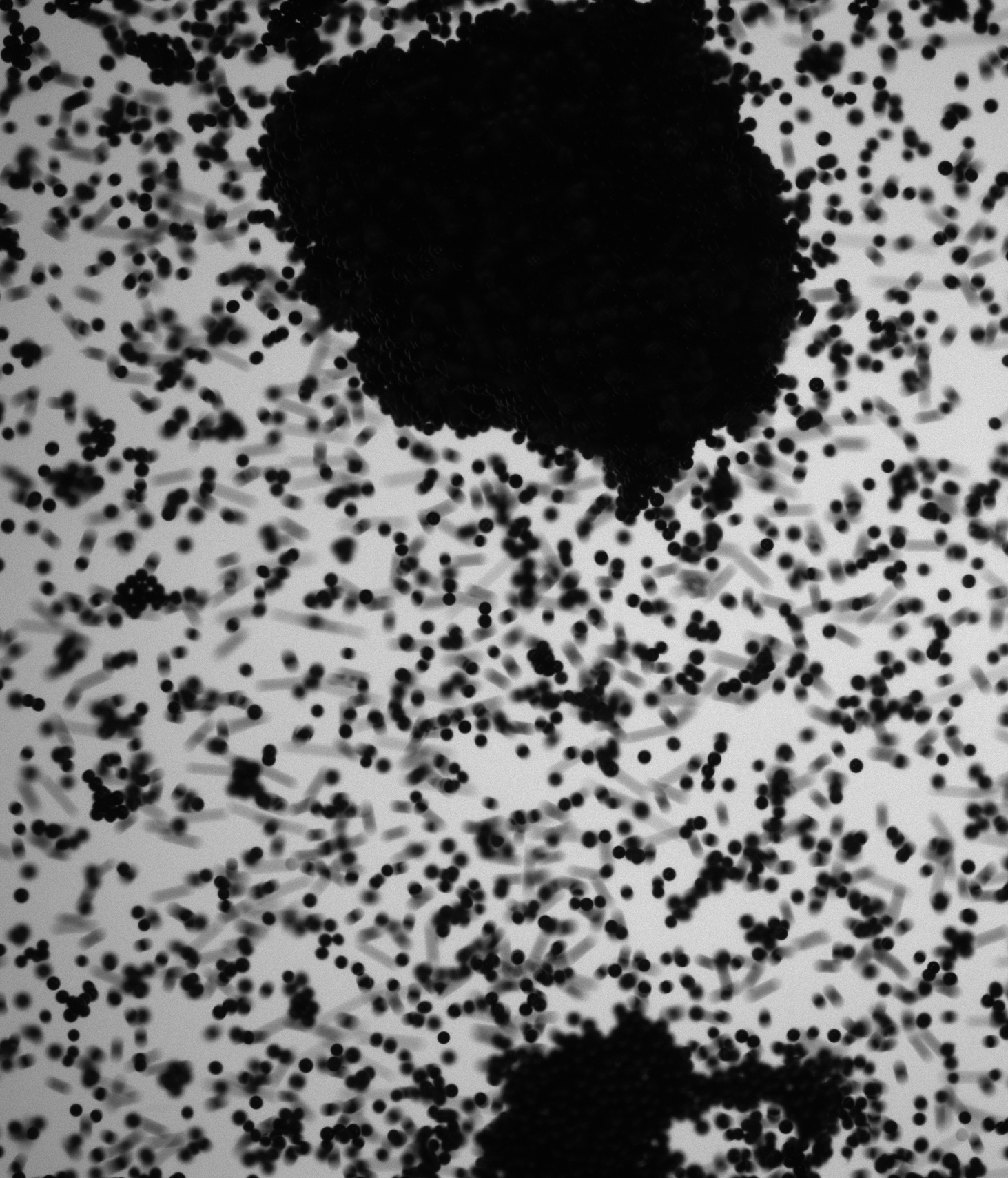}\\ 
       \hspace{0.01cm}$t= 0$\hspace{1.4cm}$t=7.500$\hspace{0.6cm}$t=8.750$
       \\
        \vspace{0.3em}
        \includegraphics[width = 0.3\linewidth, height = 0.34\linewidth]{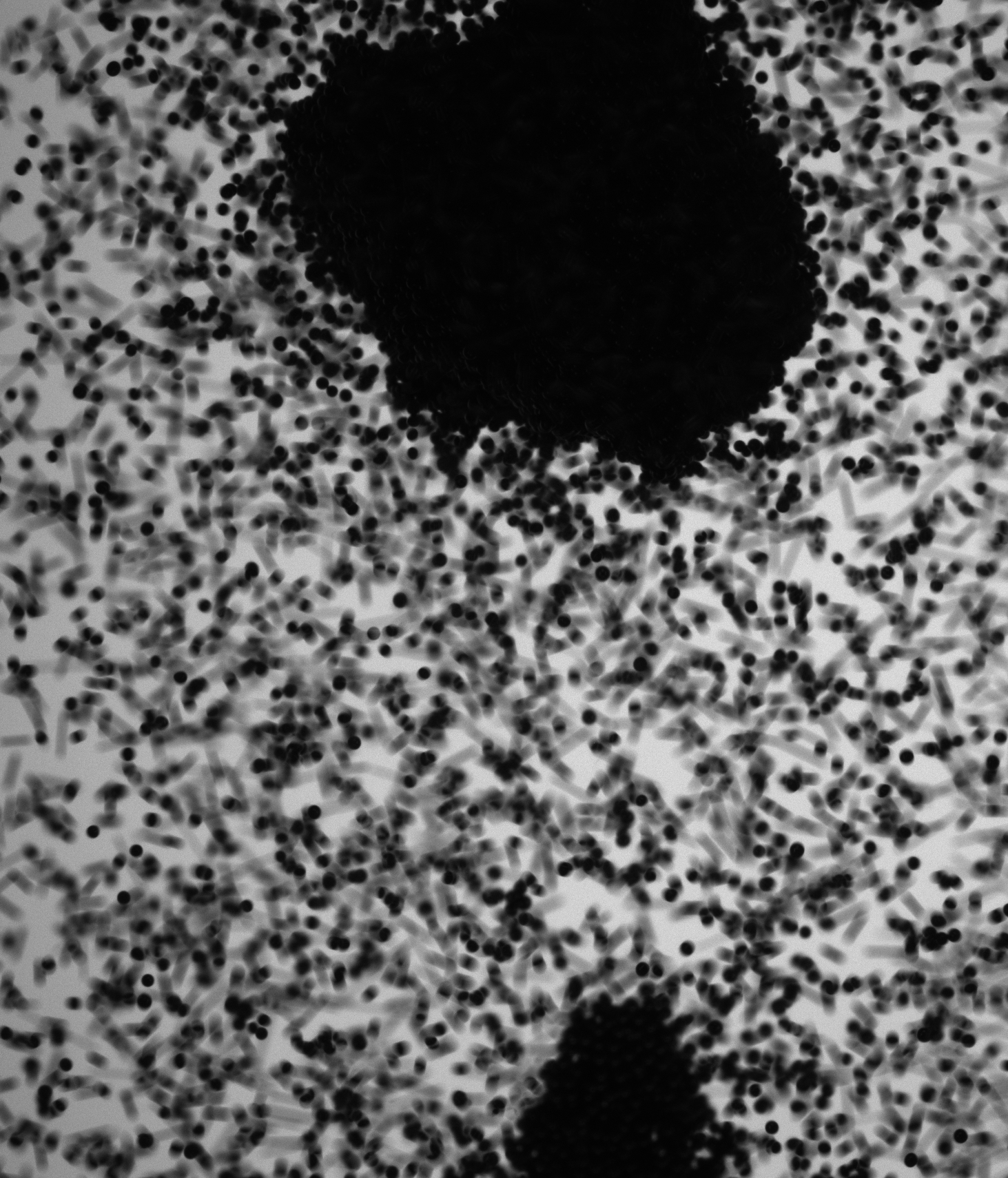} 
        \includegraphics[width = 0.3\linewidth, height = 0.34\linewidth]{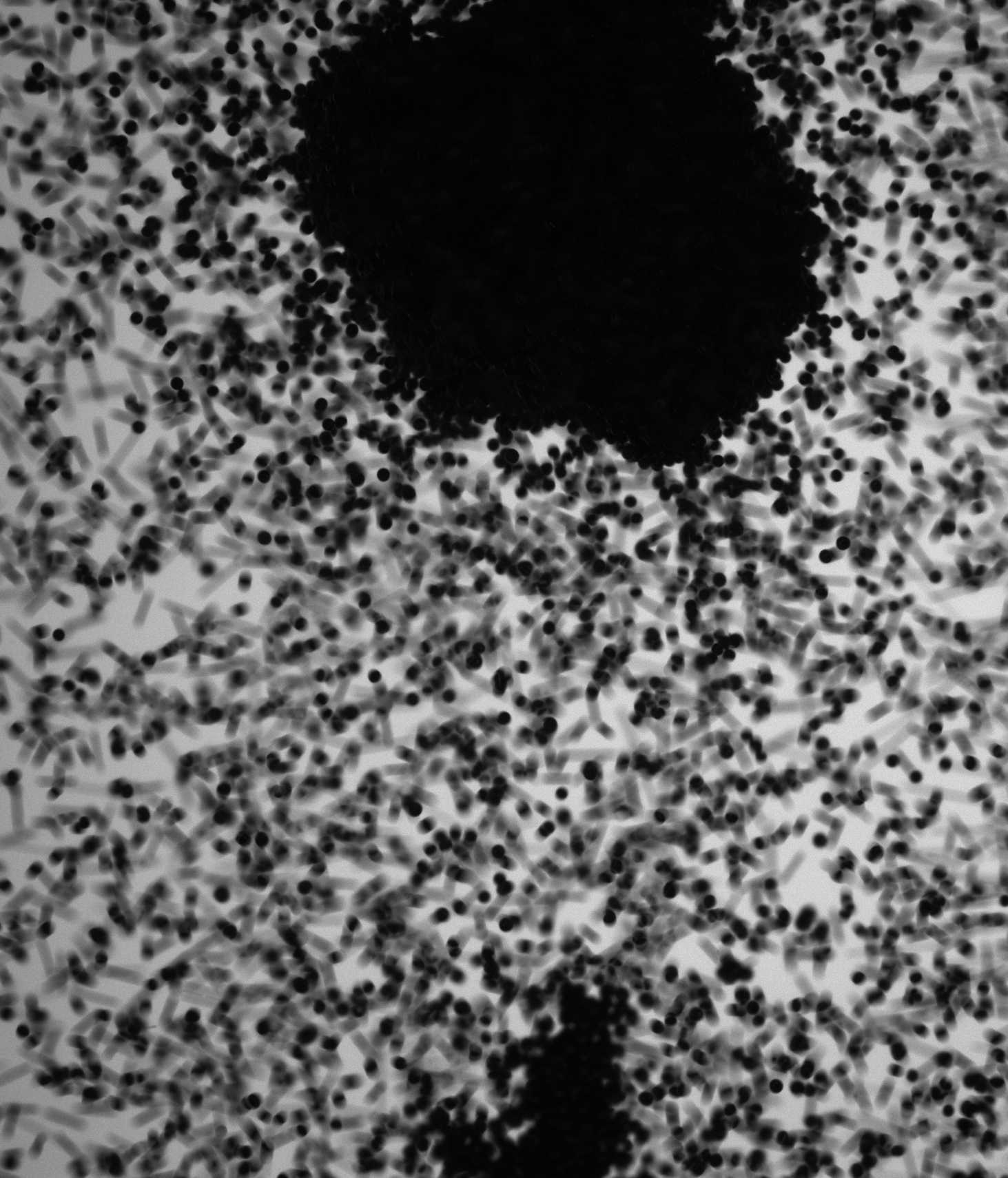} 
        \includegraphics[width = 0.3\linewidth, height = 0.34\linewidth]{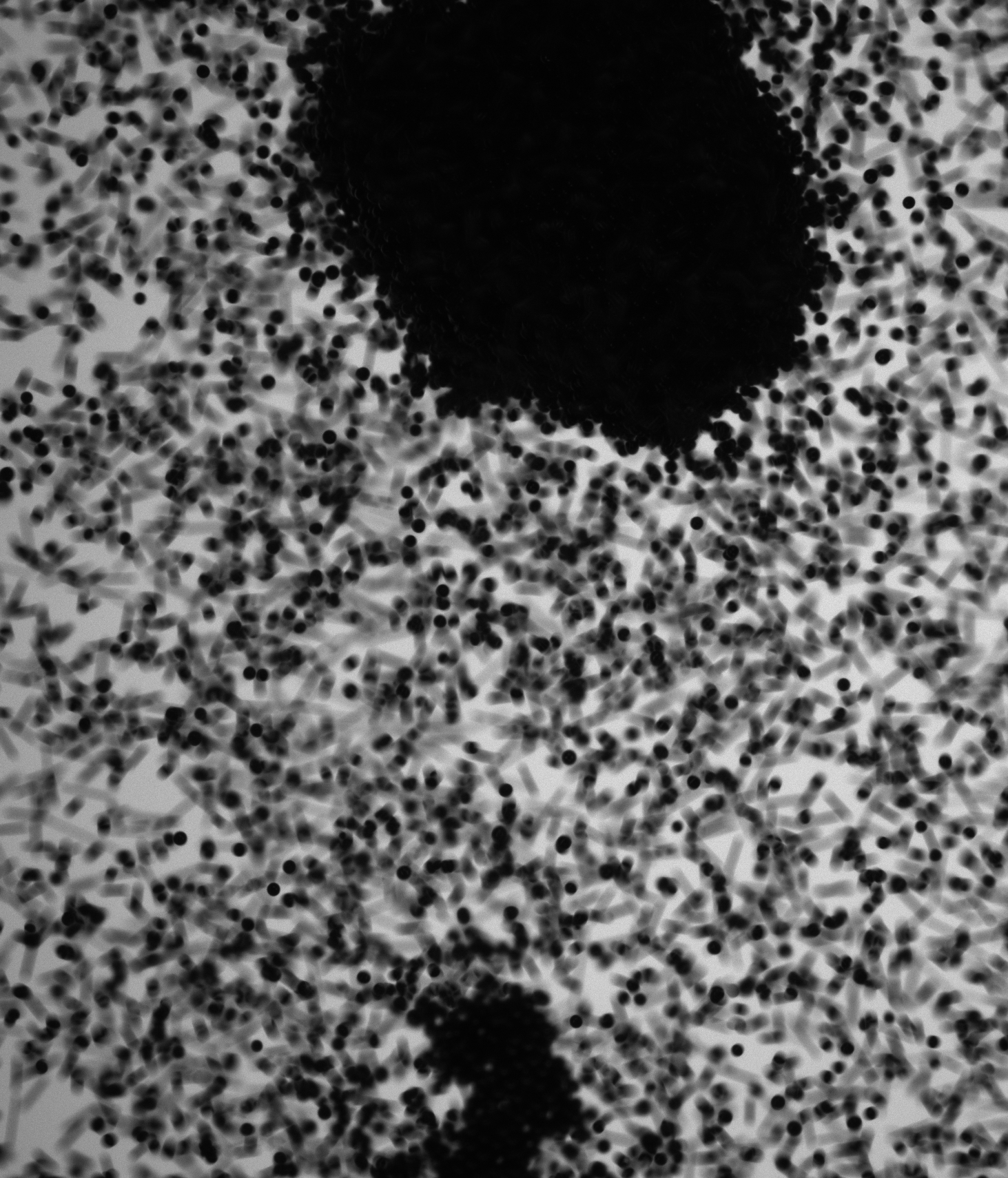}\\
        \hspace{0.01cm}$t= 9.500$\hspace{1.0cm}$t=9.750$\hspace{1.0cm}$t=10$
       \\
        \vspace{0.3em}
        \includegraphics[width = 0.3\linewidth, height = 0.34\linewidth]{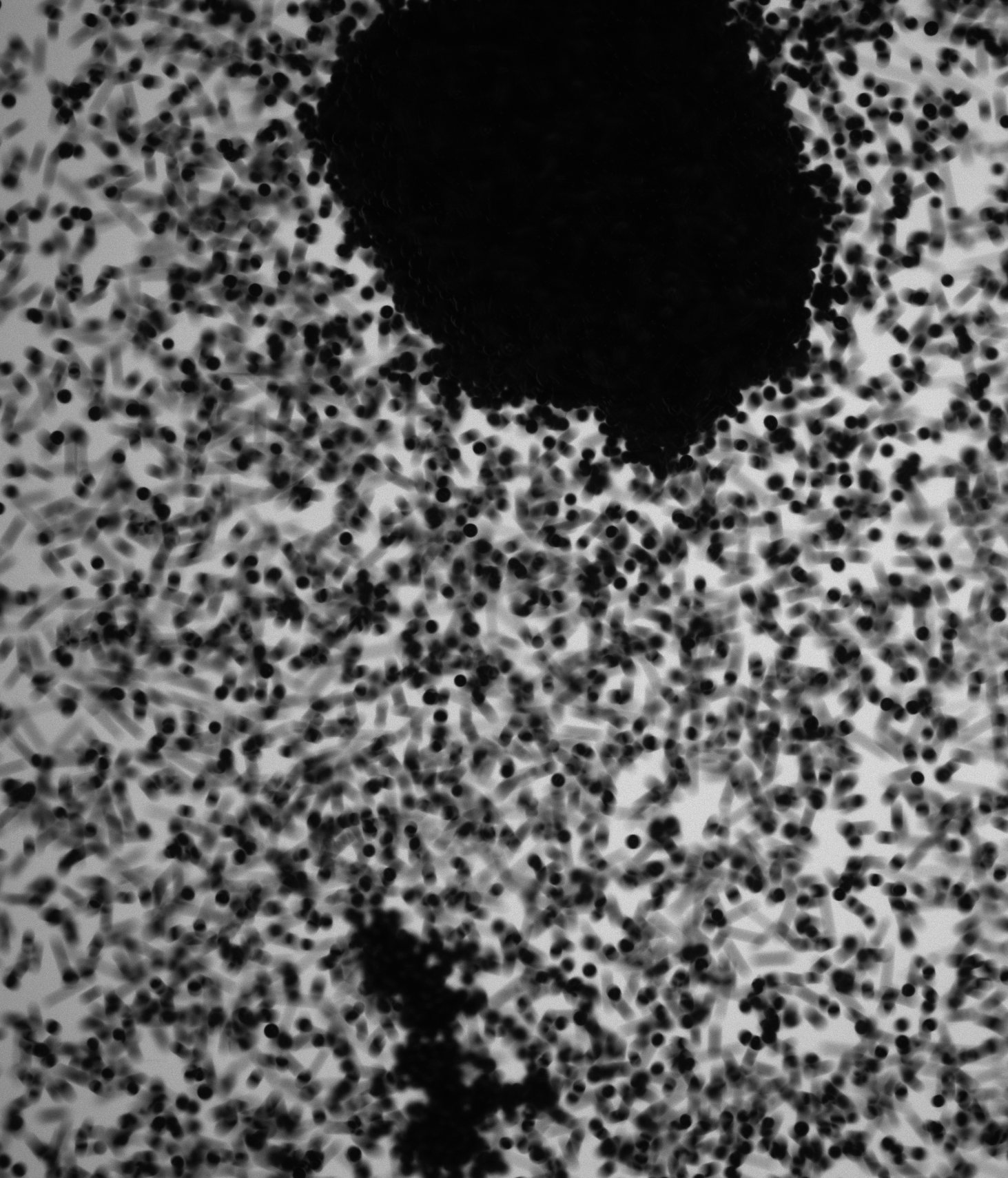} 
        \includegraphics[width = 0.3\linewidth, height = 0.34\linewidth]{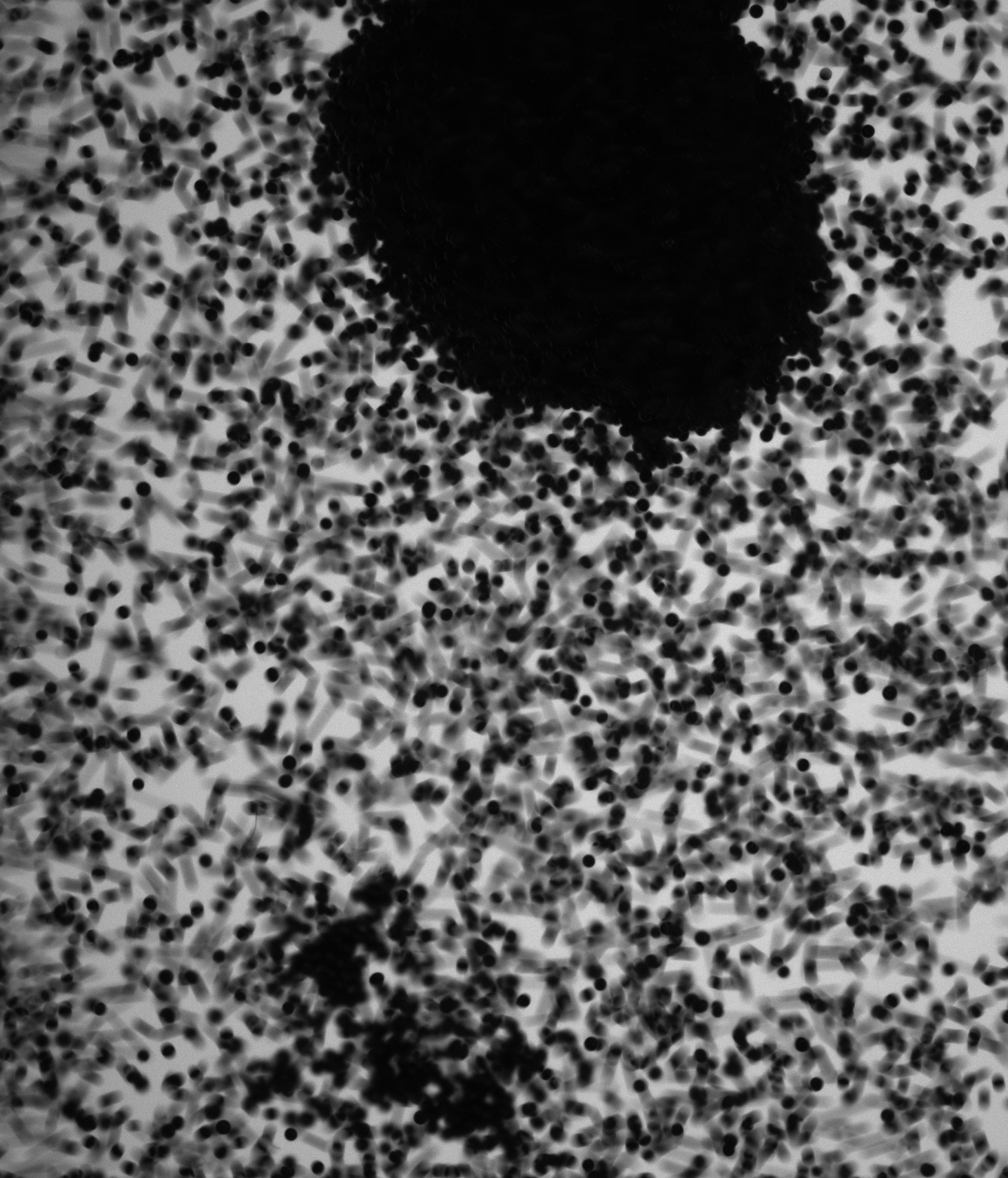} 
        \includegraphics[width = 0.3\linewidth, height = 0.34\linewidth]{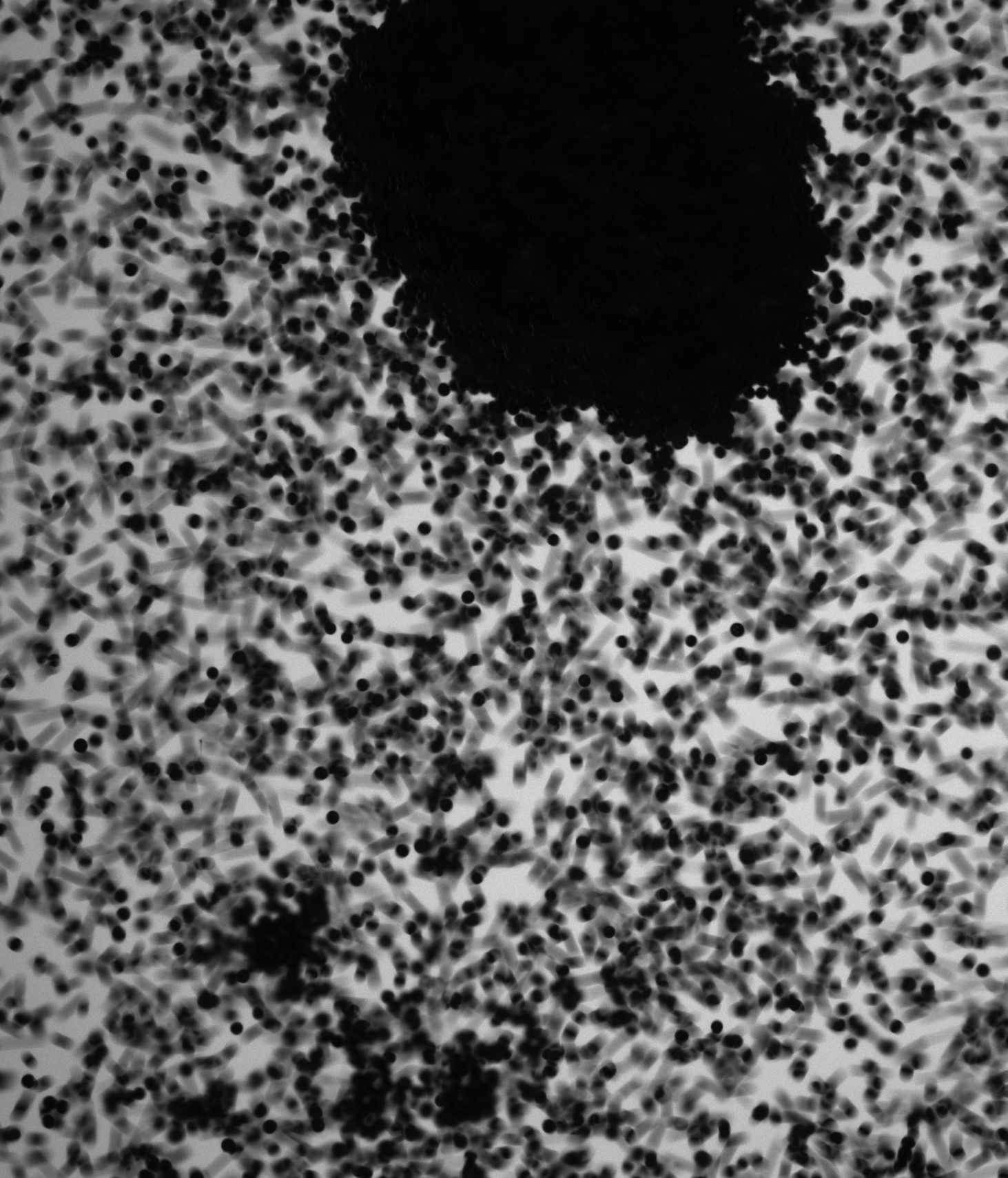}\\
        \hspace{0.02cm}$t= 10.250$\hspace{0.13cm}$t=10.375$\hspace{0.13cm}$t=10.450$
       \\
        \vspace{0.3em}
        \includegraphics[width = 0.3\linewidth, height = 0.34\linewidth]{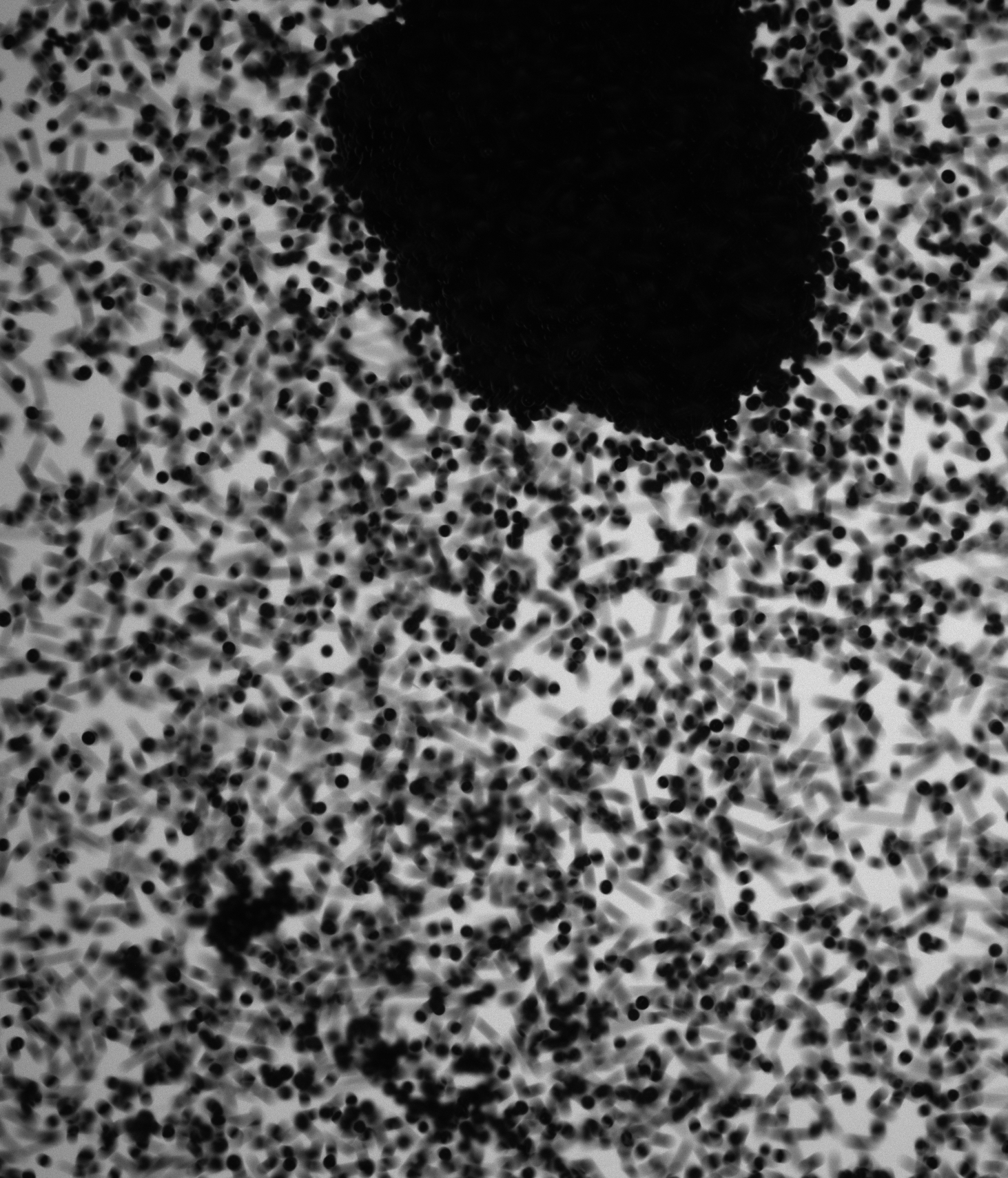} 
        \includegraphics[width = 0.3\linewidth, height = 0.34\linewidth]{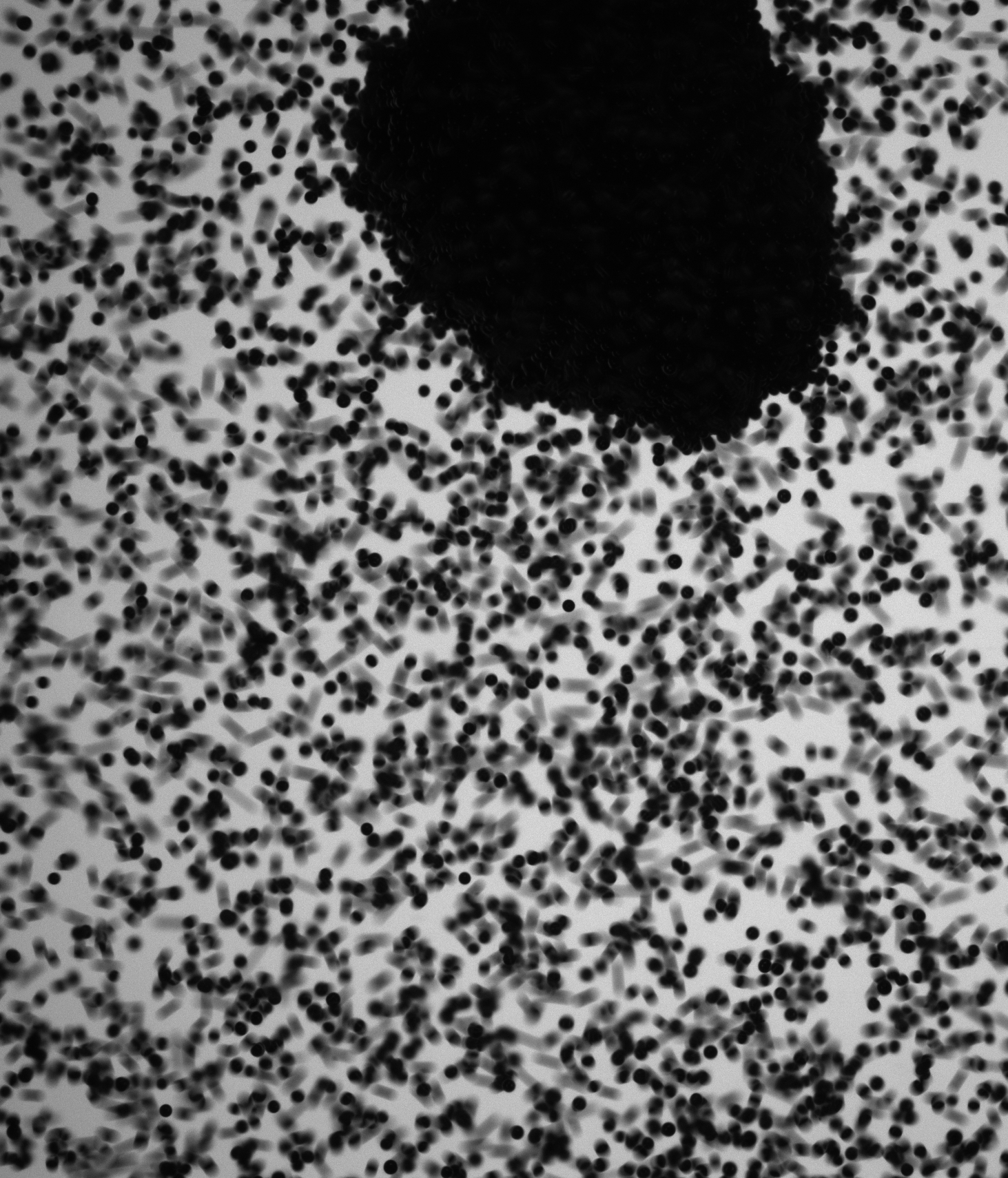} 
        \includegraphics[width = 0.3\linewidth, height = 0.34\linewidth]{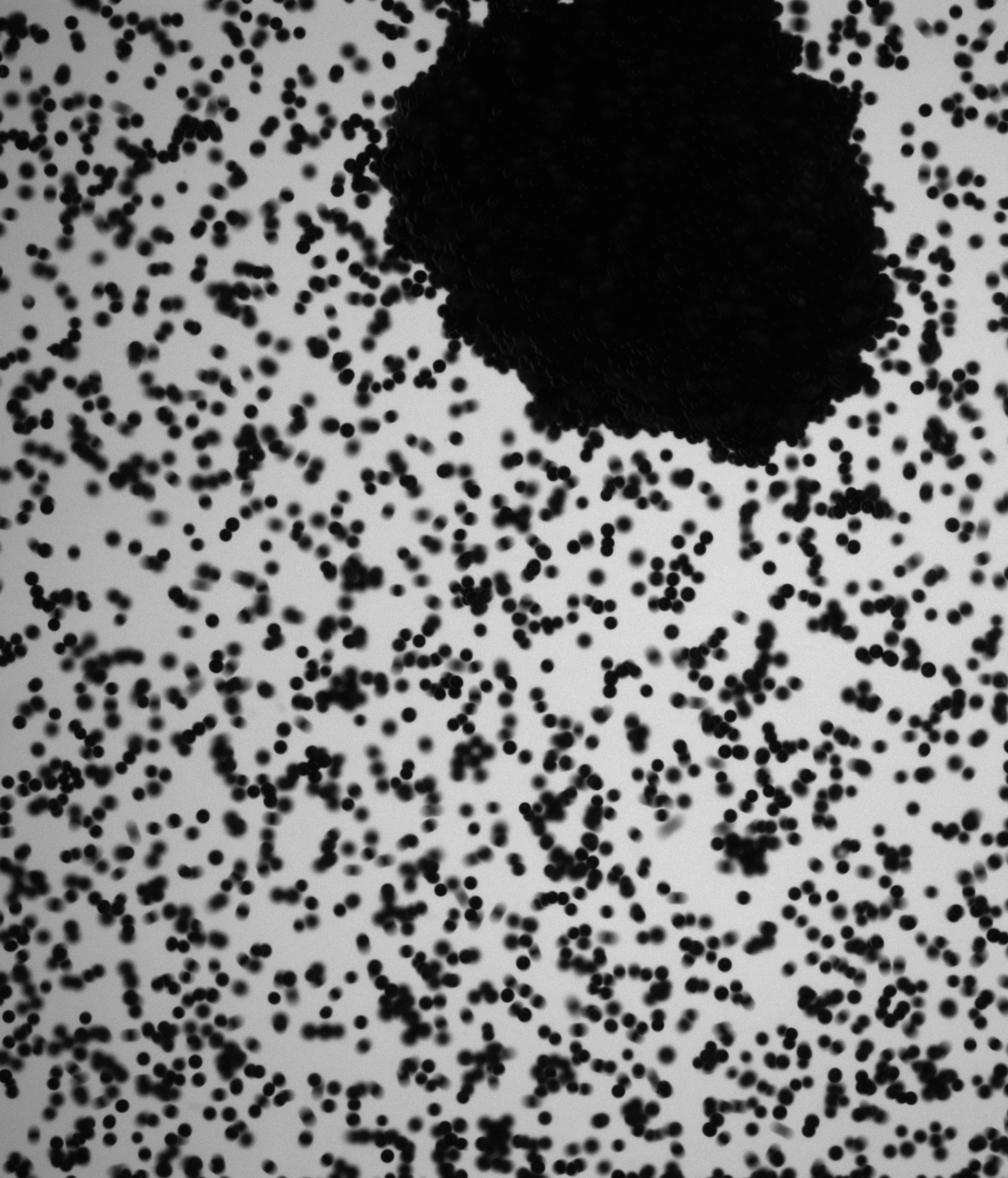}\\
        \hspace{0.01cm}$t= 10.500$\hspace{0.13cm}$t=11.250$\hspace{0.13cm}$t=12.475$
       \\
    \caption{\label{fig:Exp2}
    Sounding rocket experiment (\tt{Ser2})
    }
    \end{subfigure}
    \caption{\label{fig:Exp1-Exp2}
    The images in Figures \ref{fig:Exp1} and \ref{fig:Exp2} are selected frames out of the corresponding image sequences of two different experiments. 
    }
\end{figure}

Here, microgravity experiments are used to investigate this critical growth phase. The key questions to be answered in these experiments can be summarized as follows \cite{ref_wurm2021}:
\begin{itemize}
    \item Enable additional forces (such as electrostatic or magnetic forces) growth into the critical size range of centimeters or decimeters?
    \item Under what conditions can small dust agglomerates grow to larger sizes?
    \item What are the critical threshold velocities and / or particle sizes for growth, erosion, or catastrophic fragmentation?
    \item How large can clusters of particles grow and at which growth rate?
    \item What structures form during these processes, for example in terms of porosity, shape, and fractal dimension?
\end{itemize}
A series of microgravity experiments have been carried out to address these key questions, including experiments with short duration at the Bremen drop tower (9 s experiment time) \cite{ref_jungmann2021} and long-term experiments on a sounding rocket (6 min experiment time)\cite{ref_teiser2025}. A typical experiment consists of a test cell (volume $50\times 50\times50\,\mathrm{mm}^3$) with a sample of small basalt spheres (sphere diameter $0.5\,\mathrm{mm}$, total amount $8 - 10 \,\mathrm{g}$). The test cell can be agitated in one direction, to induce relative velocities and collisions between the particles. 


The particles moving in the test cell are recorded at 40 frames/s. A typical drop-tower experiment has a duration of 9 s, while the experiment on a sounding rocket gives an image sequence of 6 min continuous video data. This makes a detailed analysis challenging because of the amount of data and the difficulty in identifying the physical processes occurring during the experiments.
Figure \ref{fig:Exp1-Exp2} shows a typical example for image data obtained in the drop tower (left series) and in the sounding rocket (right series). We choose these data sets because they refer to two different physical situations, but otherwise are typical examples of the experiments performed. 

In Figure \ref{fig:Exp1} we observe the agglomeration process, which means that over time the particles stick together to form larger-scale structures. More specifically, in the first experiment small (single) particles are absorbed by large-scale aggregates that have formed early in the process. We refer to this setting here as {\tt Ser1}.
 
In Figure \ref{fig:Exp2} we observe the process of physical erosion of larger-scale aggregates after inducing kinetic energy into the system by shaking the test cell. The scattered particles hit the agglomerates inside the experimental cell and dissolve a larger-scale agglomerate within the observation time. Moreover, as one also might observe, towards the end of the sequence an agglomeration of scattered particles takes place forming small-size agglomerates. We refer to this setting here as {\tt Ser2}.










\section{Data Preparation and Elements of our Granulometry}

The purpose of this section is to familiarize the main elements of our granulometric analysis with respect to the data sets we consider.


Given that the data to be examined are in the form of greyscale images, whilst granulometry utilises binary images, it is necessary to perform a binarisation. This is achieved by implementing a thresholding process that determines the threshold at which grey values are assigned to the colours black and white. 
\begin{floatingfigure}[tr]{0.45\textwidth}
   \centering
   \includegraphics[width=0.22\textwidth]{2_ex2.png}
   \includegraphics[width=0.22\textwidth]{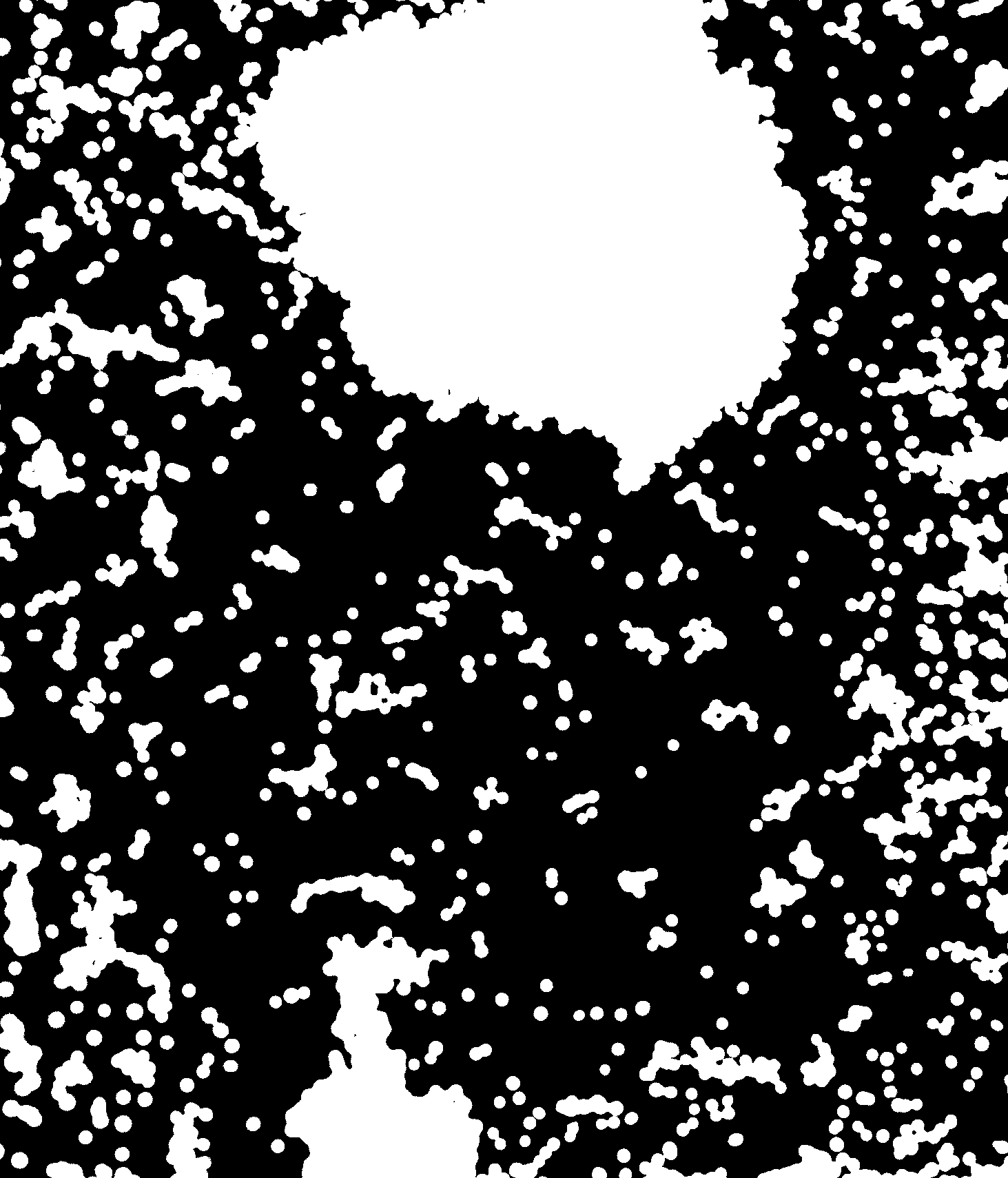} 
\caption{
     \label{fig-bin-and-inv}
     Input image and its binarised and inverted version}
\end{floatingfigure}
The approach adopted for this purpose in this paper is a method developed by Otsu \cite{ref_otsu1979}. This method involves the analysis of histograms of grey value images and the search for an optimal threshold based on the maximisation of variance between the classes of grey values that lie above and below this threshold. Otsu's method is mathematically equivalent to a globally optimal k-means clustering \cite{ref_gersh2012} applied to the image's intensity histogram. For our granulometric analysis, we then invert the images, so that {\em white} represents the particles and structures of interest and {\em black} is the background, compare Figure \ref{fig-bin-and-inv}.
\begin{figure*}[htp]
\begin{center}
\setlength{\tabcolsep}{1mm}  
\renewcommand{\arraystretch}{1}  
\begin{tabular}{llllll}
\includegraphics[width=0.9cm]{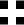} 
\hspace{2mm}
&\includegraphics[width=1.1cm]{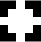} 
\hspace{2mm}
&\includegraphics[width=1.3cm]{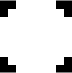}
\hspace{2mm}
&\includegraphics[width=1.5cm]{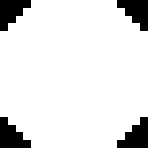}
\hspace{2mm}
&\includegraphics[width=1.7cm]{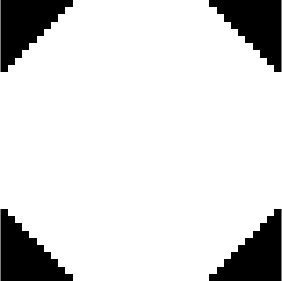}
\hspace{2mm}
&\includegraphics[width=1.9cm]{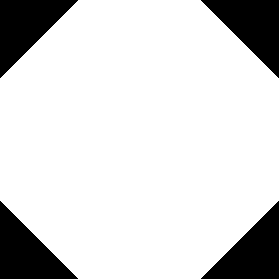} 
\end{tabular}
\vspace{-2.0ex}
\caption{\label{fig:SE} {\tt SE} visualization with  $r=1, 2, 5, 10, 20, 140$. }
\end{center}
\end{figure*}
In the implementation of the opening that generates our granulometry, we employ the disc-shaped {\tt SE} defined by the MATLAB command {\tt strel("disk",r)} where r specifies the radius, see Figure \ref{fig:SE} for a visualization.

\begin{figure*}[hp]
\begin{center}
\setlength{\tabcolsep}{1mm}  
\renewcommand{\arraystretch}{1}  
\begin{tabular}{lllllll}
\includegraphics[width=1.7cm]{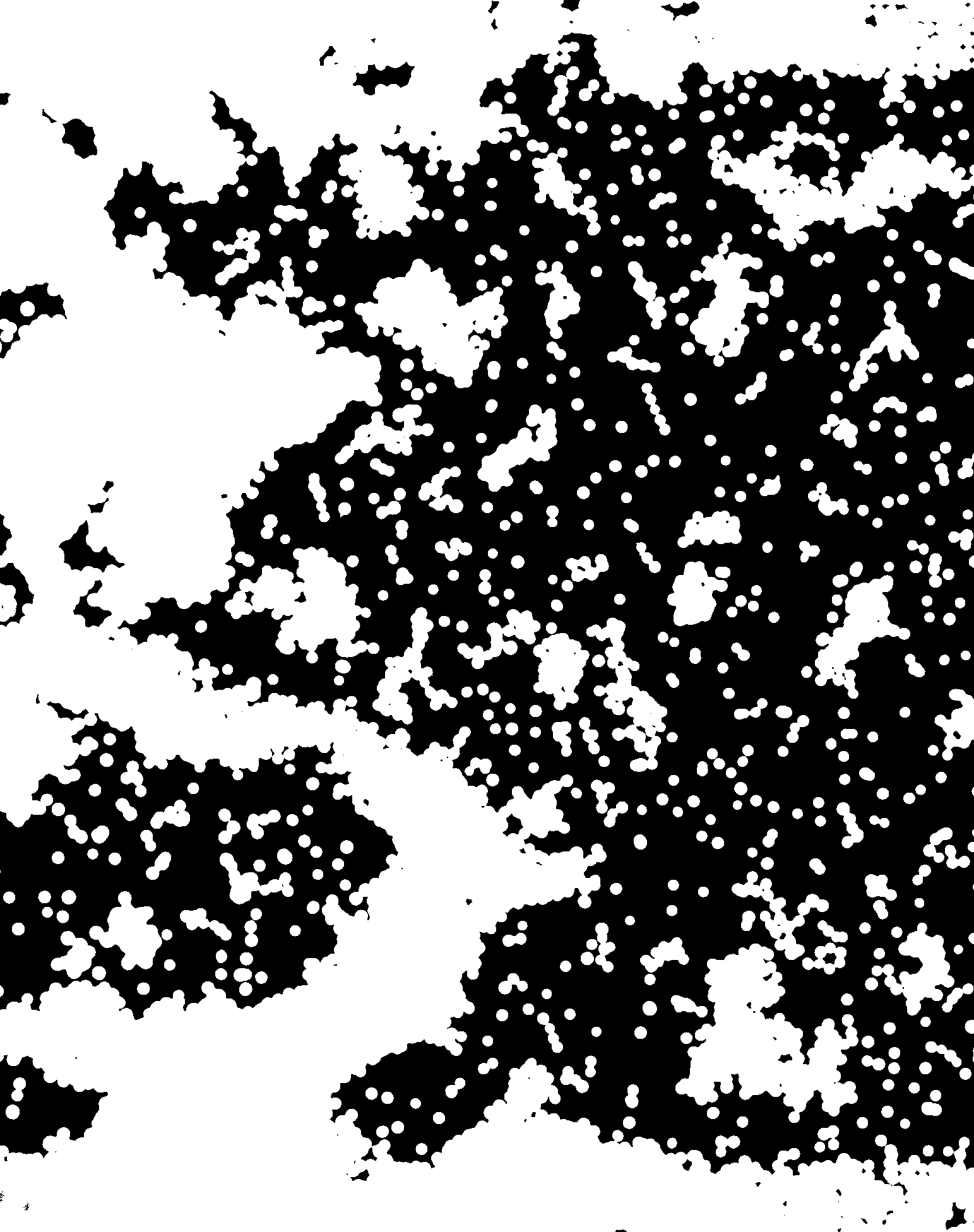} &
\includegraphics[width=1.7cm]{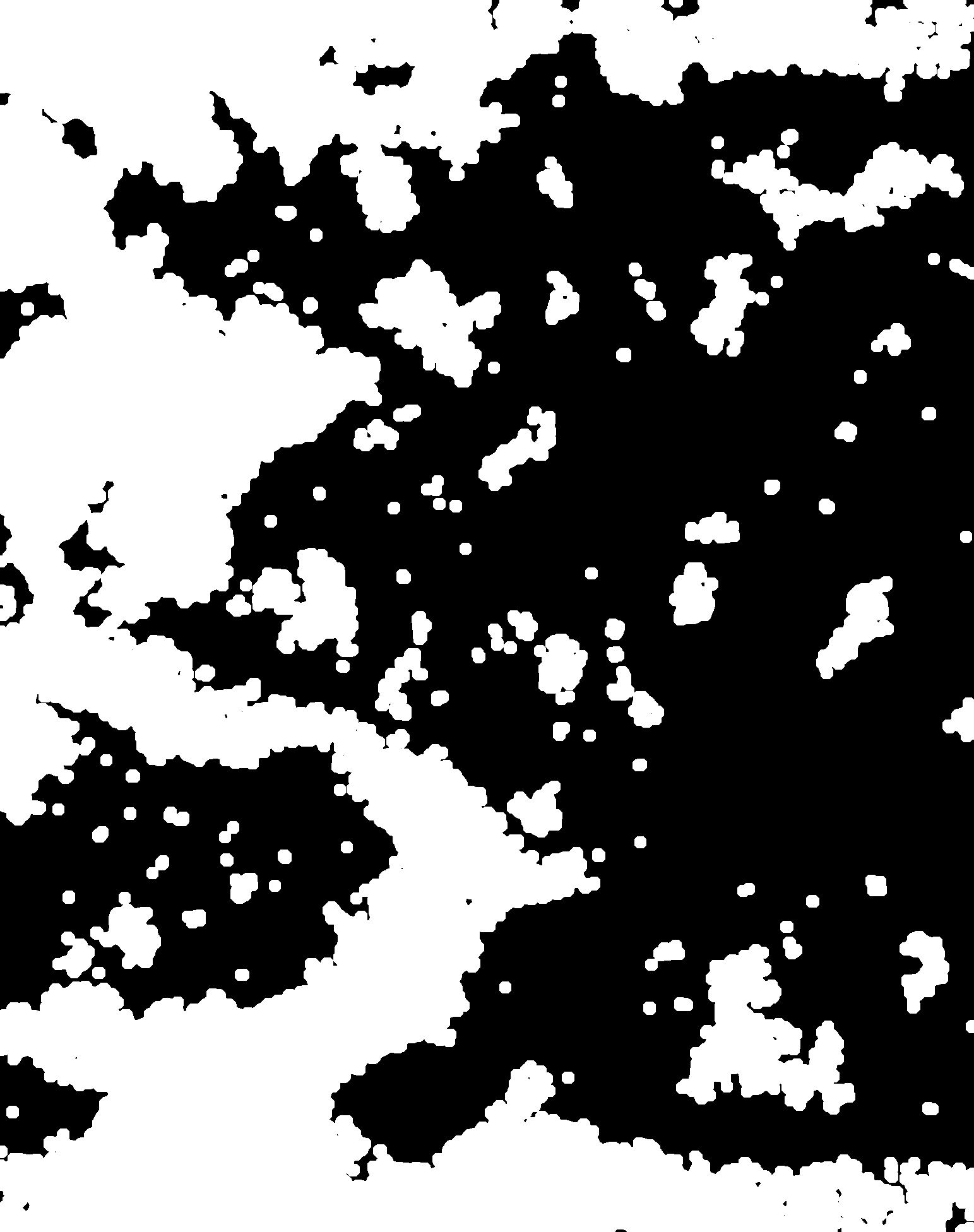} &
\includegraphics[width=1.7cm]{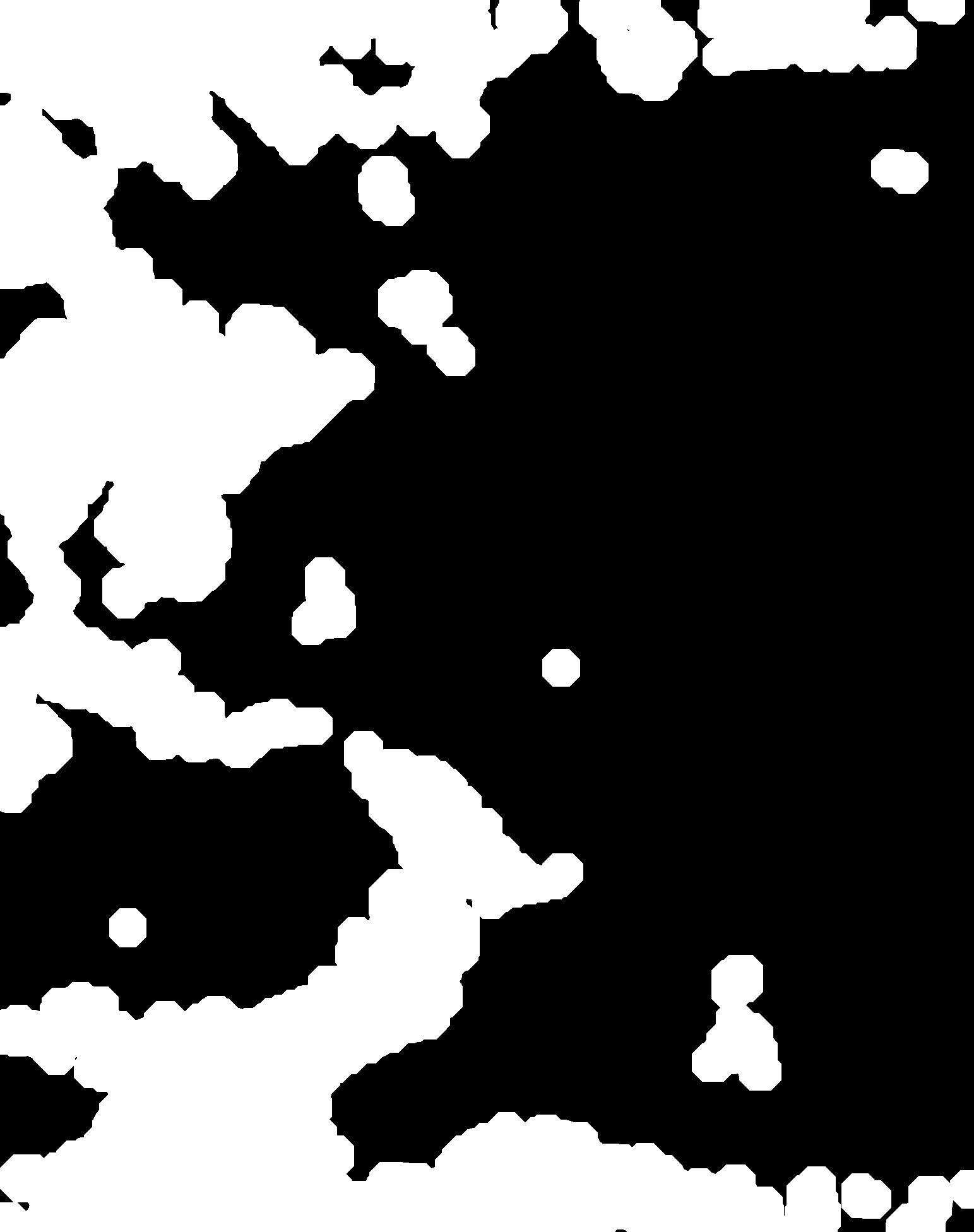} &
\includegraphics[width=1.7cm]{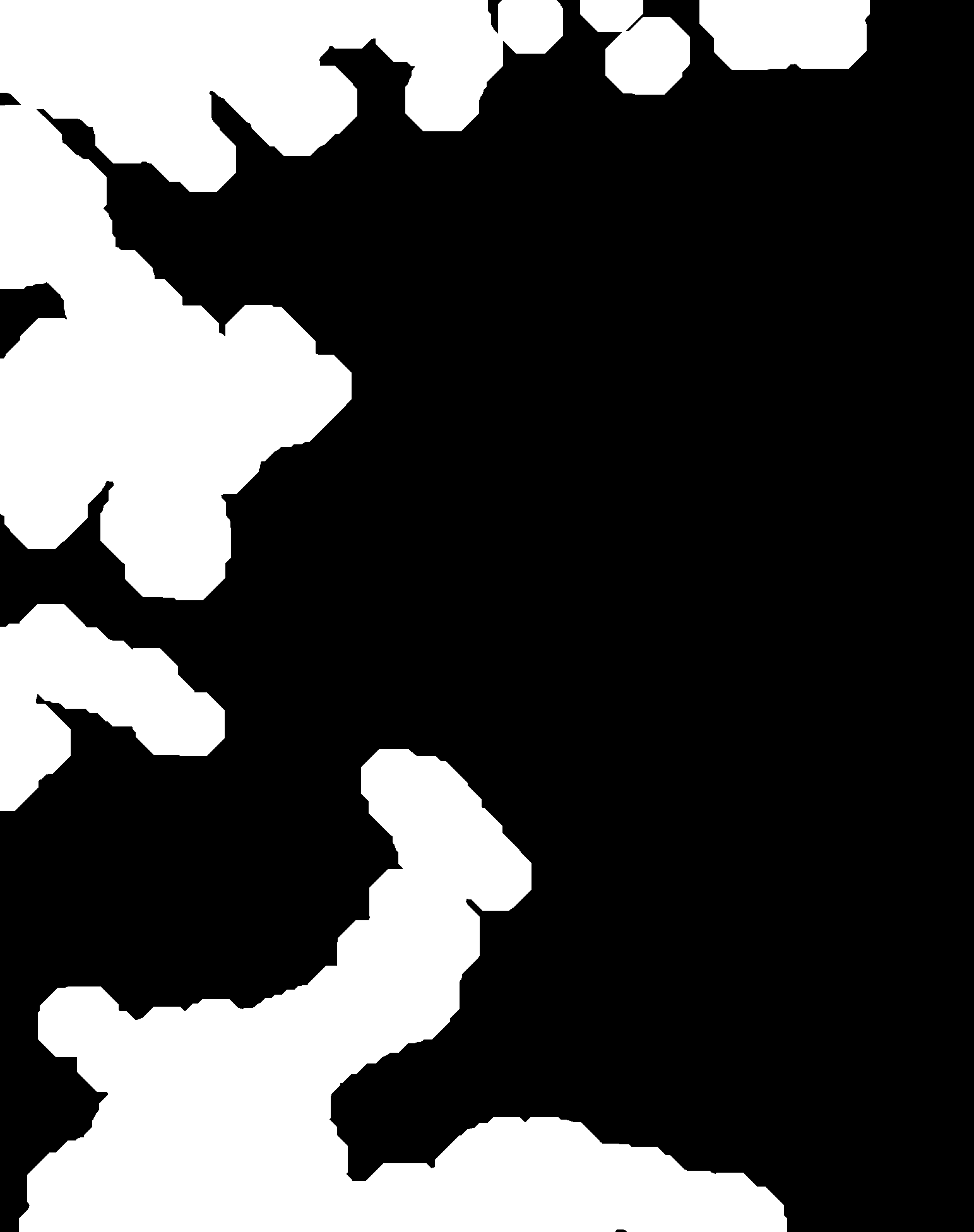} & 
\includegraphics[width=1.7cm]{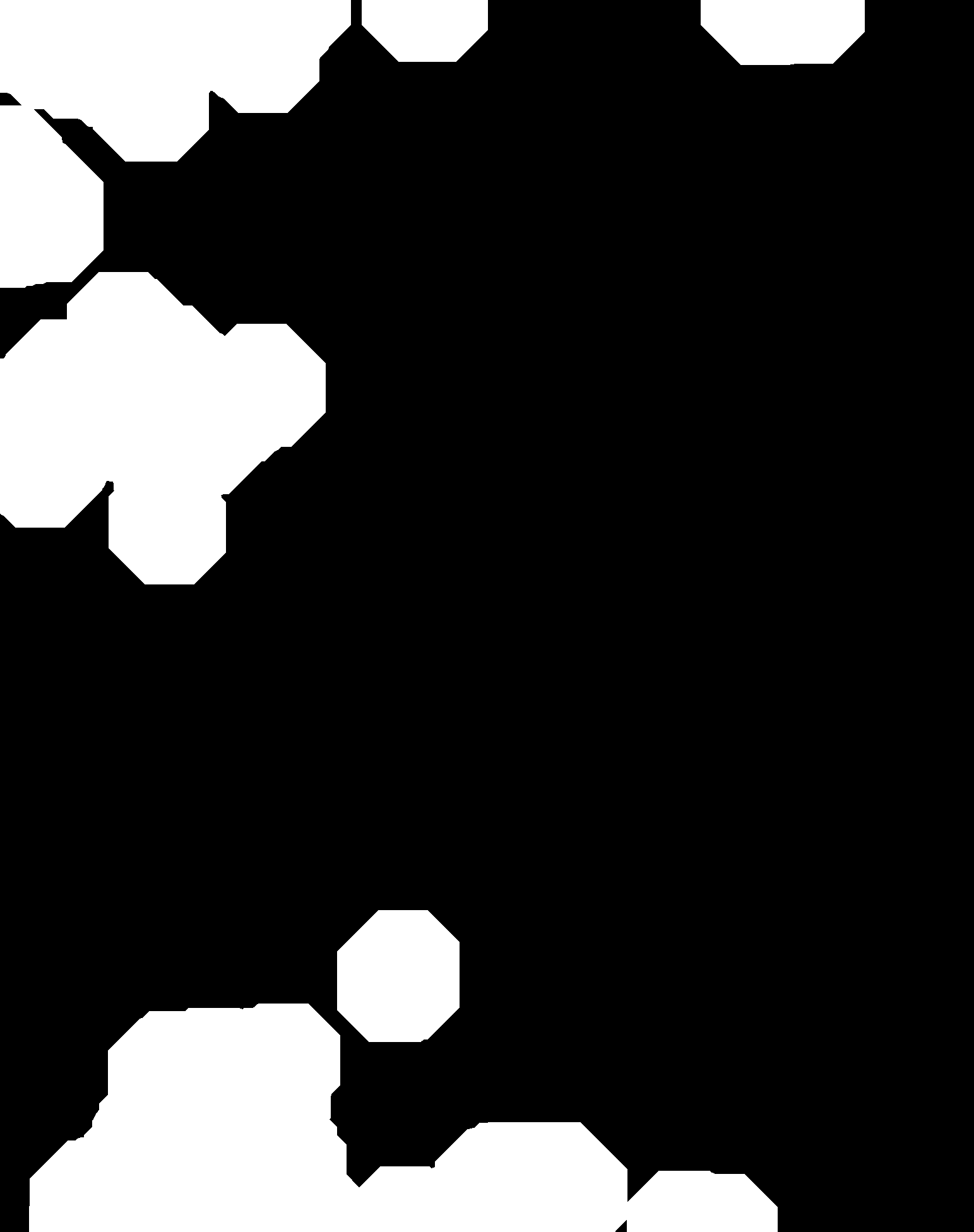} &
\includegraphics[width=1.7cm]{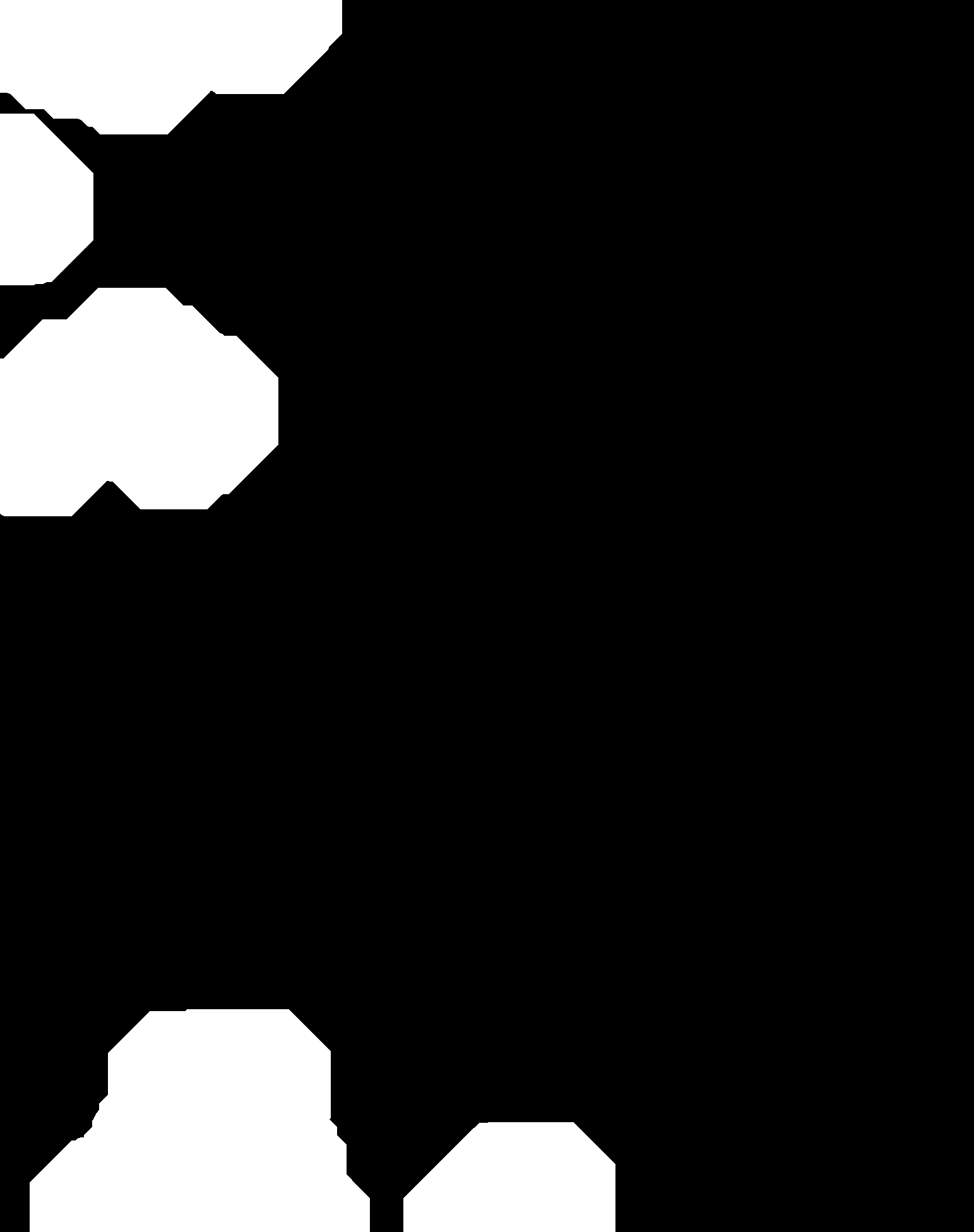}
\\
\includegraphics[width=1.7cm]{Ex2-L300-op-0.png} & 
\includegraphics[width=1.7cm]{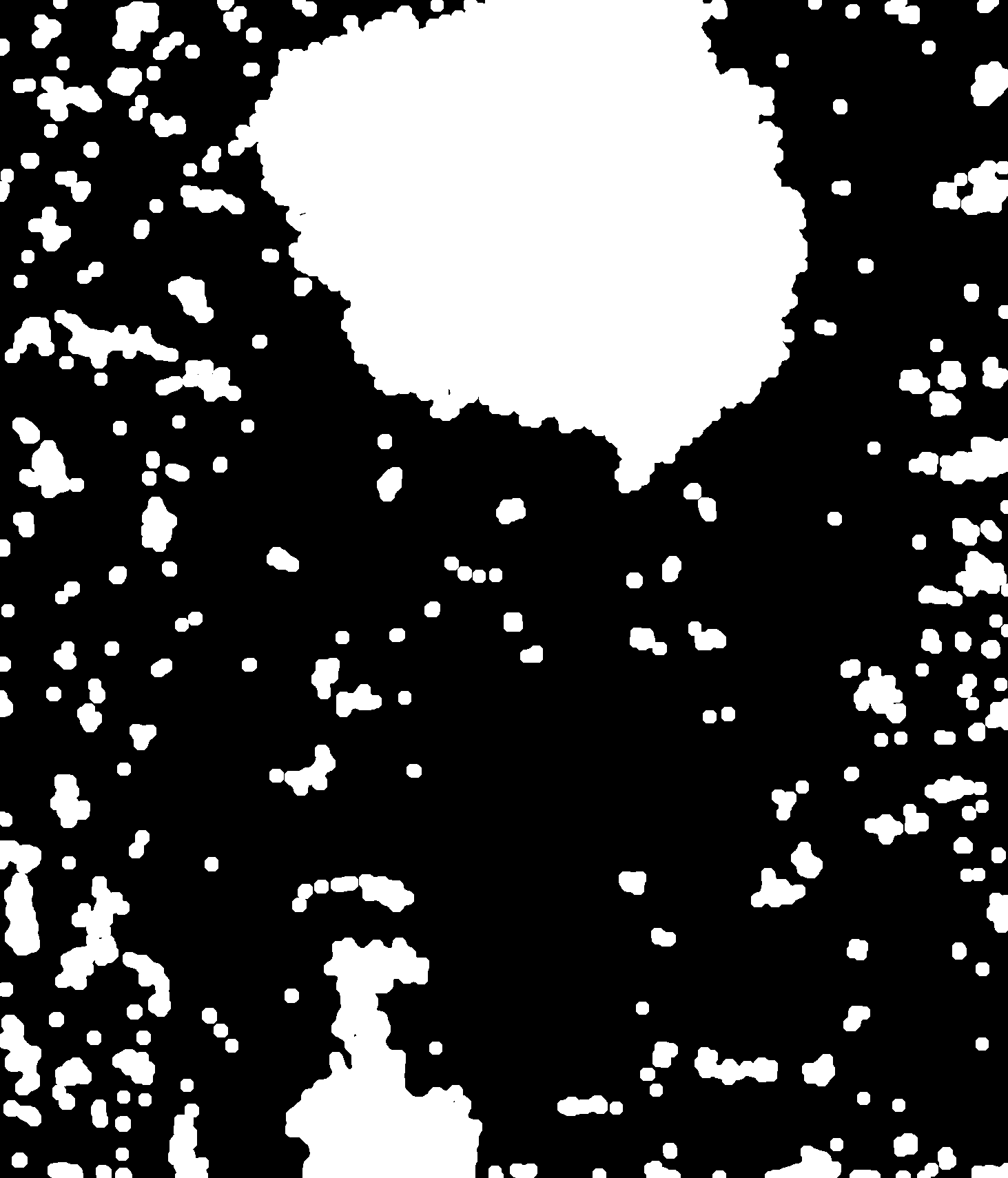} &
\includegraphics[width=1.7cm]{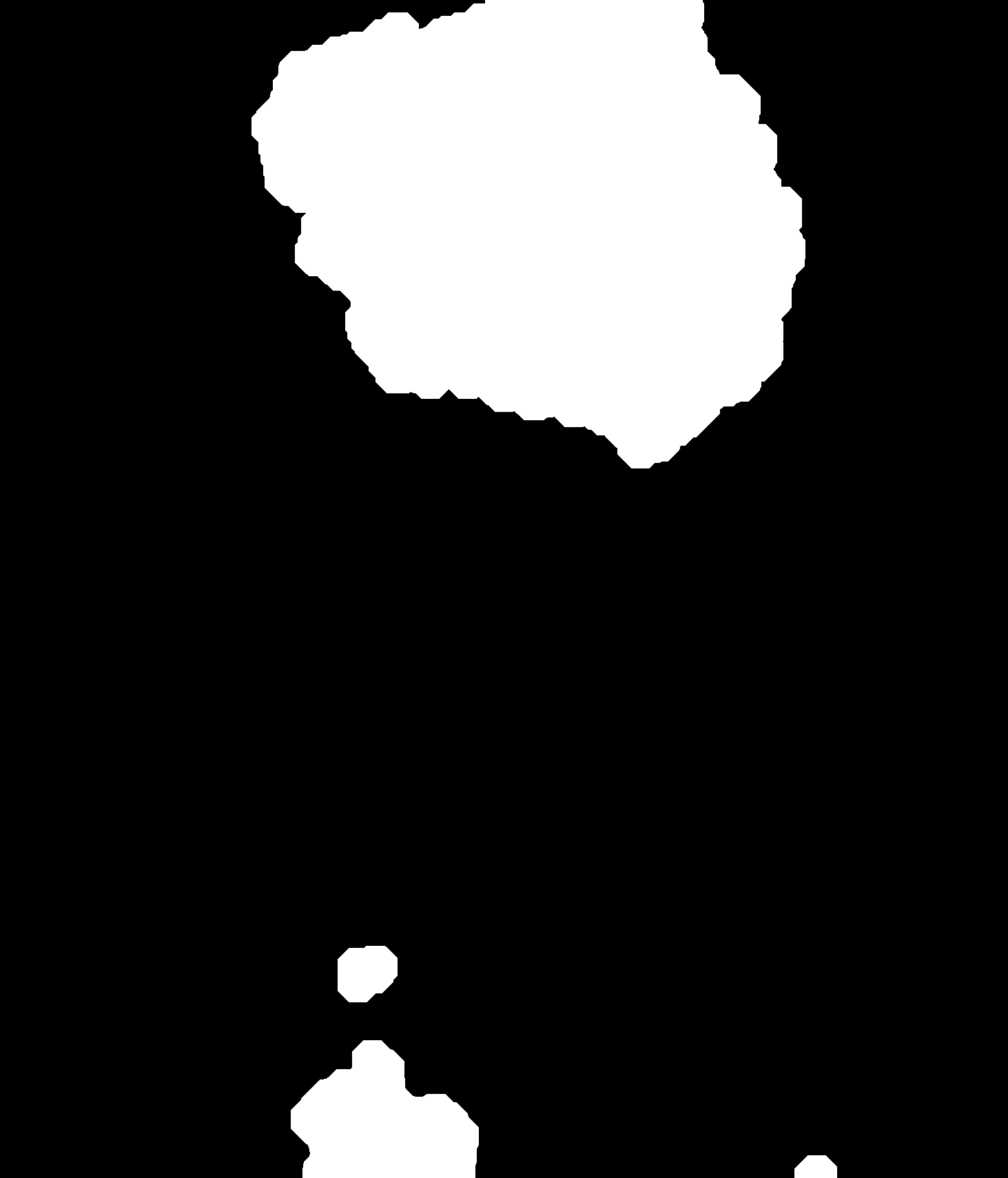} &
\includegraphics[width=1.7cm]{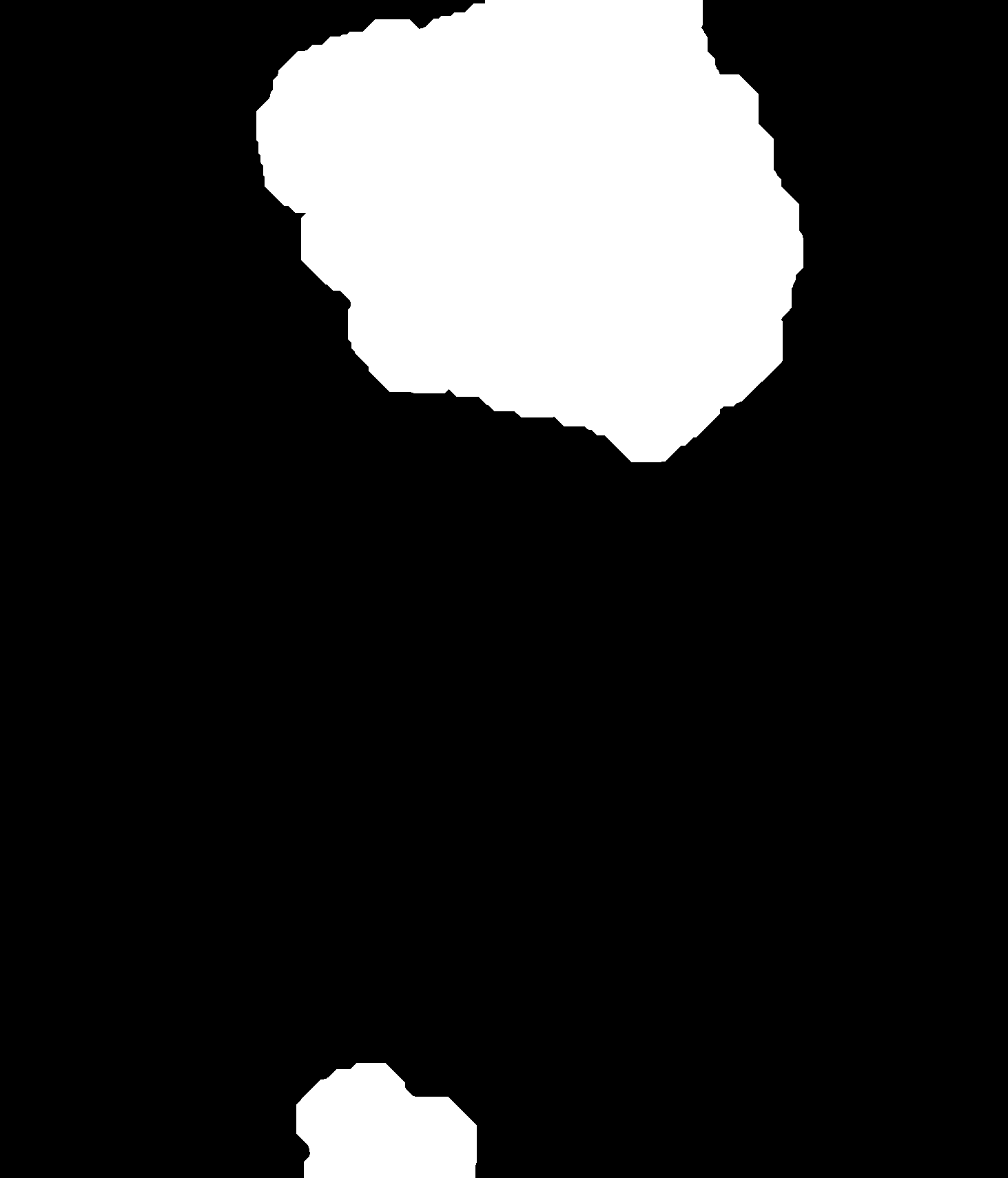} &
\includegraphics[width=1.7cm]{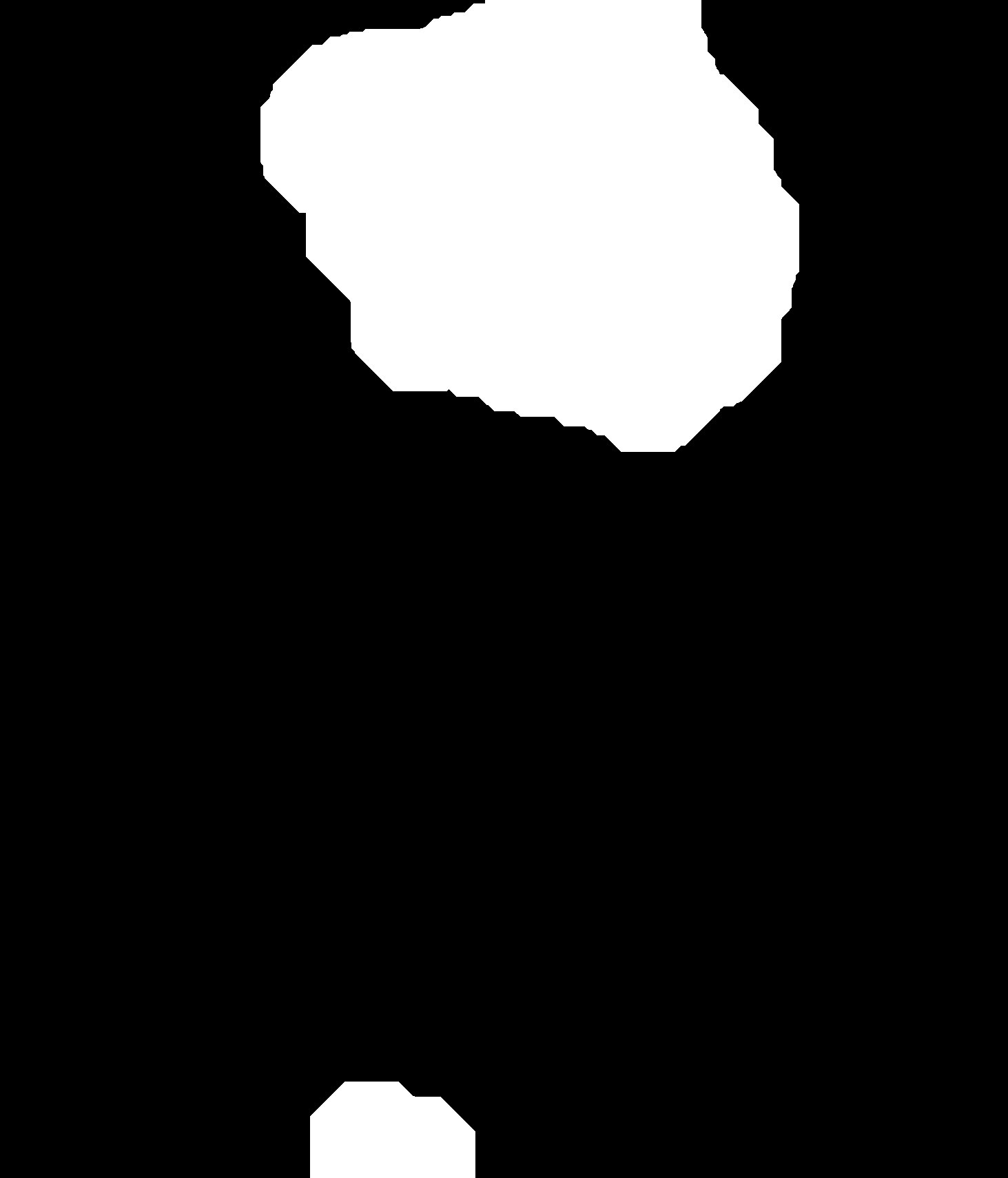} & 
\includegraphics[width=1.7cm]{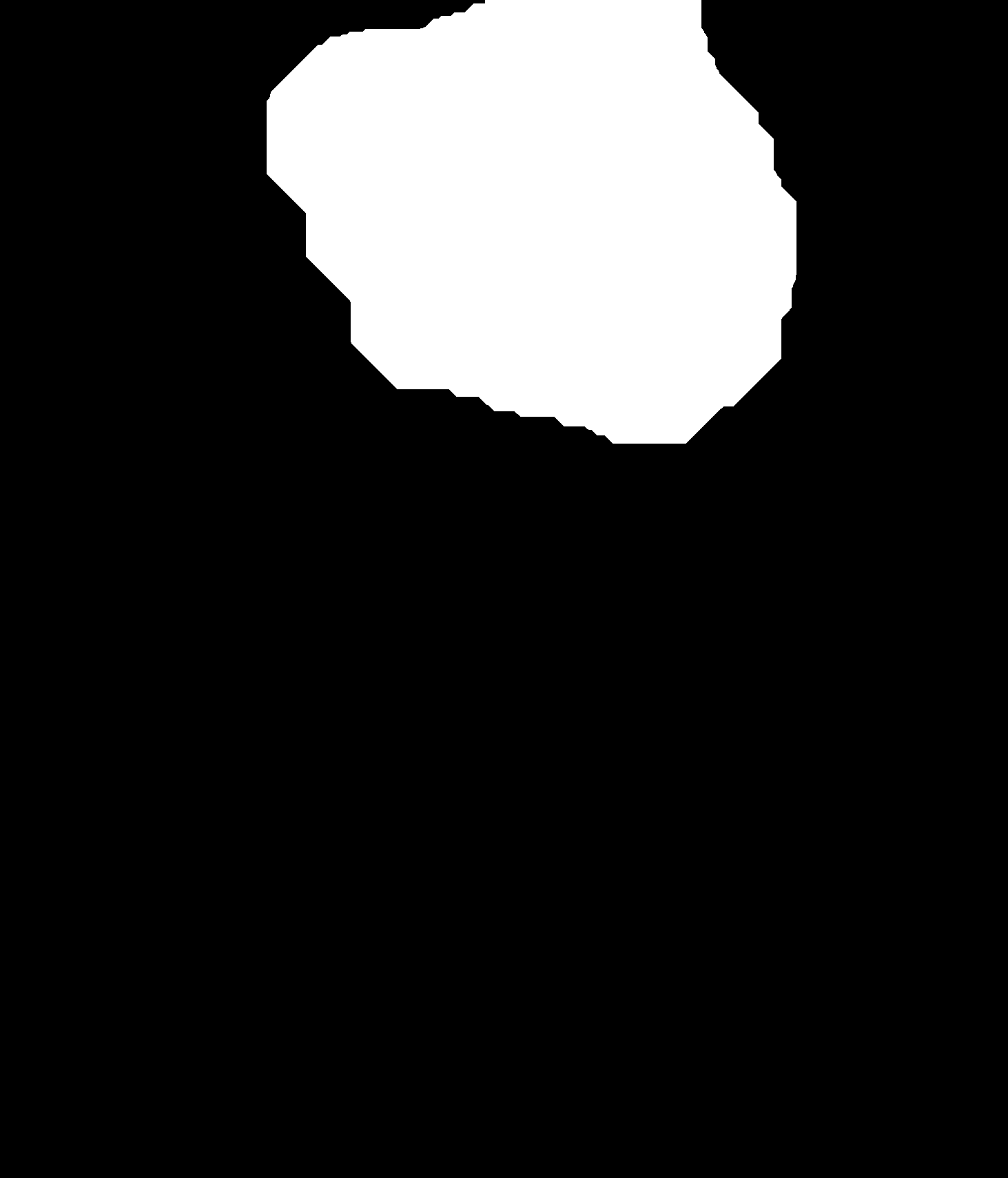}\\ 
\end{tabular}
\caption{\label{fig:Opn}Opening visualization with {\tt {SE}}$=10, 30, 50, 90, 120$. First row: frame $8$ of Figure \ref{fig:Exp1} with corresponding opened images via {\tt {SE}}. Second row: frame $2$ of Figure \ref{fig:Exp2} with corresponding opened images via {\tt {SE}}.}
\end{center}
\end{figure*}

Figure \ref{fig:Opn} gives an account of the {\em opening} of two binarised input images taken from the two image series for our application. Here we would like to remark that the size of structures related to the {\tt SE} in use is important here.
 As expected, one observes a gradual simplification of the input image. 

\begin{floatingfigure}[tr]{0.54\textwidth}
   \centering
   \includegraphics[width=0.2\textwidth]{2_ex2.png}
\includegraphics[width=0.31\textwidth]{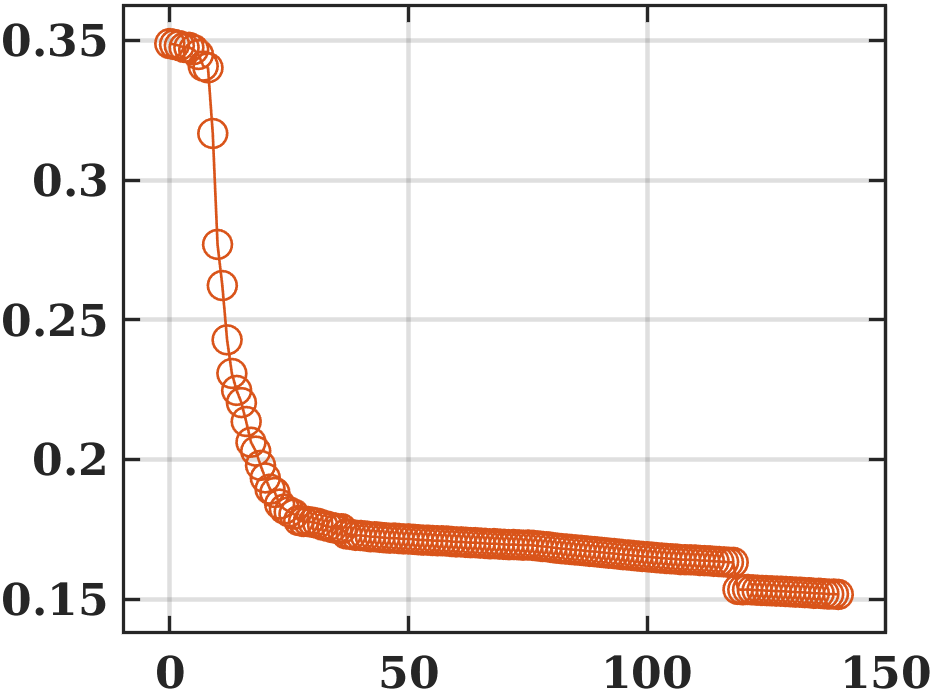}
\caption{
     \label{fig-org-and-int-plot}
     Input image and intensity distribution for opening with increasing {\tt SE} size}
\end{floatingfigure}
By image acquisition, the resolution may vary in practice from image sequence to image sequence, while for one and the same image sequence it is convenient to keep it fixed.
For an automated analysis, it is desirable to make comparable quantitative measures obtained from the imagery. Therefore, we normalize the image intensity measure which is a main quantity in granulometry so that it covers the white portion of a given image, meaning the measure is 1 if an image is completely white. 
See Figure \ref{fig-org-and-int-plot} for an account of a normalized image intensity plot that shows how the normalized intensity develops when applying an opening with increasing size of the {\tt SE}.

Concluding, our first discussion of the granulometry in our application, we turn to the filtering and identification of individual particles in an image of our series. The relative amount of such free particles and its temporal evolution is an important feature of the experiment series and allows us to assess the physical processes in an automated way. To this end, it is imperative to know at which scales of the {\tt SE} the particles are observable.

\begin{figure*}[htp]
\setlength{\tabcolsep}{1mm}  
\renewcommand{\arraystretch}{1}  
\begin{tabular}{cccc}
\includegraphics[width=2.75cm]{8_ex1.png} &\includegraphics[width=2.75cm]{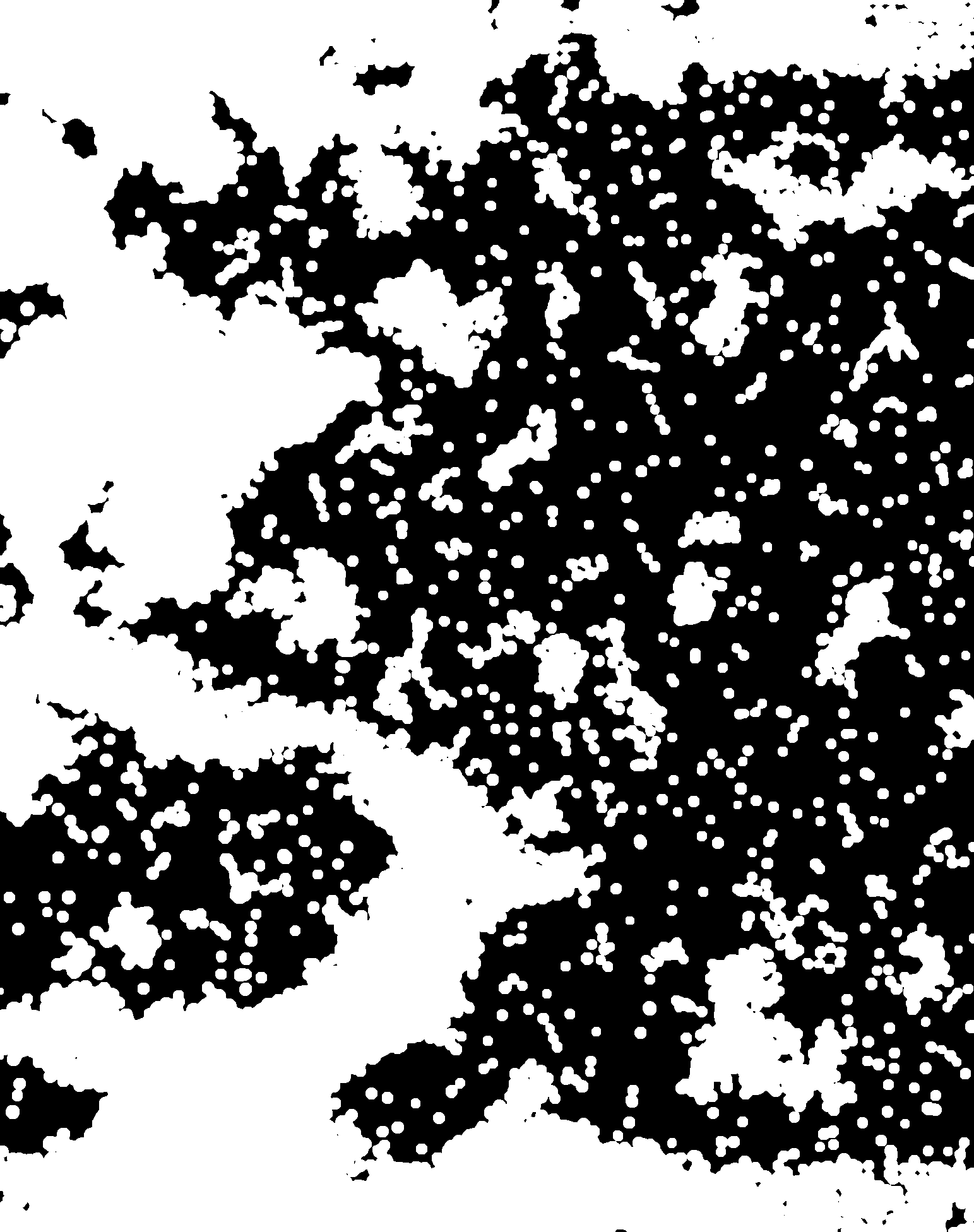} &\includegraphics[width=2.75cm]{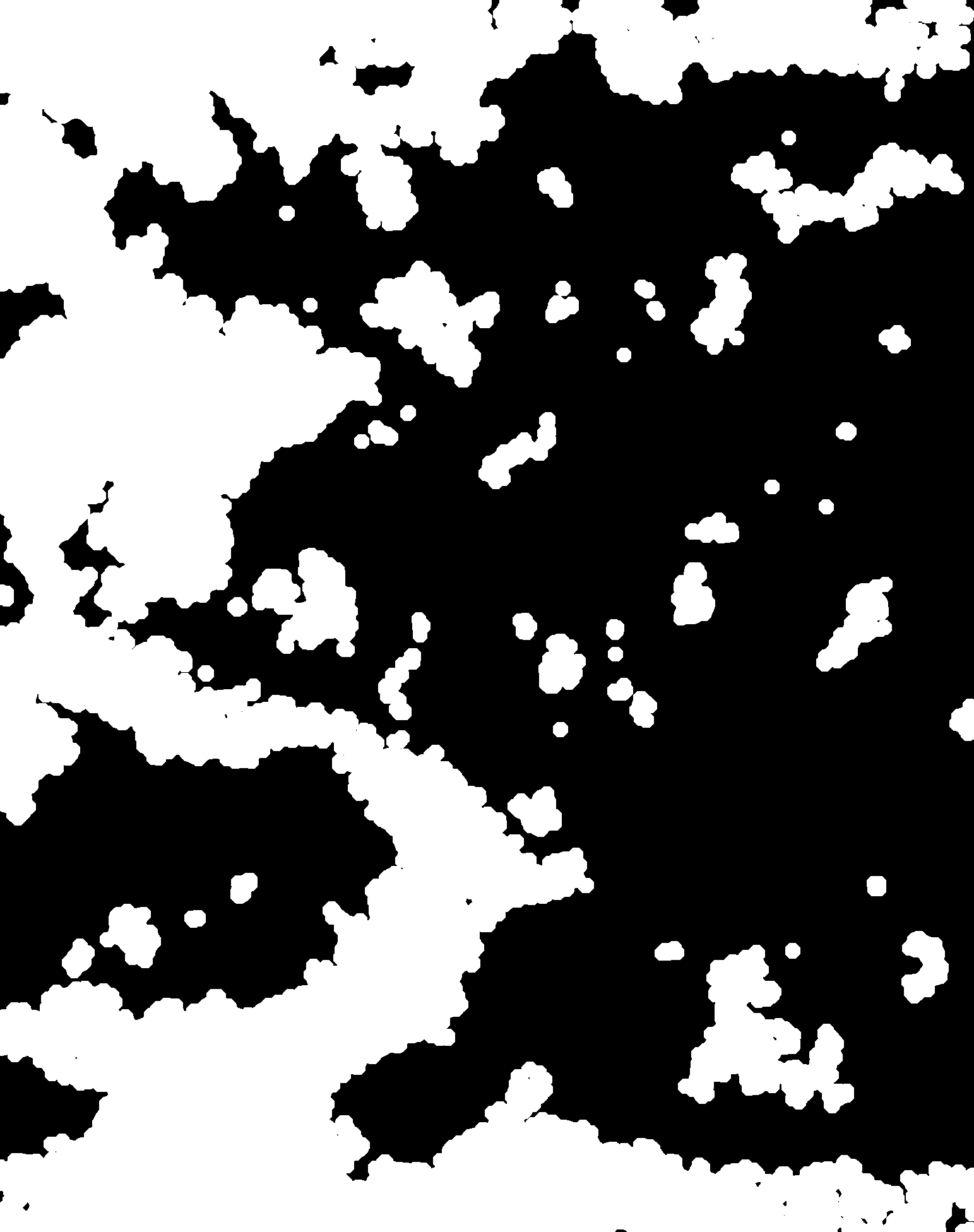}  
&\includegraphics[width=2.75cm]{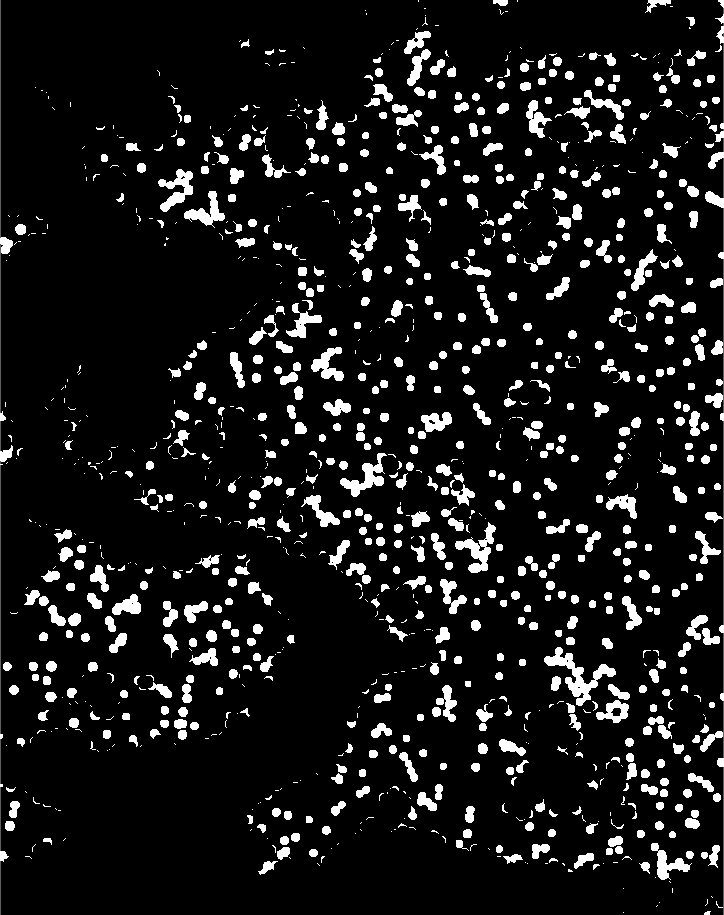}\\
\end{tabular}
\caption{\label{fig:Op-Dif} From left to right: Frame $8$ of Figure \ref{fig:Exp1}, opened images after binarisation and inversion with ${\tt {SE}}=6, 12$, respectively, and difference of these opened images. We observe that the free particles can be found to a large degree between the depicted {\tt SE} scales.} 
\end{figure*}

In Figure \ref{fig:Op-Dif} we give an account of our findings. It is important to note that, in general, the experimental cell and the camera observation point will stay the same for both of our series (and others). So, the setting for identification of the particles can be fixed for both series. As a main result for further use, the free particles can be filtered employing {\tt SE} scales from six to twelve.

\section{Experimental Results}
In this section, we present several experiments on images from both series shown in Figure \ref{fig:Exp1-Exp2}. 


\subsection{Experiment 1: Consistency with Physical Theory\label{sec:exp1}}

Observations in the Solar System (e.g. the asteroid belt) and laboratory experiments give strong empirical evidence that fragmentation processes lead to a characteristic size distribution of particles, typically following a power law. \cite{ref_deckers2014,ref_bottke2015}. The series {\tt Ser2} shows a similar fragmentation process, as small (single) particles erode larger agglomerates until they finally vanish. 
The hypothesis we follow now is that we should find similar particle statistics by 2D imagery, although the empirical evidence in the literature is found in 3D space and a more continuous scale of particle sizes. This is apparently not self-evident since the 2D imagery we evaluate is naturally subject to occlusions that hinder measurements and by the restrictions of image resolution and binarisation. If the size distribution accessible in our imagery still fits well with a power-law assumption, we could be highly confident that our granulometric analysis may convey physically meaningful interpretations. 
To this end we study if we can fit the power law 
\begin{equation}
    y=ax^b+c \label{eq:powerlaw}
\end{equation}
to our normalized image intensity distributions for increasing {\tt SE} size, an example of which we have already seen in Figure \ref{fig-org-and-int-plot}.

\begin{figure*}[hbp]
\begin{center}
\setlength{\tabcolsep}{1mm}  
\renewcommand{\arraystretch}{1}  
\begin{tabular}{lll}
\includegraphics[width=2.5cm]{2_ex2.png} &
\includegraphics[width=2.5cm]{Ex2-L300-op-0.png} &
\includegraphics[width=4.1cm]{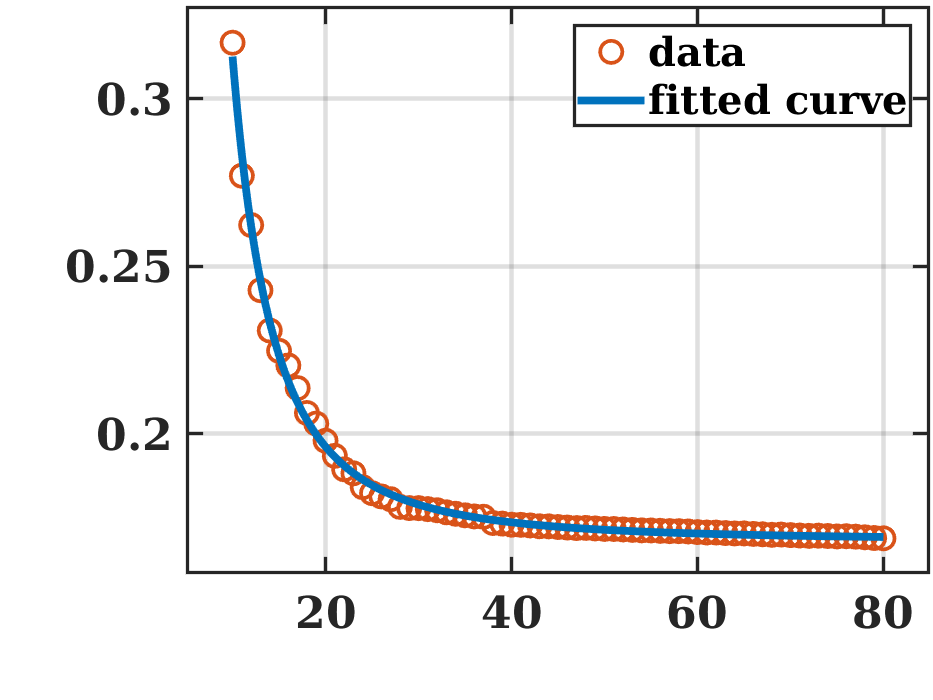} \\
\includegraphics[width=2.5cm]{8_ex2.png} &
\includegraphics[width=2.5cm]{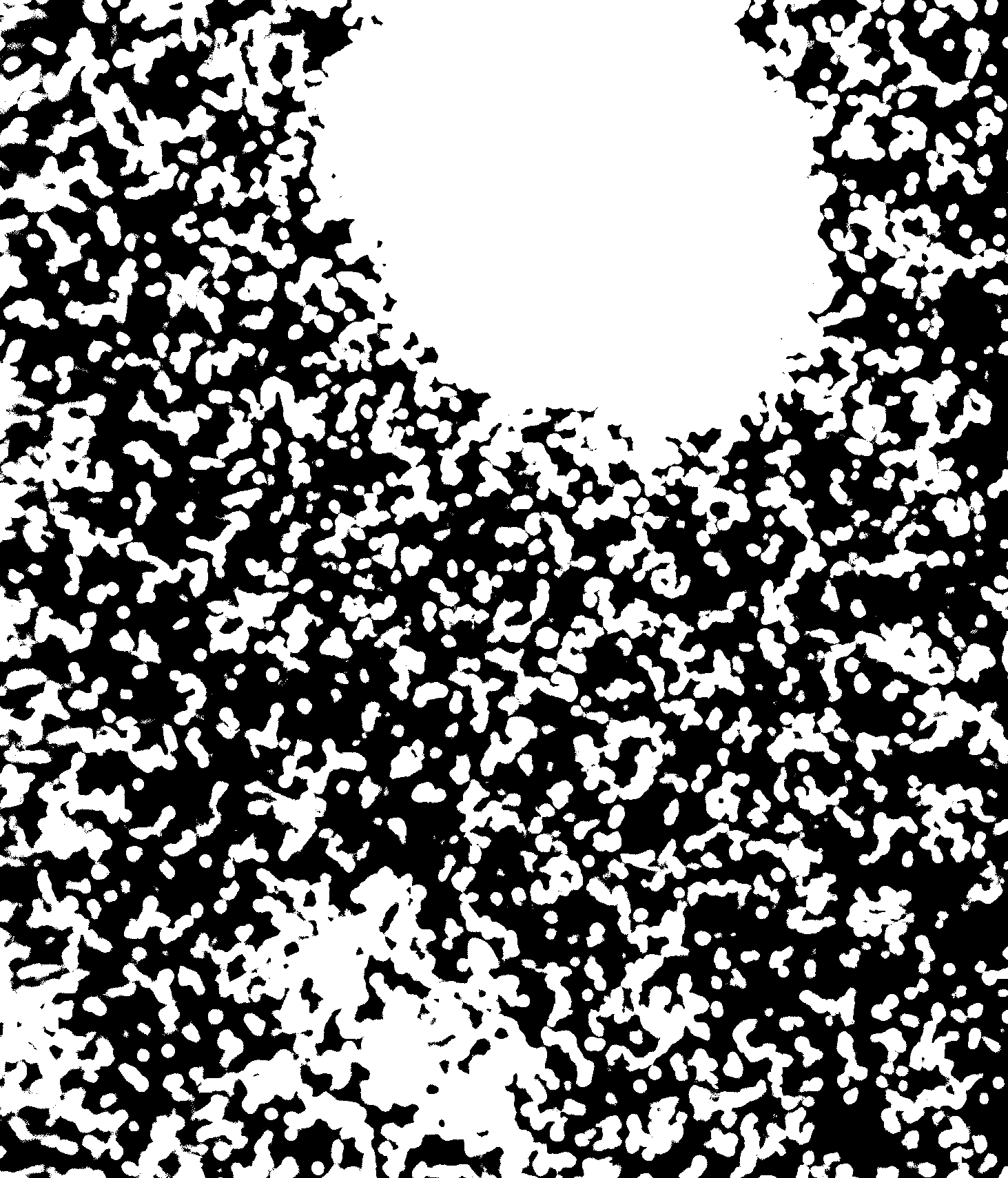} &
\includegraphics[width=4.1cm]{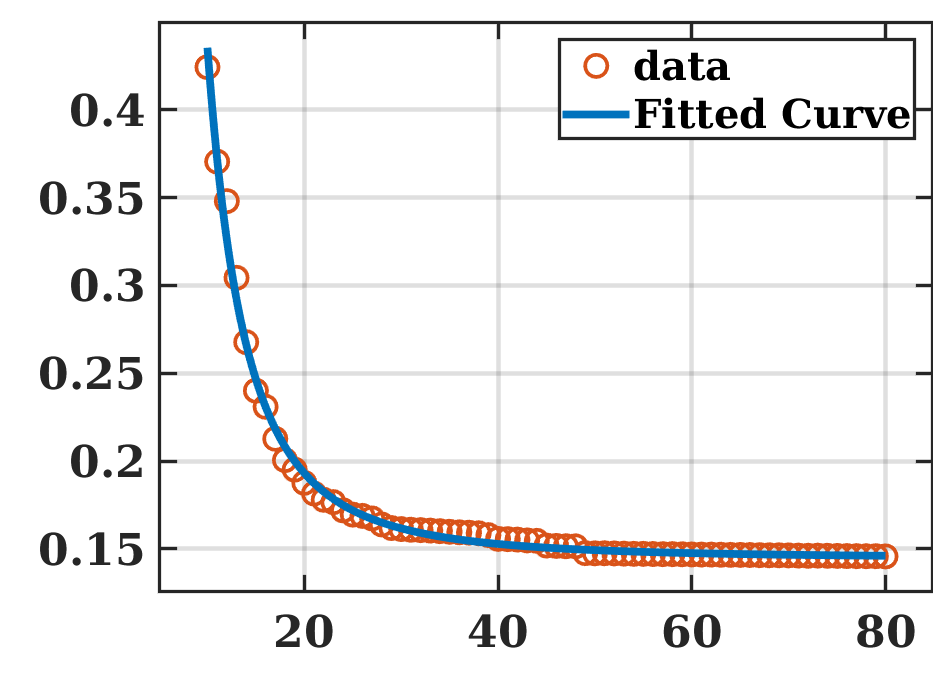}
\end{tabular}
\caption{\label{fig:PowLaw}
Frames $2, 8$ of {\tt Ser2}, first and second row, respectively, with corresponding inverted-binarized versions and intensities (red marker) and fitted power law (solid blue line) for opening process with ${\tt {SE}}=10$-$80$. 
}
\end{center}
\end{figure*}

Let us comment on the definition and usage of the power law \eqref{eq:powerlaw}, as it is an adoption of the standard power law form $y=ax^b$. As we may already observe in Figure \ref{fig-org-and-int-plot}, and also by means of Figure \ref{fig:Op-Dif}, filtering with low {\tt SE} sizes (especially of one to five) does not yield information about particle sizes. 
Therefore, for power-law fitting, the approximate plateau for very small {\tt SE} sizes bears no information and is skipped. Furthermore, in our finite setting of experimental cell and image resolution, the tail of our distribution inevitably approaches a non-zero value $c>0$ over interesting ranges of depicted agglomerations.
As we may observe in Figure \ref{fig:PowLaw},
the normalized intensities indeed appear to be fitted very well with a power law \eqref{eq:powerlaw}, analogous to empirical evidence in other studies. However, it is important to note that the exponent we find is different from the values found for fragmentation cascades in the literature \cite{ref_bottke2015,ref_deckers2014}.

\begin{figure*}[htp]
\setlength{\tabcolsep}{1mm}  
\renewcommand{\arraystretch}{1}  
\begin{tabular}{cc}
\includegraphics[width=4.5cm]{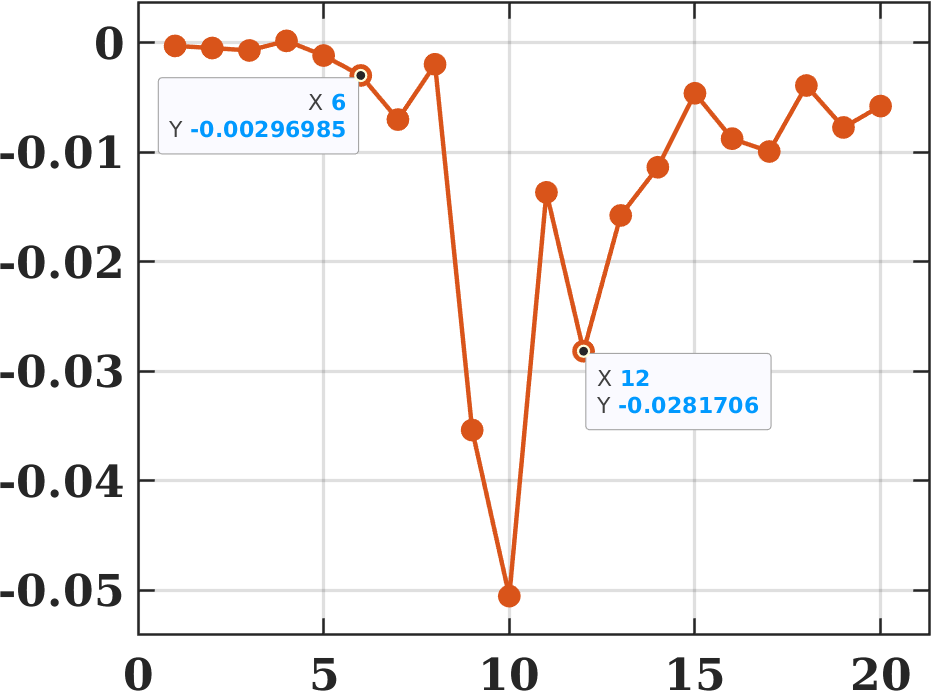} \; & \; 
\includegraphics[width=4.5cm]{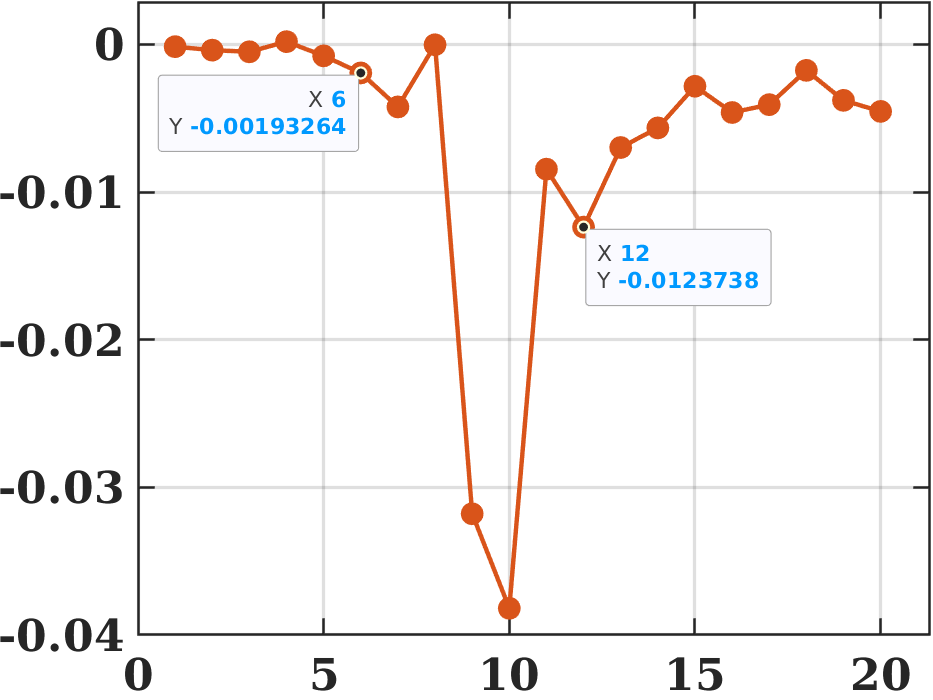}
\end{tabular}
\caption{\label{fig:Grano-1,20} First order derivatives of normalized intensities, based on (left) Frame $4$ and (right) Frame $8$ of {\tt Ser1}.}
\end{figure*}

\subsection{Experiment 2: Assessment of Free Particle Movement\label{sec:exp2}}

In this experiment we come back to the considerations leading to Figure \ref{fig:Op-Dif}. We have seen that individual particles are mainly observable making use of {\tt SE} scales six to twelve. In a first step we want to complement the discussion of Figure \ref{fig:Op-Dif} by showing the first-order derivative (computed by making use of forward differences) of the normalized intensities, compare Figure \ref{fig:Grano-1,20}.
The {\tt SE} values contributing in forming the prominent peak exactly identify a source of major loss in intensity. Evidently, the {\tt SE} range from six to twelve describes the peak, and having in mind the information carried by Figure \ref{fig:Op-Dif} we have by the peak {\em sharp} information about the free particles in an image.

\begin{figure*}[hb]
\setlength{\tabcolsep}{1mm}  
\renewcommand{\arraystretch}{1}  
\begin{tabular}{cc}
\includegraphics[width=6.0cm]{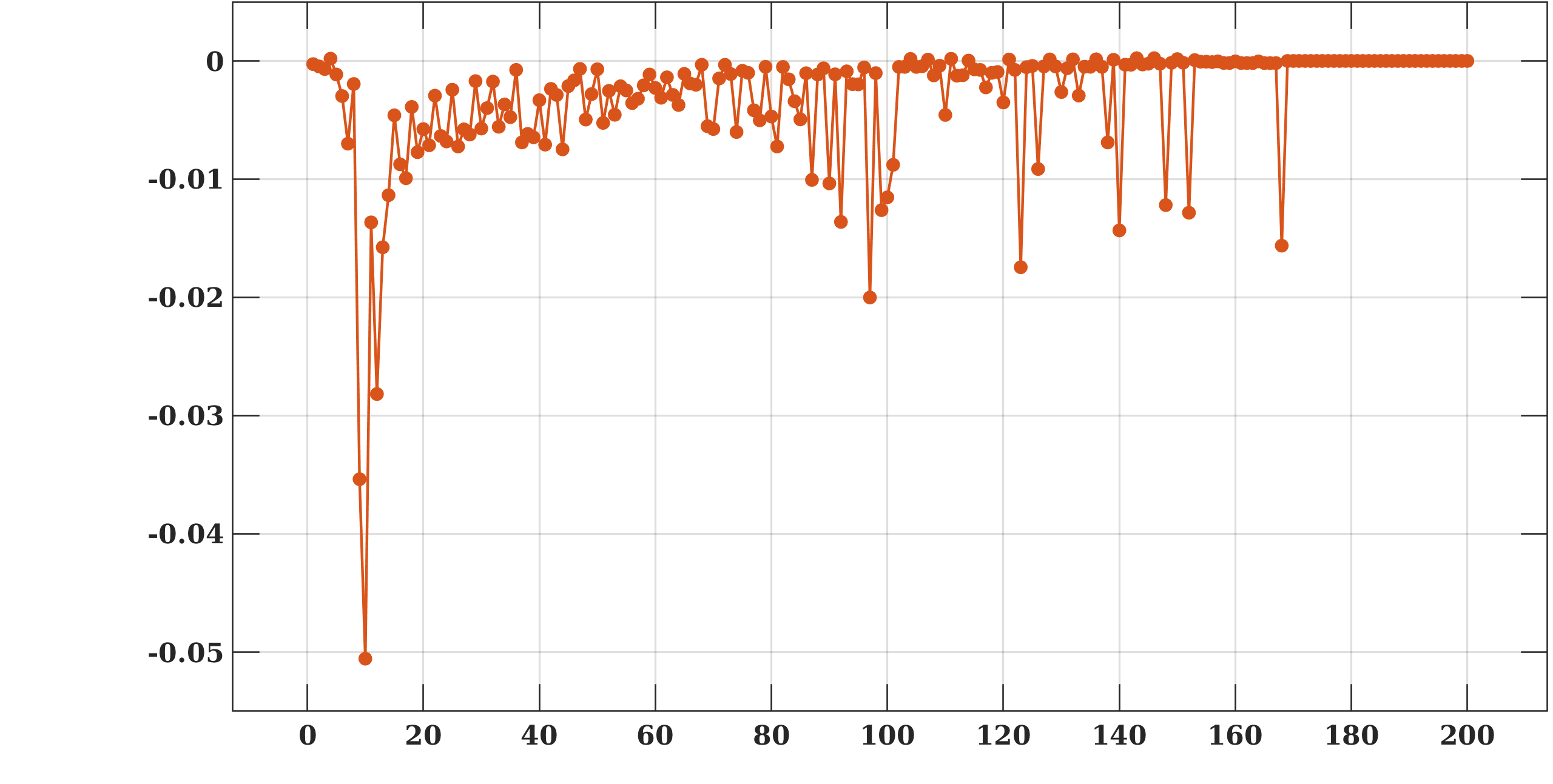} & \includegraphics[width=6.0cm]{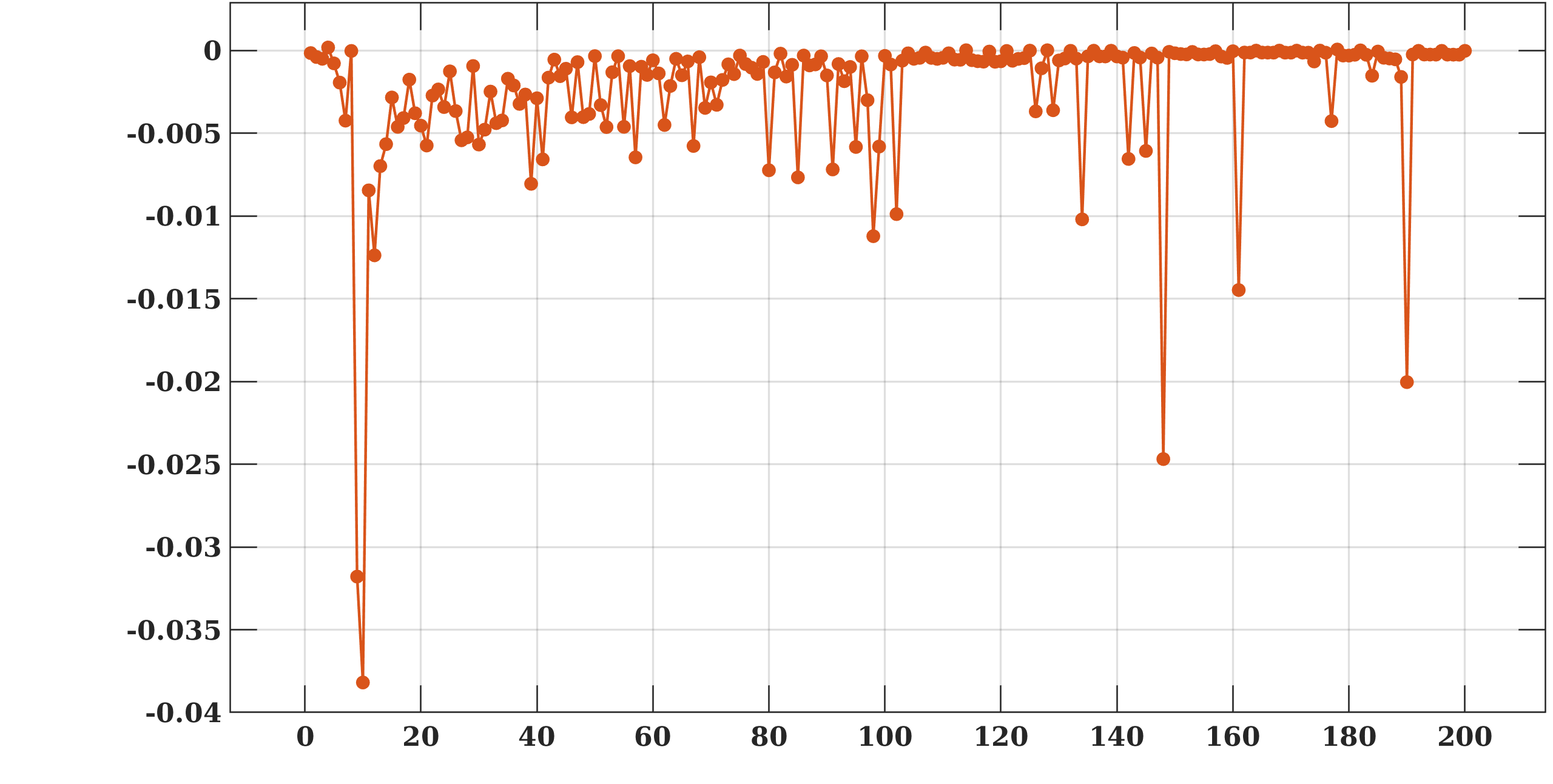}
\end{tabular}
\caption{\label{fig:Grano} First-order derivative granolumetry graphs
at a large scale of {\tt SE} sizes: (left) based on Frame $4$
and (right) based on Frame $8$ of {\tt Ser1}. Results for the other frames of {\tt Ser1} and {\tt Ser2} show analogous behaviour.}
\end{figure*}

Let us also point out that on a more global scale with respect to interesting {\tt SE} sizes, the derivative information as depicted in detail in Figure \ref{fig:Grano-1,20} contains major information in our application. This remark addresses both the range of the values of first-order derivatives as well as the issue that the largest values of the derivative can easily be assessed (for several possible uses) in an automated way, compare Figure \ref{fig:Grano}.

\begin{figure*}[htp]
\begin{center}
\setlength{\tabcolsep}{1mm}  
\renewcommand{\arraystretch}{1}  
\begin{tabular}{llll}
\includegraphics[width=5.2cm]{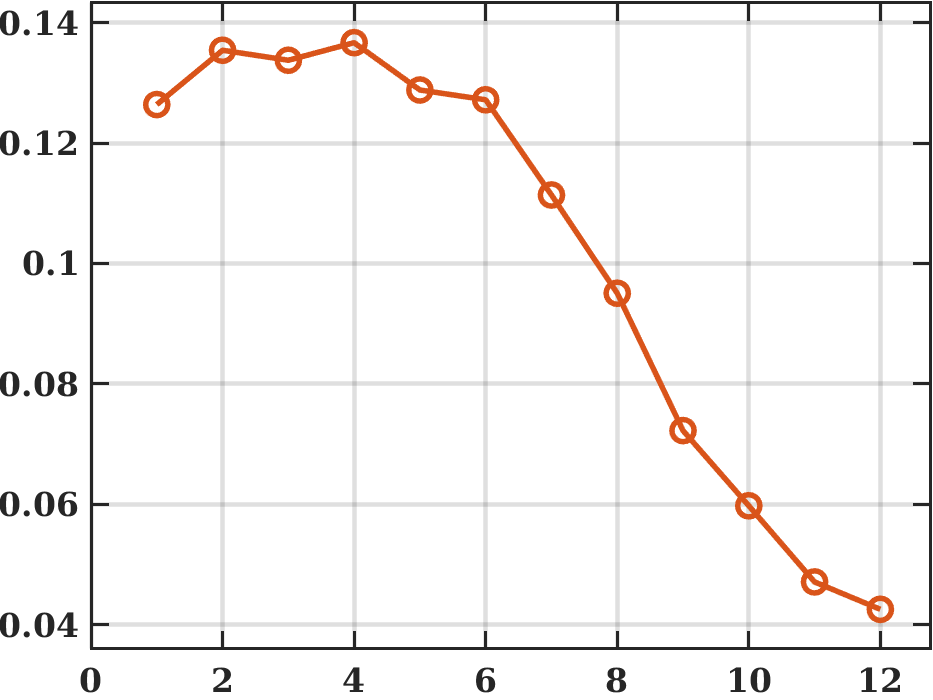} &
\includegraphics[width=5.3cm]{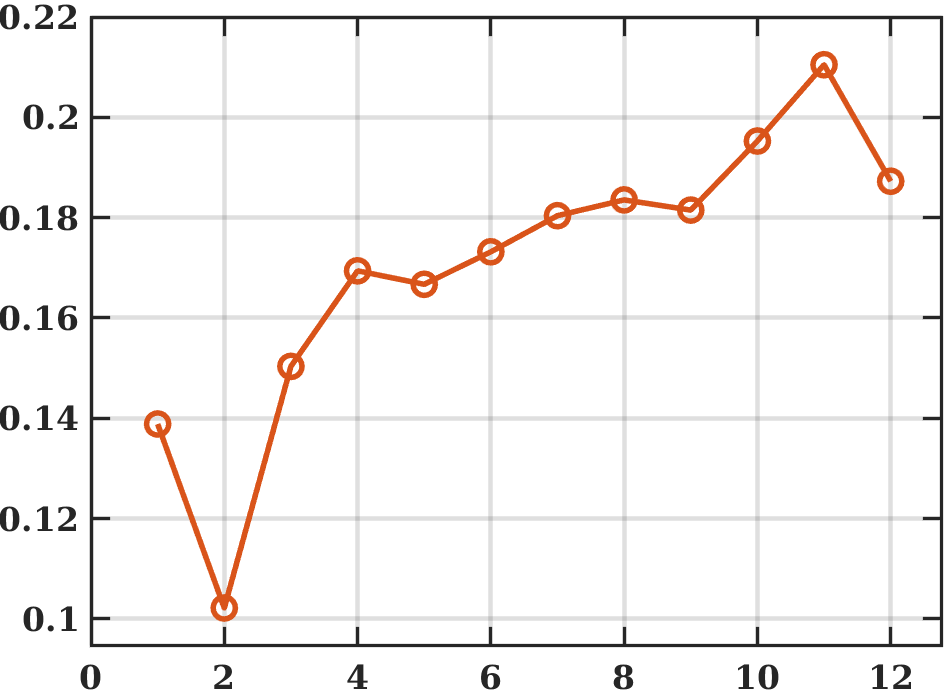} &
\end{tabular}
\caption{\label{fig:jump} Jump sizes evolving in time, for (left) {\tt Ser1} and (right) {\tt Ser2}, obtained for each frame by taking difference between normalized intensities for {\tt SE}$=6,12$.}
\end{center}
\end{figure*}

Having thus consolidated by use of granulometry that the difference between normalized image intensities after filtering with {\tt SE}$=6,12$, bear significant information about free particles in a given image, we now consider the corresponding {\em jump size} (i.e. intensity for {\tt SE}$=6$ minus intensity for {\tt SE}$=12$). 
More specifically, we are interested in the {\em temporal evolution} of jump sizes, see Figure \ref{fig:jump}, where we observe that evolution for {\tt Ser1} and {\tt Ser2}, respectively. As one may confirm visually by comparing with Figure \ref{fig:Exp1-Exp2}, we have obtained here a very characteristic, quantitative account of the significant decrease over time of the free particles in the agglomearation process given by {\tt Ser1} as well as of the increase of free particles in the fragmentation process given by {\tt Ser2}.

\subsection{Experiment 3: Assessing larger-scale structures\label{sec:exp3}}

We now aim to keep track of larger-scale image structures, such as close to the lower image boundaries in {\tt Ser2}. In that fragmentation process, the mentioned structure is dissolved over time, which is an important physical process.

\begin{figure*}[hbp]
\begin{center}
\setlength{\tabcolsep}{1mm}  
\renewcommand{\arraystretch}{1}  
\begin{tabular}{lllllll}
\includegraphics[width=1.5cm]{Ex2-L300-op-30.png} &
\includegraphics[width=1.5cm]{Ex2-L300-op-30.png} &
\includegraphics[width=1.5cm]{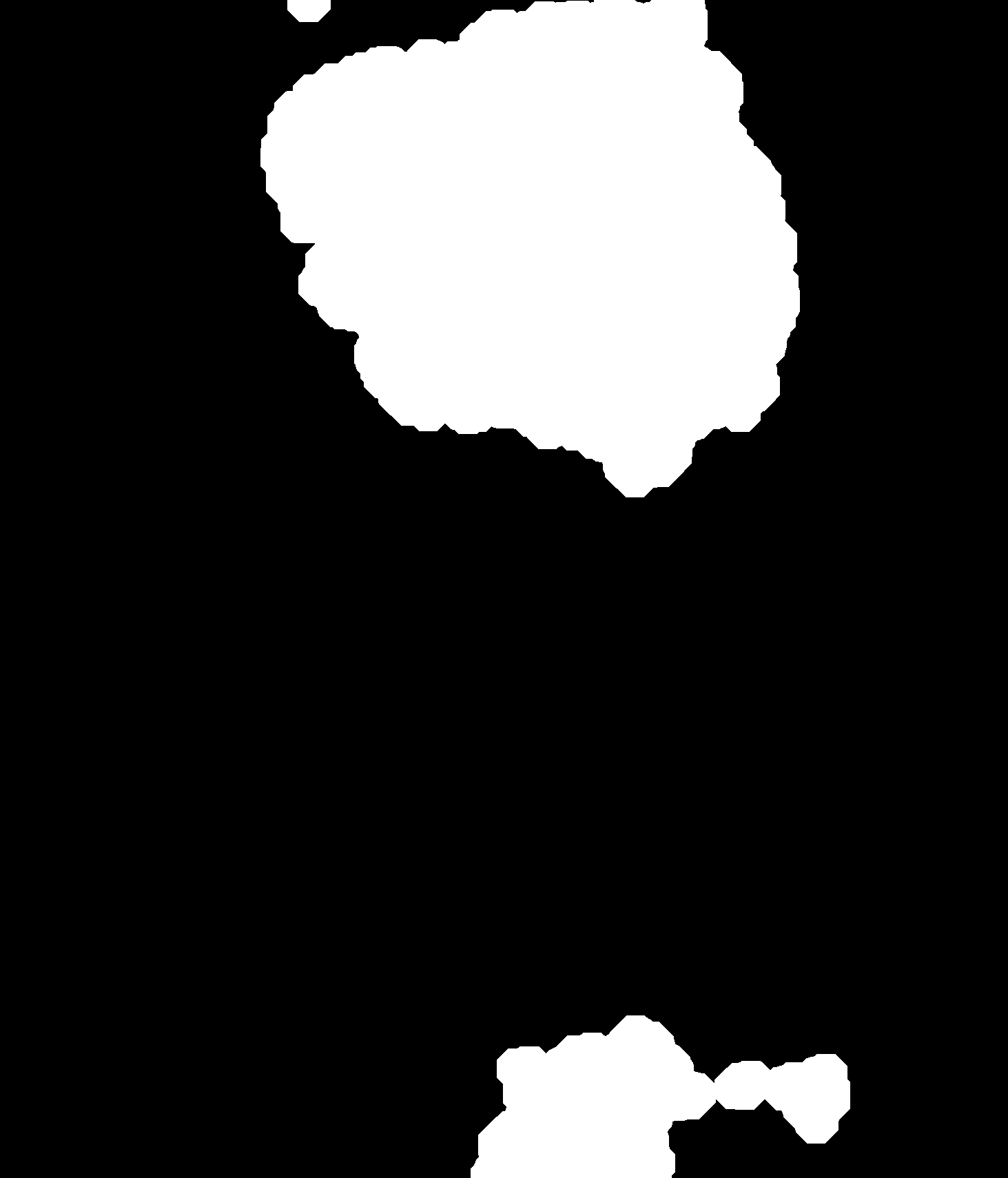} &
\includegraphics[width=1.5cm]{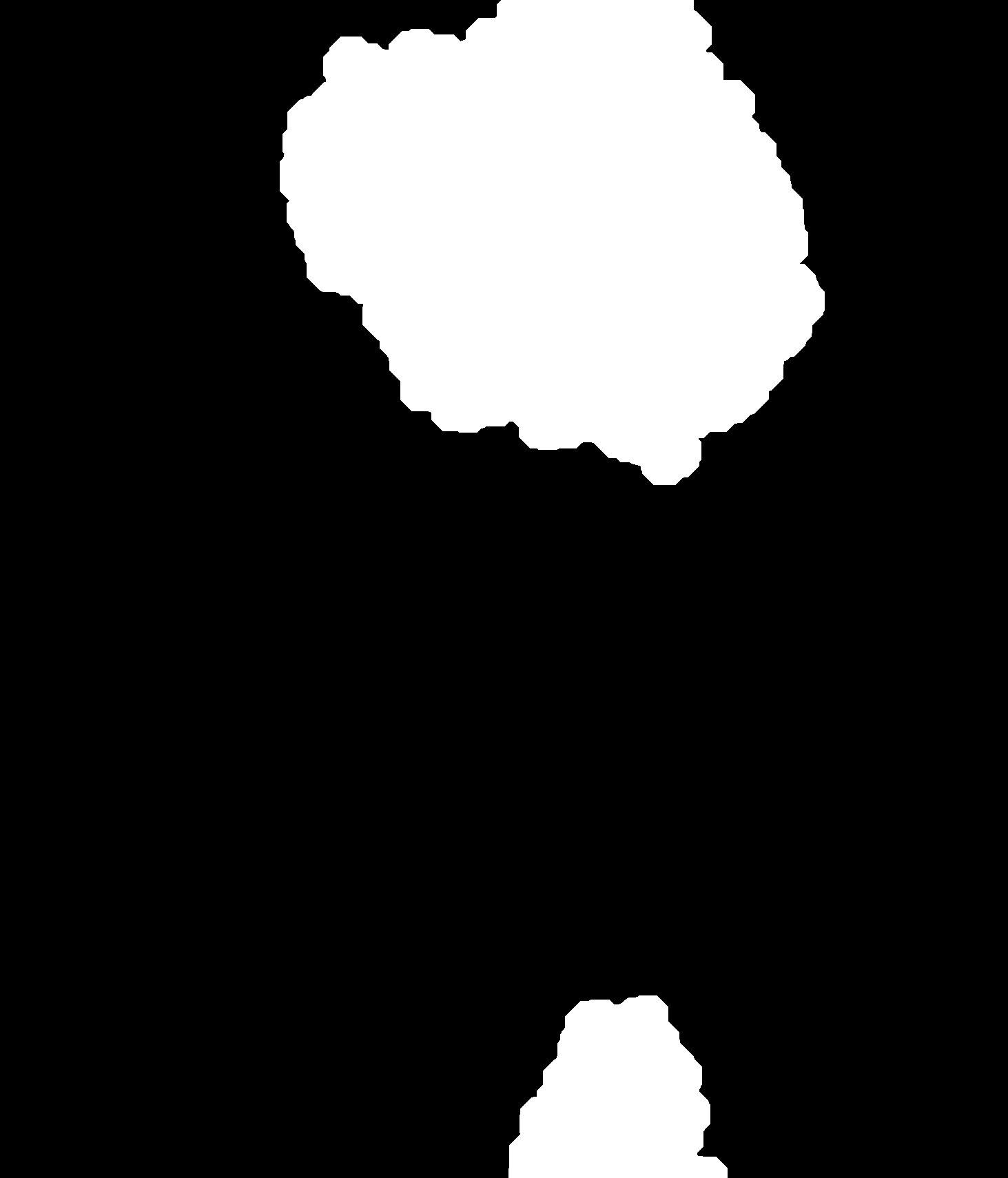} & 
\includegraphics[width=1.5cm]{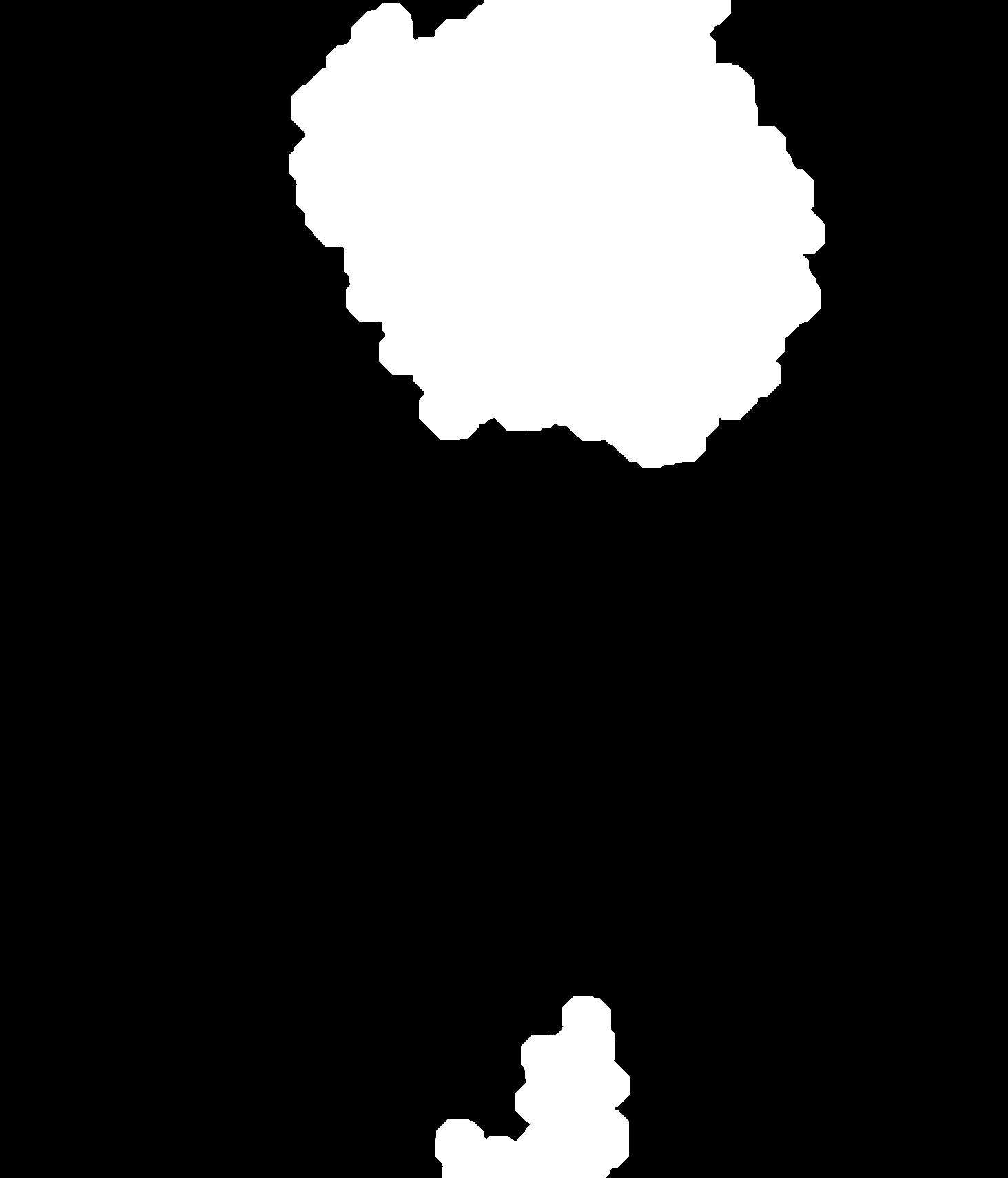}&
\includegraphics[width=1.5cm]{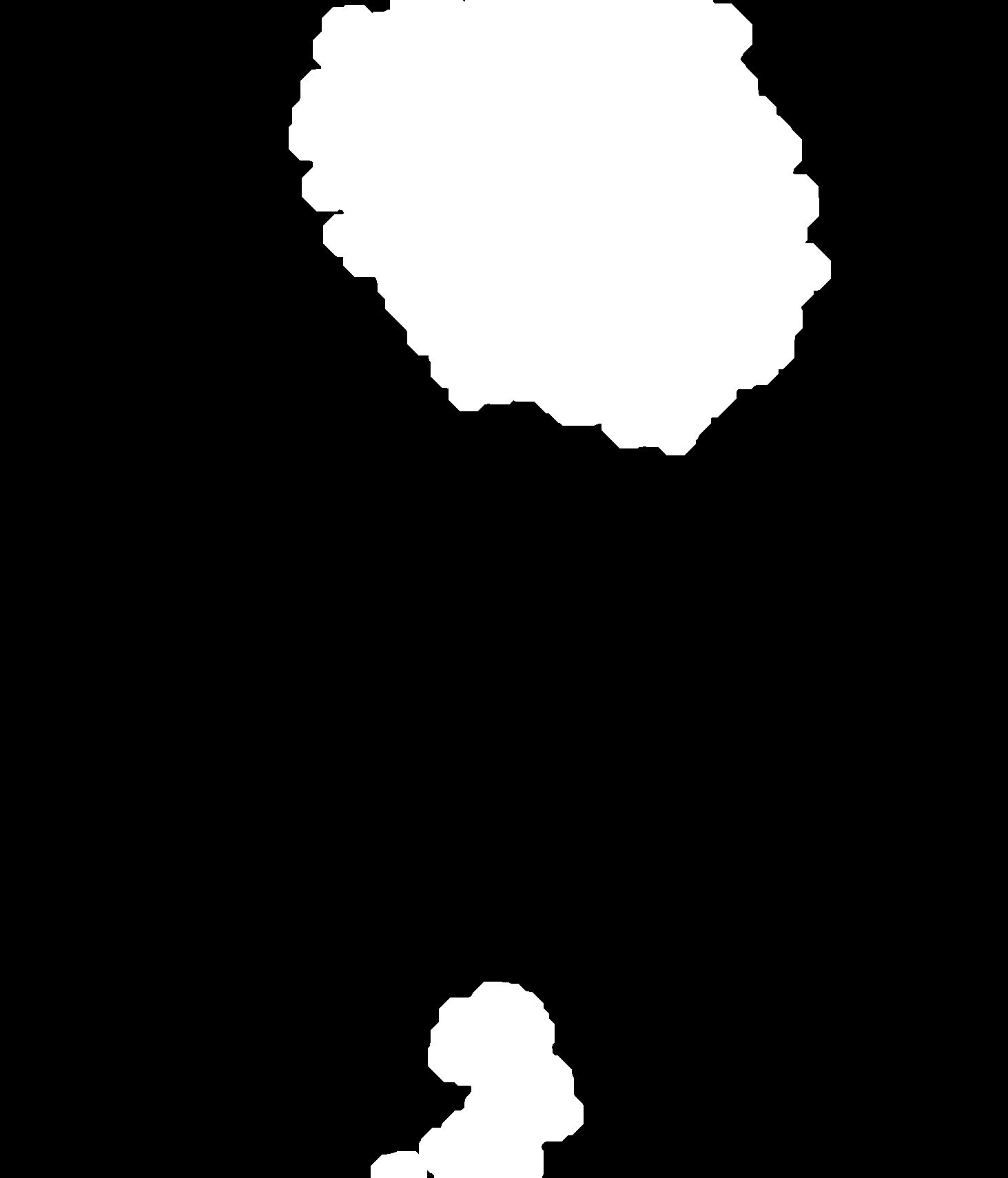} &\\ 
\includegraphics[width=1.5cm]{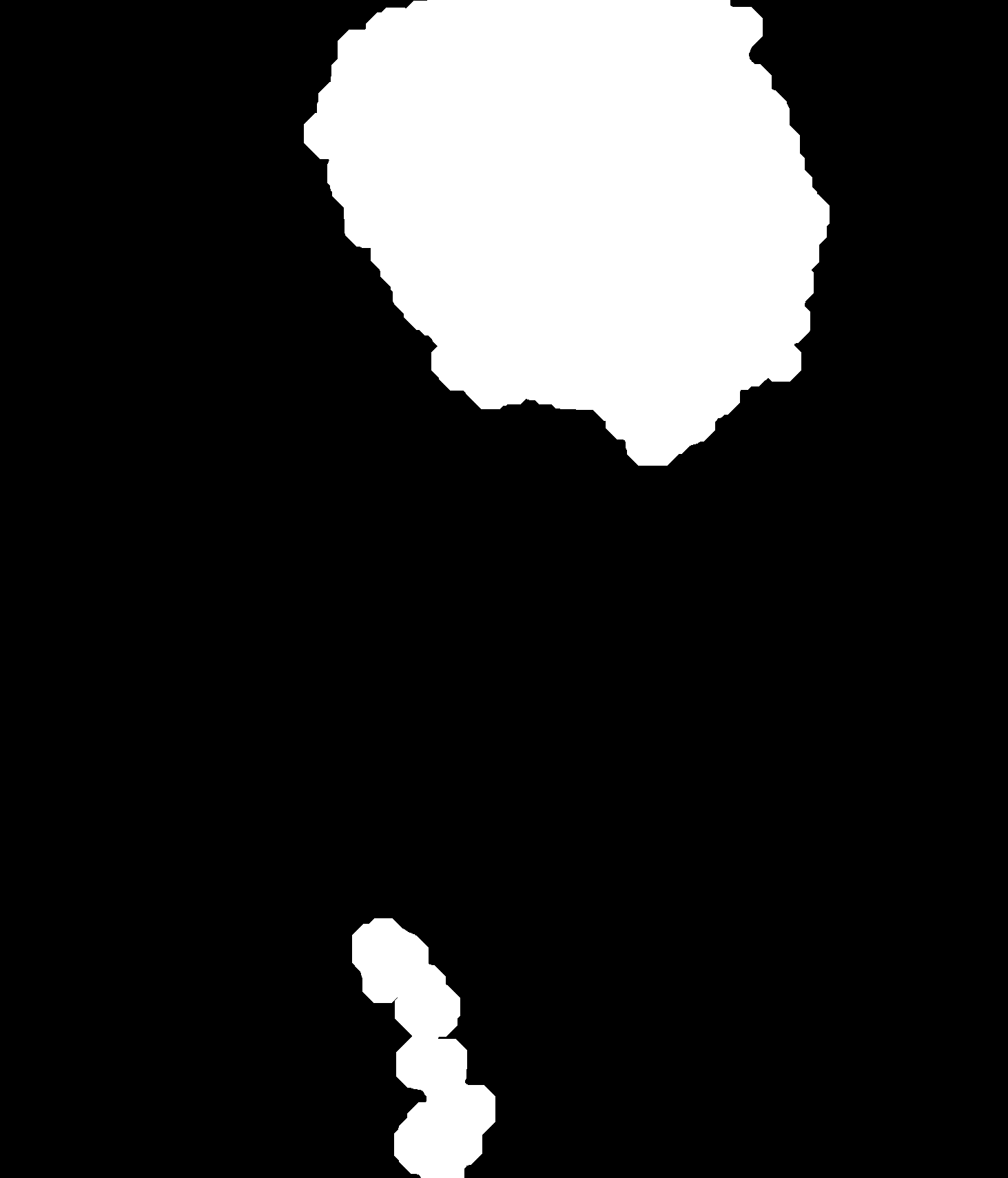} &
\includegraphics[width=1.5cm]{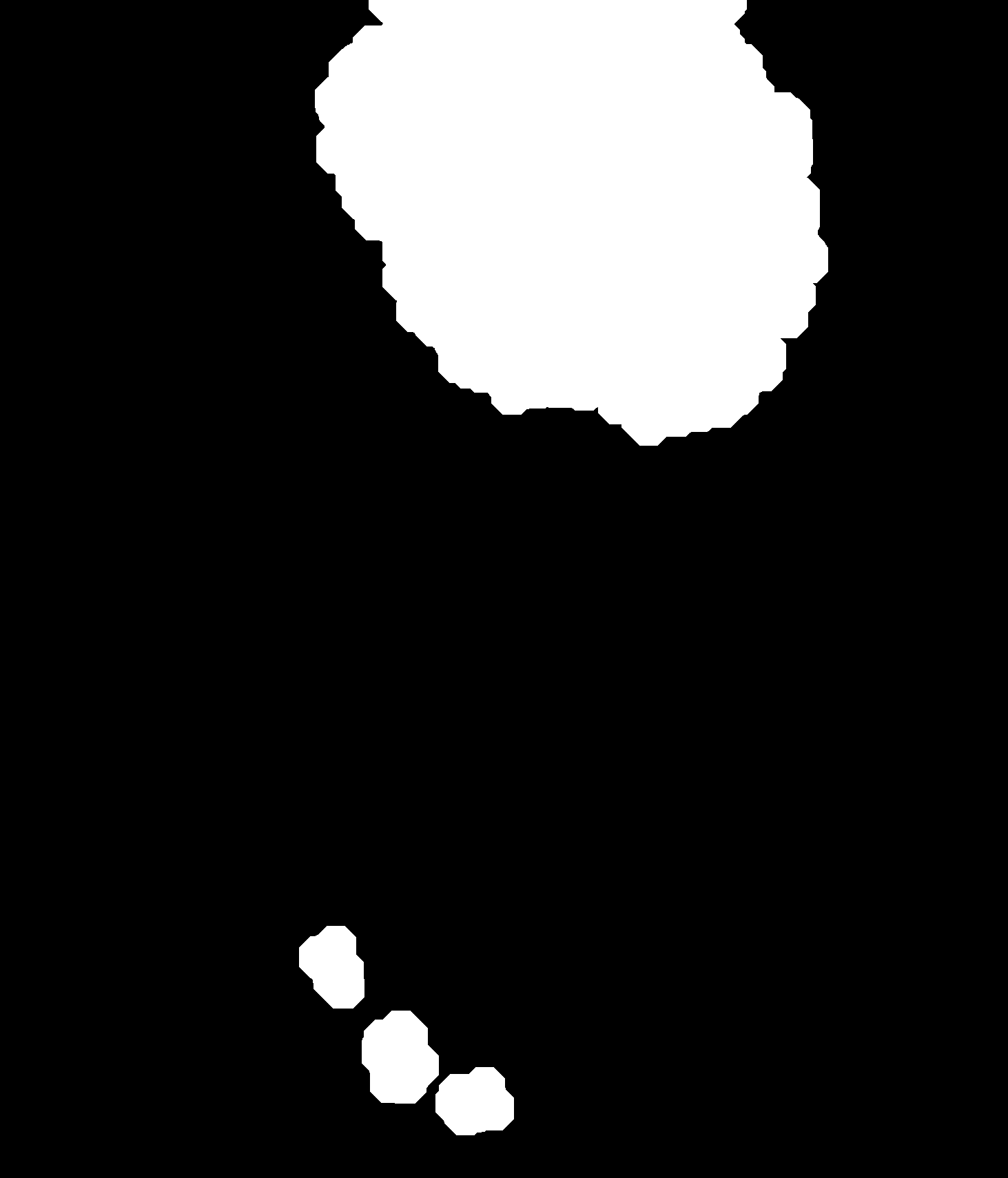} &
\includegraphics[width=1.5cm]{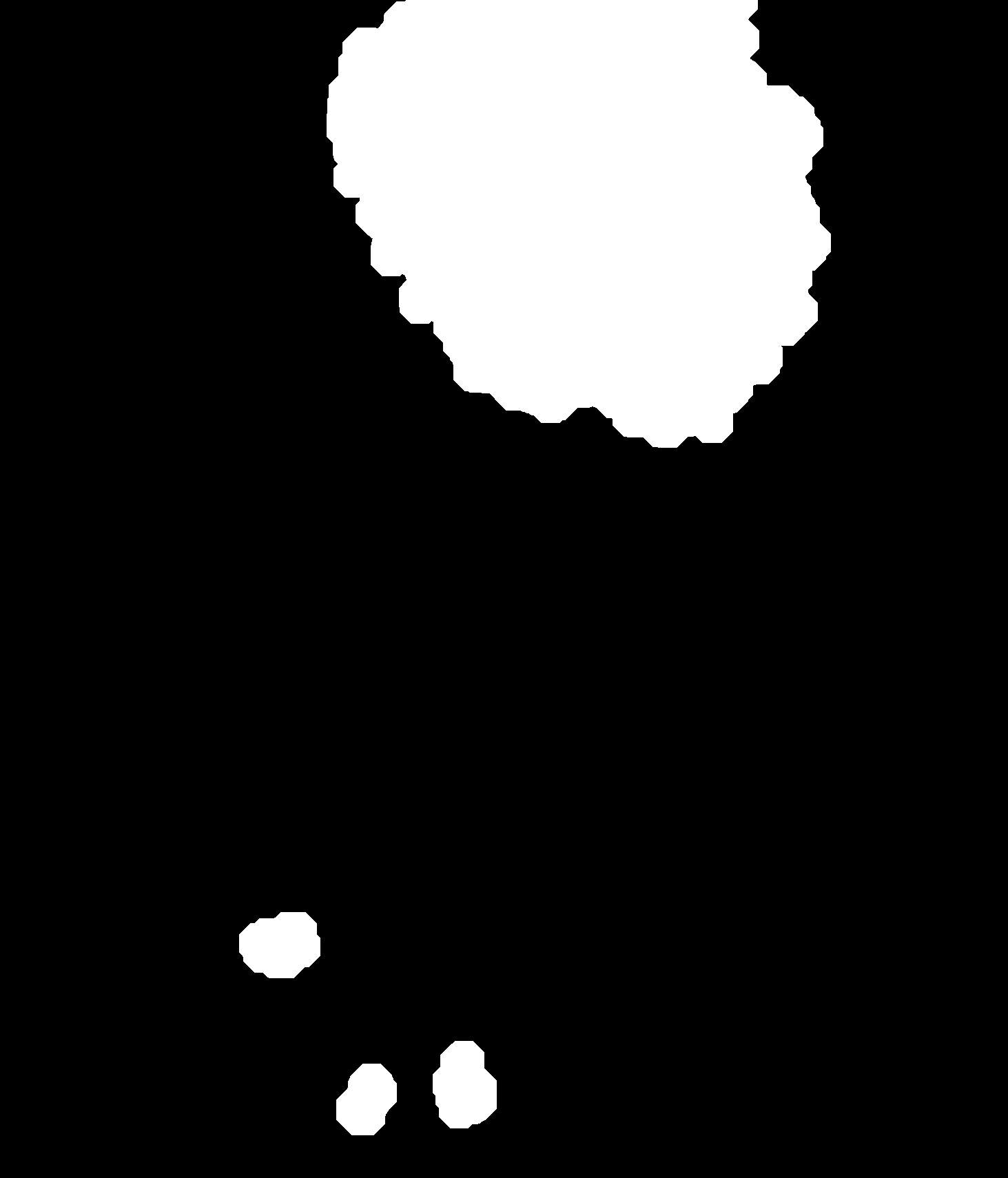} &
\includegraphics[width=1.5cm]{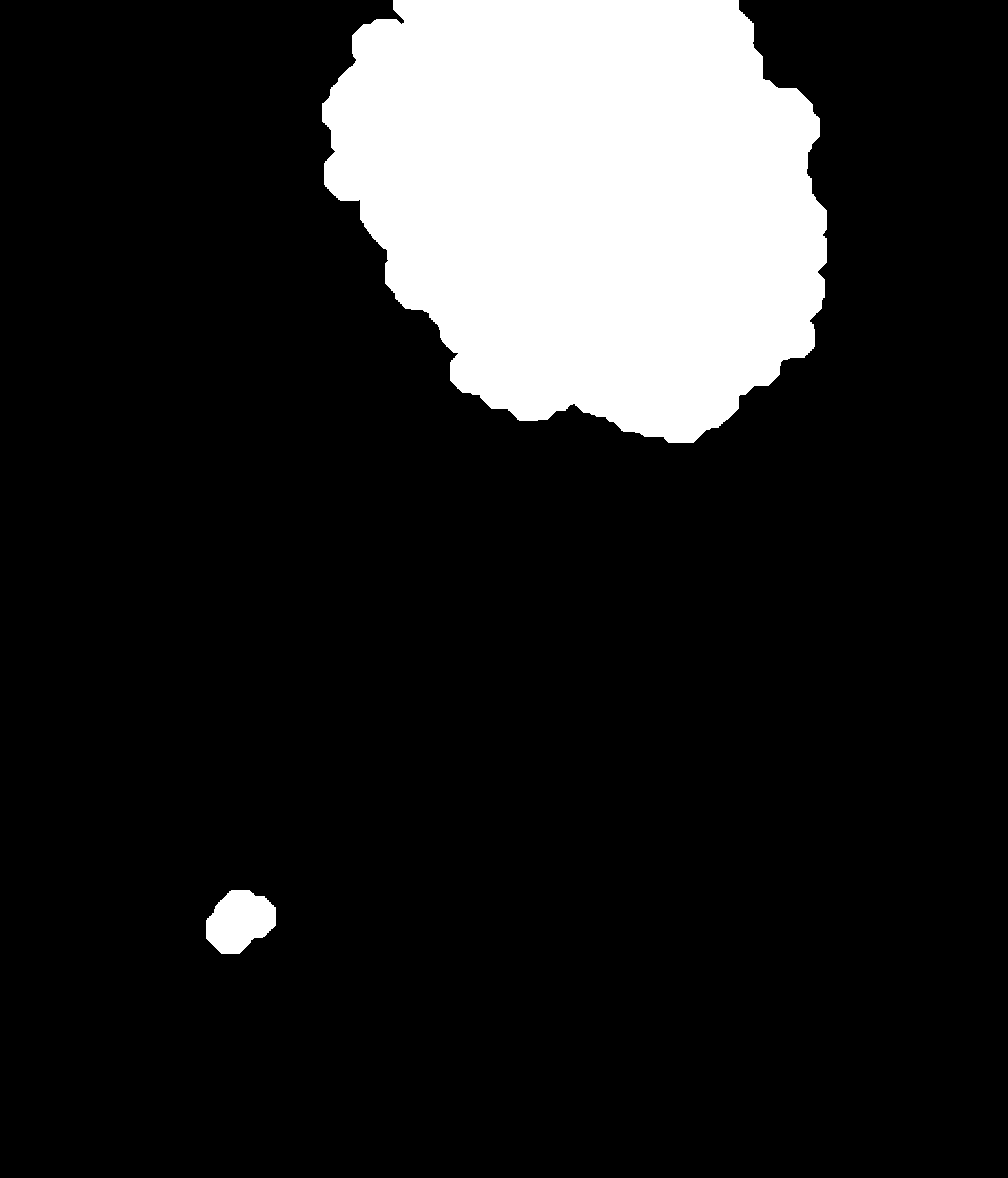} &
\includegraphics[width=1.5cm]{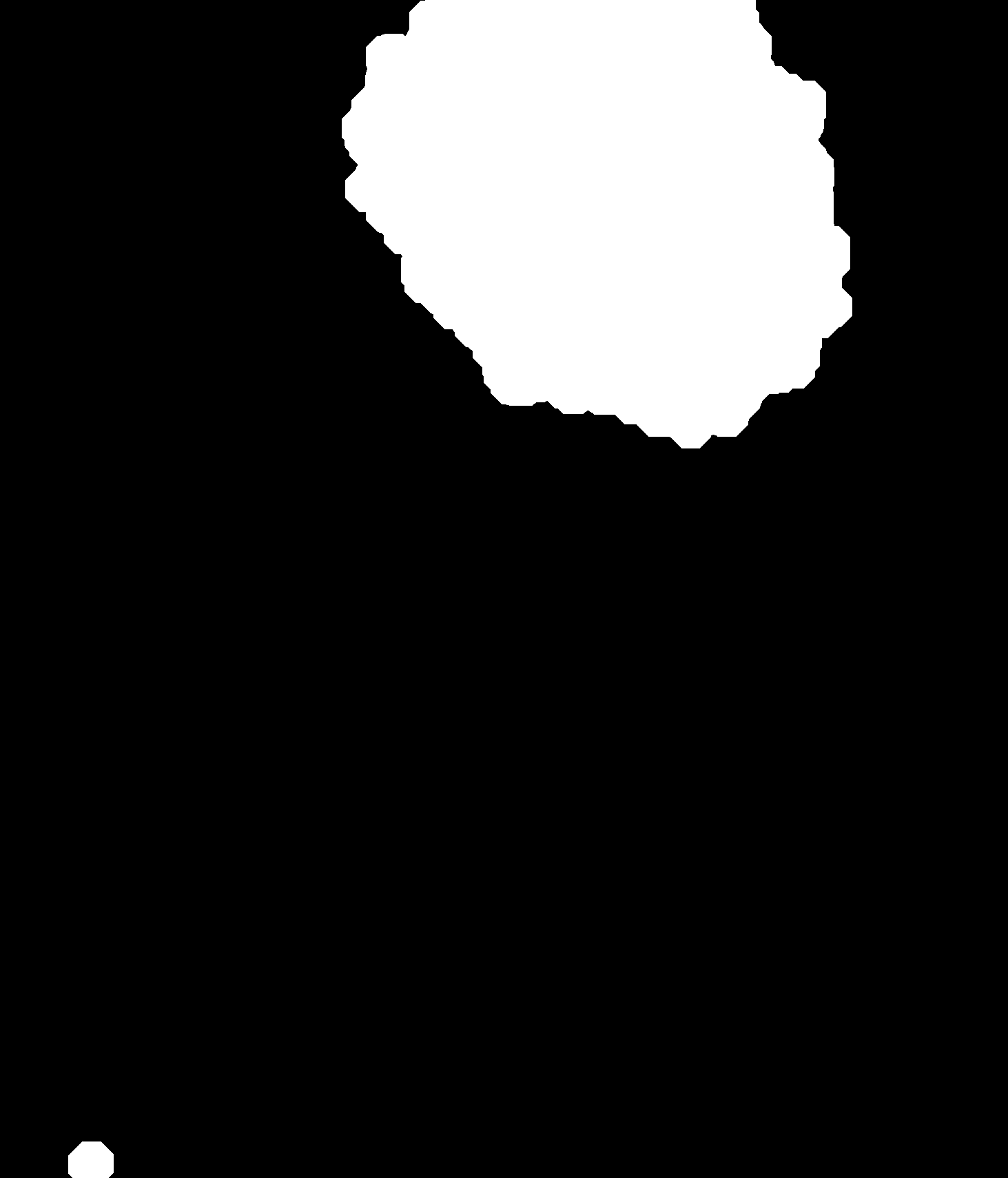} & 
\includegraphics[width=1.5cm]{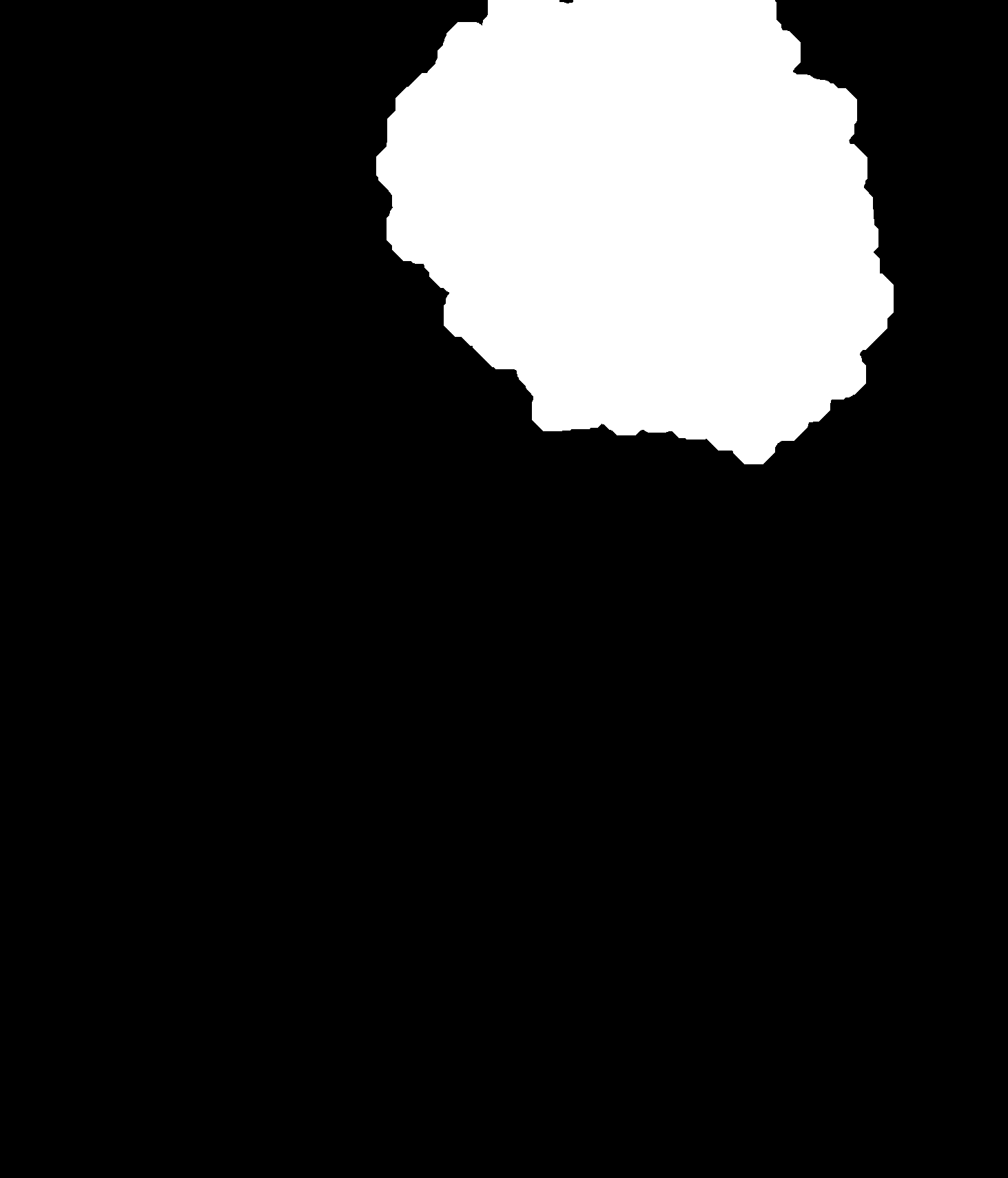} 
\end{tabular}
\caption{\label{fig:Op-Tim-Ser-30}Opening with {\tt {SE}}$=30$ to show the way of vanishing mid-sized structure in the process, (first row) frame $1$-$6$ and (second row) frame $7$-$12$ of {\tt Ser2}.}
\end{center}
\end{figure*}

\begin{figure*}[htp]
\setlength{\tabcolsep}{1mm}  
\renewcommand{\arraystretch}{1}  
\begin{tabular}{cccc}
\includegraphics[width=3.3cm]{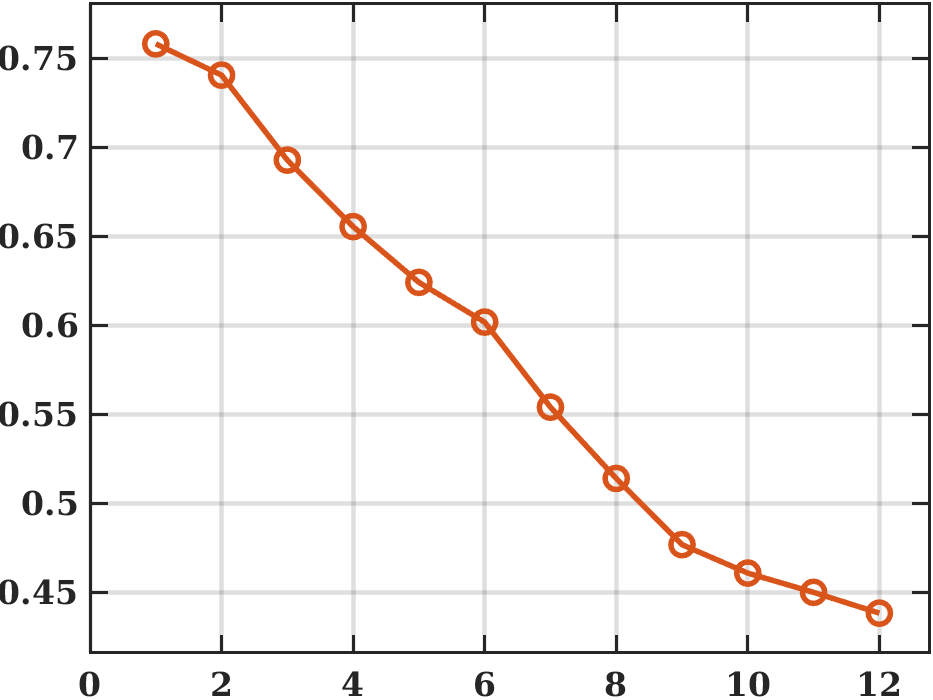} &
\includegraphics[width=3.3cm]{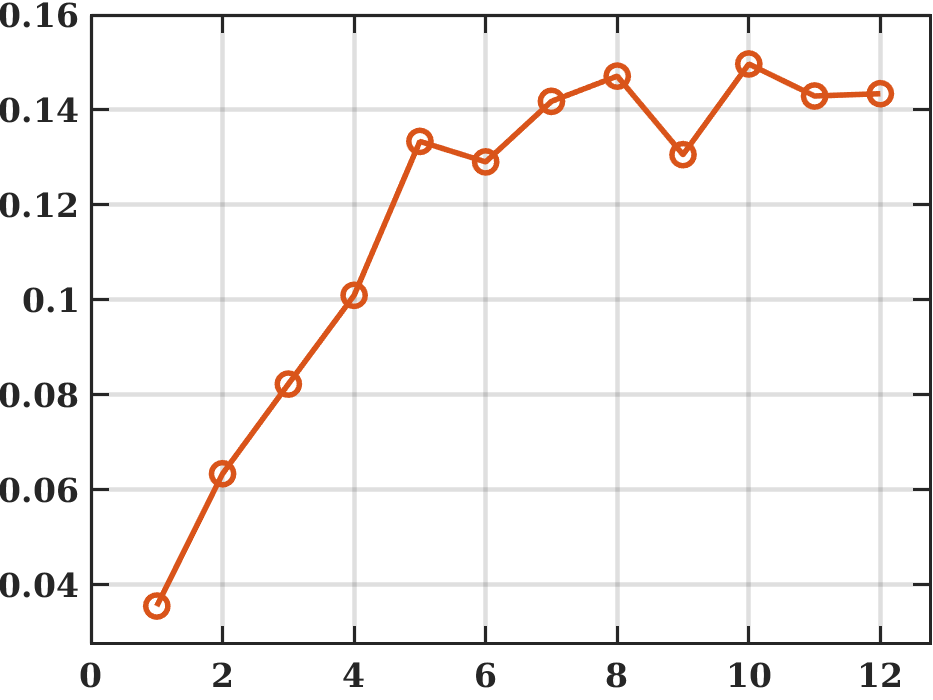} &
\includegraphics[width=3.3cm]{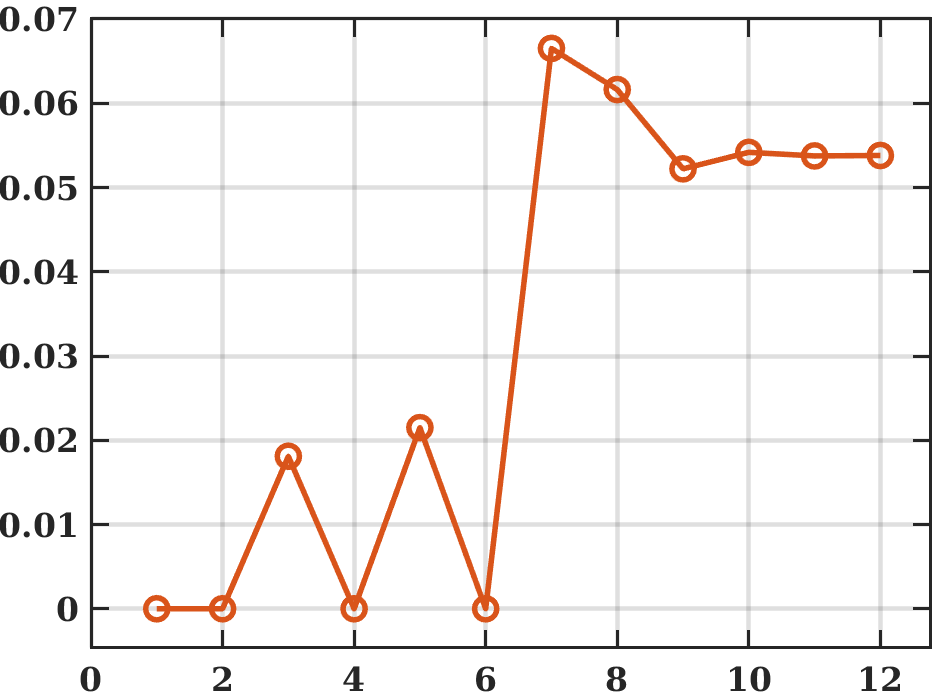} &
\\
\includegraphics[width=3.3cm]{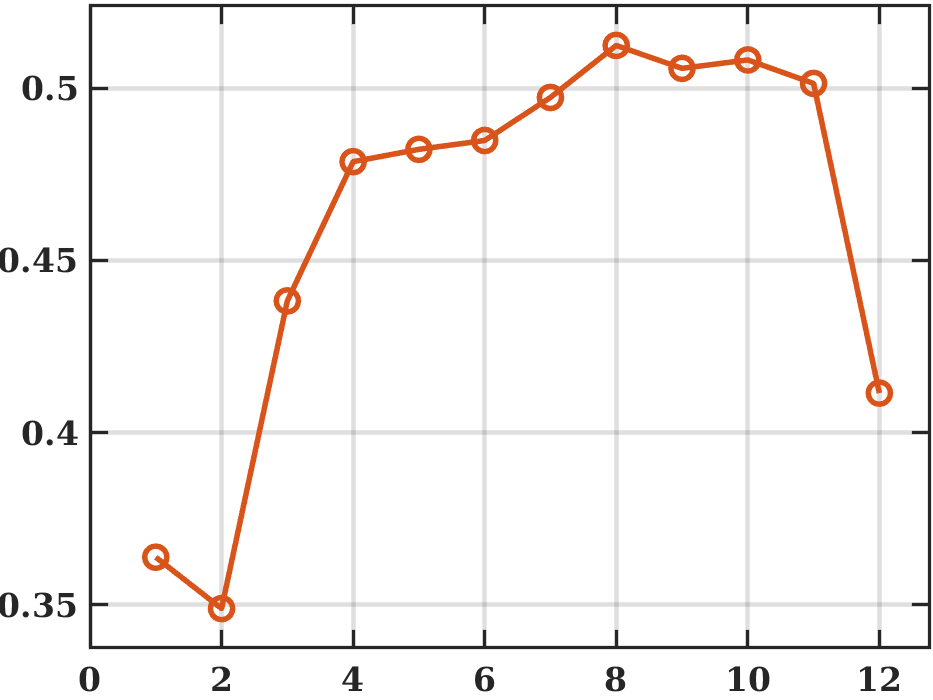} &
\includegraphics[width=3.3cm]{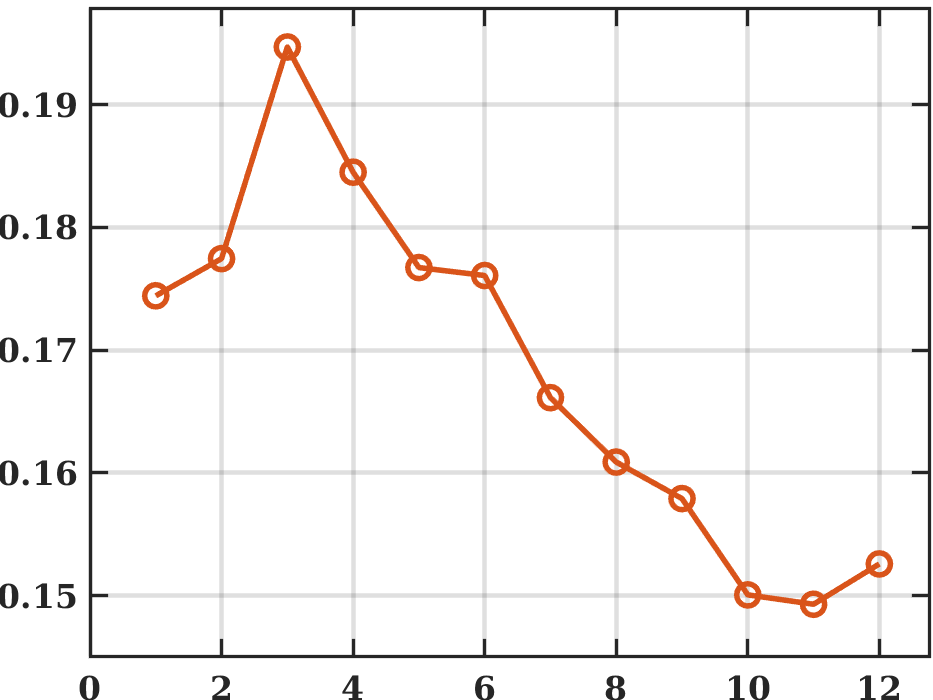} &
\includegraphics[width=3.3cm]{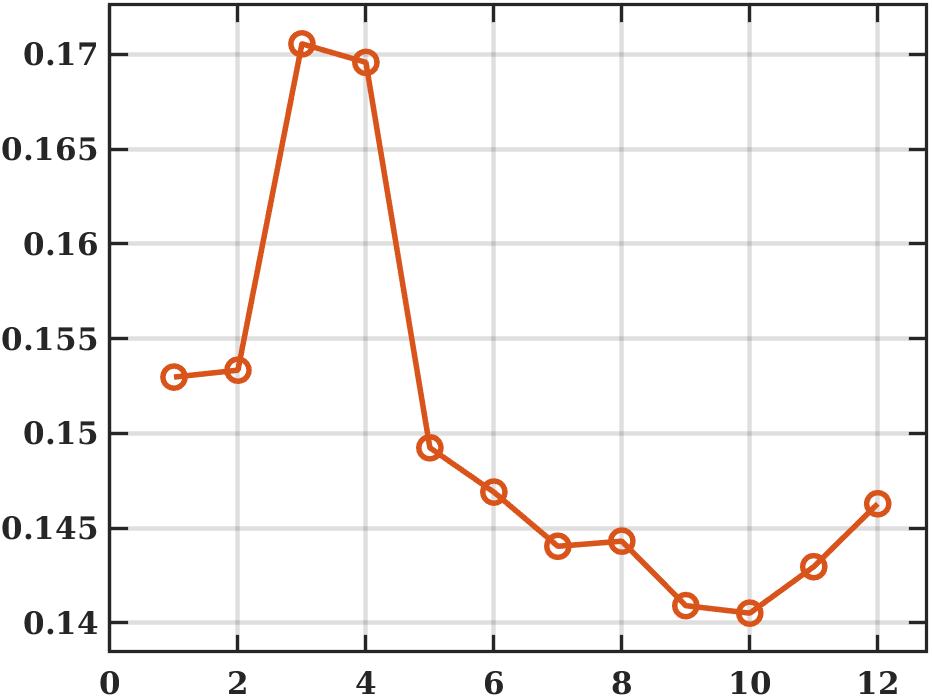}&
\end{tabular}
\caption{\label{fig:Time_Series} 
Time series intensity plots with different {\tt SE} sizes. First Row:  {\tt Ser1} with ${\tt {SE}}=0, 120, 180$. We clearly observe that the size of structures at larger scales grow over time. Second row:  {\tt Ser2} with ${\tt {SE}}=0, 30, 120$. We can clearly observe the fragmentation, the slight growth for large {\tt SE} values in last two frames is due to movement of the big agglomeration in the sequence. Let us note that {\tt SE}$=0$ denotes intensity directly.} 
\end{figure*}

In Figure \ref{fig:Op-Tim-Ser-30} we give a visual account of the time evolution in {\tt Ser2} using a medium-size structuring element {\tt SE}$=30$. Comparing with {\tt Ser2} as depicted in Figure \ref{fig:Exp1-Exp2}, one can clearly observe when the medium-size structure close to the lower image boundaries dissolves. Making use of a sample of {\tt SE} sizes and studying time evolution of normalized intensities, one can clearly distinguish agglomeration processes as in {\tt Ser1} and fragmentation taking place in certain scales, see Figure \ref{fig:Time_Series}.

\section{Summary and Conclusion}

To interpret the date of the image in the astrophysical context, it is crucial to identify the physical processes during the different phases of the experiment. In Section \ref{sec:exp2}, we show that granulometry is a powerful tool to distinguish different phases of the experiment from each other, and identify growth and erosion phases in the experiment. This is already an important tool for existing data, but it is absolutely crucial for long-term experiments, as they are currently planned for the International Space Station. 

In Section \ref{sec:exp1} we show that granulometry is a suitable tool to derive particle statistics from the image data. In case of erosion processes as shown in {\tt Ser2}, the intensity percentage follows a power law statistics. This follows the empirical observation in astronomical observations and laboratory experiments. Fragmentation processes typically result in a size distribution following a power law, although the exponent typically found in literature ($x^{-3.5}$) is larger than the one derived from the granulometry \cite{ref_bottke2015,ref_deckers2014}.

To analyze image sequences efficiently, it is important to follow the motion and evolution of larger-scale agglomerates. This means, that a reliable tool is required to identify and locate these agglomerates automatically. In Section \ref{sec:exp3} we show that granulometry is able to identify such agglomerates and simultaneously gives a first estimate for the agglomerate size.


\begin{credits}
\subsubsection{\ackname} 
This project is supported by DLR Space Administration with
funds provided by the Federal Ministry for Economic Affairs and Climate Action (BMWK) under grant numbers 50WK2270F, 50WK2270C, and 50WM2142. Access to the drop tower ande sounding rocket was provided by the European Space Agency under grant number ESA-HRE-RS-LE-0214.
\end{credits}

\end{document}